\documentclass[12pt]{article}
\usepackage{amssymb,amsmath}
\usepackage{epsfig,psfrag}

\catcode`\@=11
\def \Im {\mathop{\rm Im}\nolimits}

\def \e  {\mathop{\rm e}\nolimits}
\newcommand\lr[1]{{\left({#1}\right)}}

\newcommand \ket [1] {|{#1}\rangle}
\newcommand \bra [1] {\langle {#1}|}
\newcommand\re[1]{(\ref{#1})}
\def \qqquad {\qquad\quad}
\def \qqqquad {\qquad\qquad}

\newcommand{\pa}{\partial}
\newcommand{\bet}{\bar\eta}

\newcommand{\be}{\begin{equation}}
\newcommand{\ee}{\end{equation}}
\newcommand{\bea}{\begin{eqnarray}}
\newcommand{\eaa}{\end{eqnarray}}
\newcommand{\nn}{\nonumber}

\renewcommand{\a}{\alpha}

\newcommand{\da}{{\dot\alpha}}
\newcommand{\db}{{\dot\beta}}
\newcommand{\bl}{{\tilde\lambda}}

\renewcommand{\b}{\beta}

\renewcommand{\tt}{\tilde\ell}
\newcommand{\la}{\lambda}
\renewcommand{\t}{\ell}

\newcommand{\q}{\theta}

\newcommand{\ep}{\epsilon}

\newcommand{\cN}{{\cal N}}
\newcommand{\cA}{{\cal A}}

\newcommand{\p}[1]{(\ref{#1})}
\newcommand{\bt}[1]{{\bar t}}

\newcommand \vev [1] {\langle{#1}\rangle}
\newcommand \ran [1] {|{#1}\rangle}
\newcommand \lan [1] {\langle{#1}|}

\def\numberbysection{\@addtoreset{equation}{section}
                     \def\theequation{\thesection.\arabic{equation}}}
\numberbysection

\begin{document}
\thispagestyle{empty}
\null\vskip-12pt \hfill IPhT-T09/076~~~ \\
\null\vskip-12pt \hfill  LAPTH-1331/09 \\
\null\vskip-12pt \hfill NSF-KITP-09-95
\vskip2.2truecm
\begin{center}
\vskip 0.2truecm {\Large\bf
 Symmetries and analytic properties of\\[3mm]  scattering amplitudes in   $\cN=4$ SYM theory}

\vskip 1truecm
{\bf  G.P. Korchemsky$^{*}$\footnote{On leave from Laboratoire de Physique Th\'eorique, Universit\'e de Paris XI, 91405 Orsay C\'edex, France}
 and E. Sokatchev$^{**}$ \\
}

\vskip 0.4truecm
$^{*}$ {\it Institut de Physique Th\'eorique%
\footnote{Unit\'e de Recherche Associ\'ee au CNRS URA 2306},
CEA Saclay, \\
91191 Gif-sur-Yvette C\'edex, France\\
\vskip .2truecm $^{**}$ {\it LAPTH\footnote[3]{Laboratoire d'Annecy-le-Vieux de Physique Th\'{e}orique, UMR 5108},   Universit\'{e} de Savoie, CNRS, \\
B.P. 110,  F-74941 Annecy-le-Vieux, France
                       }
  } \\
\end{center}

\vskip 1truecm 
\centerline{\bf Abstract} 

\medskip

 \noindent
In addition to the superconformal symmetry of the underlying Lagrangian, the scattering amplitudes in planar $\mathcal{N}=4$ super-Yang-Mills theory exhibit a new, dual superconformal symmetry.
We address the question of how powerful these symmetries are to completely determine the scattering amplitudes.

We use the example  of the NMHV superamplitudes to show that the combined action of conventional and dual superconformal symmetries is not sufficient to fix all the freedom in the tree-level amplitudes. We argue that the additional information needed comes from the study of the analytic properties of the amplitudes. The requirement of absence of  spurious singularities, together with the correct multi-particle singular behavior, determines the unique linear combination of superinvariants corresponding to the $n-$particle NMHV superamplitude. The same result can be obtained recursively, by relating the $n-$ and $(n-1)-$particle amplitudes in the  singular collinear limit. We also formulate constraints on the loop corrections to the superamplitudes, following from the analytic behavior in the above limits.

We then show that,
at one-loop level,  the holomorphic anomaly of the tree amplitudes leads to the breakdown of dual Poincar\'e supersymmetry (equivalent to ordinary special conformal supersymmetry) of the ratio of the NMHV and MHV superamplitudes, but this anomaly does not affect dual conformal symmetry.

\newpage

\thispagestyle{empty}

{\small \tableofcontents}

\newpage
\setcounter{page}{1}\setcounter{footnote}{0}

\section{Introduction}

Recently it has been  realized that the scattering amplitudes in planar
$\mathcal{N}=4$ super-Yang-Mills theory exhibit a new remarkable symmetry, called dual superconformal symmetry. It appears in addition to the known classical symmetries of this gauge theory and it is believed to be related to the hidden integrability of the $\mathcal{N}=4$ theory. The dual superconformal symmetry was first identified at weak coupling by examining
the properties of the tree and one-loop MHV and NMHV scattering amplitudes \cite{Drummond:2008vq}. It was subsequently shown that at
strong coupling this new symmetry is related, through the AdS/CFT correspondence, to the invariance of the string sigma model in $\rm AdS_5\times S^5$ under bosonic and fermionic T-duality \cite{Berkovits:2008ic,Beisert:2008iq}.

In principle, the combination of this new symmetry with the ordinary $\mathcal{N}=4$ superconformal symmetry of the underlying Lagrangian may have far reaching consequences for the scattering amplitudes and might eventually lead to their exact determination for arbitrary values of the 't Hooft coupling  in planar  $\mathcal{N}=4$ SYM theory. However, a lot of work still needs to be done before we fully understand all the implications of these symmetries. In the present paper we try to answer some of the open questions. We first consider the tree-level MHV and NMHV superamplitudes, and show that the combined action of the two superconformal symmetries is not sufficient to completely fix all the freedom in the amplitudes. We demonstrate that the additional information needed comes from the study of the physical and spurious singularities, or, alternatively, of the collinear singularities of the amplitudes. We then show that  the holomorphic anomaly of the tree amplitudes leads to the breakdown of dual supersymmetry (which is equivalent to ordinary special conformal supersymmetry) at one-loop level, but this anomaly does not affect dual conformal symmetry.

\subsection{Tree-level superamplitudes}

A convenient framework for discussing the symmetries of scattering amplitudes in
$\mathcal{N}=4$ SYM theory is on-shell superspace \cite{Nair:1988bq,Witten:2003nn,Drummond:2008vq,ArkaniHamed:2008gz}, closely related to the light-cone superspace formalism \cite{Mandelstam:1982cb,Brink:1982pd}. In this approach, all asymptotic states in $\mathcal{N}=4$ SYM theory,  gluons ($G^\pm$), gluinos ($\Gamma_A, \bar \Gamma^A$) and scalars ($S_{AB}$), are combined into a single on-shell superstate,
\begin{align} \notag
\Phi(p,\eta) = G^+(p) +& \eta^A \Gamma_A(p) +\frac12 \eta^A\eta^B S_{AB}(p)
\\ \label{superstate}
& +\frac1{3!} \eta^A\eta^B\eta^C \epsilon_{ABCD} \bar\Gamma^D(p)
+\frac1{4!} \eta^A\eta^B\eta^C\eta^D \epsilon_{ABCD} G^-(p)\,,
\end{align}
with the help of Grassmann variables $\eta^A$ carrying helicity $1/2$ and an $SU(4)$ index $A=1\ldots4$.  The coefficients in the expansion \p{superstate} describe the on-shell states of particles with a light-like momentum $p_\mu$ (with $p^2=0$) and helicities ranging from $+1$ ($G^+$) to $-1$ ($G^-$). Making use of such superstates, we can combine all $n-$particle color-ordered scattering amplitudes in the $\mathcal{N}=4$ SYM theory into a single on-shell object, the superamplitude $\mathcal{A}_n(p_1,\eta_1; \ldots; p_n,\eta_n)$. It depends on
the supermomenta $(p_i,\eta_i)$ of the particles and its expansion in  powers of  $\eta$'s generates the scattering amplitudes for the various types of particles from \re{superstate}.

Another useful tool for discussing the symmetries of the superamplitudes is the spinor-helicity formalism \cite{Xu:1986xb}. In it one resolves the on-shell
conditions on the particle momenta $p_i^2=0$ by introducing a pair of commuting spinors for each particle,
\begin{align}\label{p=ll}
p_i^{\a\da} = p^\mu (\sigma_\mu)^{\a\da} = \la_i^\a \bl_i^\da \,,
\end{align}
or equivalently $p_i =  \ket{i} [i|$. In Minkowski space-time the two spinors are complex conjugate, $\bl = (\la)^*$, and are defined up to a phase factor (helicity).

The $SU(4)$ invariance of $\mathcal{A}_n(p_1,\eta_1; \ldots; p_n,\eta_n)$ implies that it is expanded in powers of $(\eta)^{4k}$ with $k=0,1,\ldots,n$. On-shell Poincar\'e supersymmetry requires the absence of the terms with $k=0,1$. So,  the first term in the expansion of the superamplitude is of degree 8. It  generates all $n-$particle MHV amplitudes and we shall denote it by $\mathcal{A}_n^{\rm MHV}$. The next term of the expansion, $\mathcal{A}_n^{\rm NMHV}$, has degree $12$ in $\eta$ and it generates the NMHV amplitudes, etc.
The simplest, tree-level $n-$particle MHV superamplitude takes the form \cite{Nair:1988bq}
 \begin{equation}\label{MHVsua}
    {\cA}_n^{\rm MHV;0} = i\frac{ \delta^{(4)}(\sum_{i=1}^n \la_i^\a\bl_i^{\da})\ \delta^{(8)}(\sum_{i=1}^n \la_i^\a \eta_{i}^A)}{\vev{12}\vev{23}\ldots\vev{n1}}\,,
\end{equation}
where the superscript `$0$' in the left-hand side denotes the tree approximation and we use the standard notation for the Lorentz invariant spinor contractions $\vev{ij} = \la_i^\a\la_{j,\a} = \la^\a_{i}\epsilon_{\a\b}\la^\b_{j}$.

At tree level, the superamplitudes  have neither infrared nor ultraviolet divergences, so they inherit all the classical symmetries of the $\mathcal{N}=4$ Lagrangian. Indeed, it can be shown directly that ${\cA}_n^{\rm MHV;0}$ is invariant under $\cN=4$ superconformal transformations, realized {\it non-locally} on the particle momenta \cite{Witten:2003nn}.

In addition to the ordinary superconformal symmetry, ${\cA}_n^{\rm MHV;0}$ has another, {\it dual $\cN=4$ superconformal symmetry}. To make it manifest, one introduces new dual variables \cite{Broadhurst,Drummond:2006rz,Drummond:2008vq}
 related to the supermomenta
$(p_i,\eta_i)$ as follows
\begin{align}
& p^{\a\da}_i = \la_i^\a \bl_i^\da = x^{\a\da}_i - x^{\a\da}_{i+1}\,,
\qqqquad 
\la_i^\a \eta_i^A = \theta_i^{\a A}-\theta_{i+1}^{\a A}\,, 
\label{super-mom}
\end{align}
with the periodicity conditions $x_{n+1} \equiv x_1$  and $\theta_{n+1}  \equiv \theta_{1}$.
The dual superconformal symmetry acts {\it locally} on the dual variables $x_i$ and $\theta_i$ (and, as a consequence, on the spinor variables $\la,\bl$) as if they were coordinates in some dual superspace. Most remarkably, the superamplitude \p{MHVsua}, rewritten in terms of the dual coordinates, transforms covariantly under dual superconformal symmetry with dual conformal weight $(+1)$ at each point, equal to the helicity
of the superstate \re{superstate}. We wish to stress that this symmetry is by no means obvious. Its dynamical origin in perturbation theory is still not completely clear. At strong coupling, it can be interpreted via the AdS/CFT correspondence as a T-duality symmetry of the string sigma model \cite{Berkovits:2008ic,Beisert:2008iq}.

The $n$-particle  NMHV tree superamplitude $\mathcal{A}_n^{\rm NMHV;0}$ has a considerably more complicated form. It was first found in \cite{Drummond:2008vq} by comparing its gluon components to the known NMHV gluon amplitudes from \cite{Bern:2004ky,Bern:2004bt}, and later on rederived by the supersymmetric versions of the generalized unitarity method \cite{Britto:2004nc,Drummond:2008bq} and of the BCFW recursion relations \cite{Britto:2005fq,Drummond:2008cr}.
It can be represented in a factorized form involving the MHV tree superamplitude
\re{MHVsua} as a prefactor,
\begin{align}\label{NMHV-tree0}
\mathcal{A}_n^{\rm NMHV;0} & = \mathcal{A}_n^{\rm MHV;0} {  \sum_{4\le s+1< t\le n }   R_{1st}}\,.
\end{align}
{Here the MHV prefactor has Grassmann degree 8 and it is covariant under dual superconformal symmetry. The new feature of Eq.~\p{NMHV-tree0} is the appearance of a sum of {\it dual superconformal invariants} (or `superinvariants' in short) $R_{1st}$ of Grassmann degree 4 defined in Eq.~\re{R} below.

\subsection{Fixing the freedom in the NMHV tree superamplitudes}

A remarkable property of \re{NMHV-tree0} is that all the
superinvariants enter the sum with equal (unit) coefficients. What properties of the superamplitude are responsible for this? The natural question arises: Is the combined action of ordinary and dual superconformal symmetries powerful enough to completely determine the form of the tree superamplitudes? \footnote{Recently it has been shown that the closure of the two superconformal symmetries is infinite dimensional and has a Yangian structure \cite{Drummond:2009fd}. This has renewed \cite{Beisert:2009cs} the hope  that the scattering amplitudes in $\cN=4$ planar SYM may be integrable in some sense \cite{Berkovits:2008ic,Beisert:2008iq,Gorsky:2009ew}. }

In the present paper we show that this is not the case since each term in the sum \re{NMHV-tree0} is separately invariant under {\it both} symmetries and, therefore, any linear combination of $R$'s is invariant. Thus, the complete determination of the tree-level superamplitudes does not rely on symmetries alone, further physical input is needed.

In this paper we argue that the values
of the coefficients in \re{NMHV-tree0} are unambiguously fixed by the requirement that the scattering amplitude $\mathcal{A}_n^{\rm NMHV;0}$  should have correct analytic properties as a function of the Mandelstam invariants. More precisely, the NMHV amplitude should be free from the so-called spurious singularities \cite{Bern:2004ky,Bern:2004bt}.%
\footnote{The role of the spurious singularities has recently been discussed in \cite{Hodges:2009hk}, in the context of a renewed interest \cite{Mason:2009sa,ArkaniHamed:2009si} in Witten's twistor transform \cite{Witten:2003nn} of scattering amplitudes.} These are poles which do not correspond to the vanishing of {multi-particle} invariant masses. They are present however in each superinvariant $R_{1st}$.  We show that demanding their cancellation fixes all the relative coefficients  in \p{NMHV-tree0}. Further, the amplitude should have the necessary poles {in multi-particle
invariant masses}
leading to its factorization into MHV amplitudes. This requirement fixes the overall factor in \p{NMHV-tree0}.

Alternatively, we can make use of the so-called collinear (or two-particle) singularities of the amplitude. They relate {amplitudes with different numbers of particles}, $\cA_n$ to $\cA_{n-1}$, and, hence, allow us to derive linear equations for the coefficients in the two amplitudes. We show that these equations fix all the coefficients in \re{NMHV-tree0} recursively, by relating $\mathcal{A}_n^{\rm NMHV;0}$ (for $n=6,7,8$) to the simplest NMHV-like `googly' five-particle amplitude $\mathcal{A}_5^{\rm \overline{MHV};0}$.

Comparing the two methods of fixing the freedom in the NMHV tree superamplitudes, we may say that the spurious singularities operate `horizontally', i.e. they establishes a property of the $n-$particle NMHV amplitude, without reference to other amplitudes (putting aside the use of the multi-particle factorization, needed to fix the overall normalization). In this sense, $\mathcal{A}_n^{\rm NMHV;0}$ is the unique $n-$point invariant of the two superconformal symmetries, free from spurious singularities. At the same time, the method based on collinear singularities operates `vertically', i.e. it recursively relates the $n-$particle amplitude to the amplitudes with fewer particles. Since in both cases one arrives at the same, unique expression for the NMHV tree amplitude, built out of ordinary+dual superconformal invariants, the two approaches are equivalent
at tree level. However, at loop level this may not be the case anymore (see Section \ref{spcvcf}).

A comment is due here about the BCFW recursion procedure \cite{Britto:2004ap,Britto:2005fq}. It provides a powerful mechanism for constructing tree-level scattering amplitudes from elementary blocks, the MHV amplitudes (in fact, the starting point of the recursion is the simplest, 3-particle MHV amplitude). It consists in deforming two of the particle momenta by a complex parameter $z$ and then exploiting the resulting poles in $z$ to factorize the amplitude into simpler blocks. An important ingredient in the construction is the assumption about the regular behavior at $z\to\infty$. Although this approach cannot directly explain the origin of dual conformal symmetry, it can reduce it to the properties of the elementary building blocks, the MHV amplitudes, as shown in \cite{Brandhuber:2008pf}. Recently, the complete tree-level superamplitude has been found in
\cite{Drummond:2008cr}
through the supersymmetric generalization of the BCFW recursion~\cite{Bianchi:2008pu,Brandhuber:2008pf,ArkaniHamed:2008gz,Elvang:2008na}. The reason why we wish to reexamine the tree superamplitudes is not to find an alternative construction (the BCFW one is very efficient), {but to understand to  what extent the amplitudes are determined by their symmetries (ordinary and dual superconformal).}

\subsection{Anomalous symmetries at loop level}

Let us now turn to the situation at loop level, where the massless particle scattering amplitudes are infrared divergent and
require regularization. This inevitably breaks part of the classical symmetries and renders problematic the way they constrain the all-order amplitudes. In particular, to make use of the ordinary and dual superconformal symmetries we should be able to control their breakdown at loop
level.

Important progress in this direction has been made after the discovery of the MHV amplitude/Wilson loop duality. It stipulates the equivalence of the loop corrections to the MHV amplitude ${\cA}_n^{\rm MHV}$, on the one hand,  with a Wilson loop in {\it dual space-time} evaluated along a  light-like polygon contour with cusps at the points $x_i$ from \p{super-mom}, on the other hand. In the $\mathcal{N}=4$ SYM theory,
this relation was first observed at strong coupling 
\footnote{The duality between scattering amplitudes and light-like Wilson loops was first noticed in QCD in the high-energy (Regge) limit \cite{Korchemskaya:1996je}.}
 using the
string description of the scattering amplitudes in AdS/CFT \cite{Alday:2007hr} and it was later confirmed
at weak coupling by matching the perturbative corrections to both quantities
\cite{Drummond:2007aua,Brandhuber:2007yx,Drummond:2007cf,Drummond:2007au,Drummond:2008aq,Bern:2008ap,Anastasiou:2009kn,Gorsky:2009nv}. The scattering amplitude/Wilson loop duality implies that the dual conformal symmetry of the MHV superamplitude is equivalent to the conformal symmetry of the light-like Wilson loops in $\mathcal{N}=4$ SYM. The latter is broken {\it locally} by the ultraviolet cusp singularities~\cite{Korchemskaya:1992je}. This allows us to determine the dual conformal anomaly of the all-loop MHV superamplitudes from an anomalous conformal Ward identity for the light-like Wilson loop \cite{Drummond:2007au}. This Ward identity is a powerful constraint on the form of the MHV amplitude. In particular, it fixes the finite part of the $n=4$ and $n=5$ amplitudes in accord with the BDS conjecture \cite{Anastasiou:2003kj,Bern:2005iz}, while for $n\ge 6$ it reduces the problem to finding a single function of the conformally invariant cross-ratios of the dual coordinates~\cite{Alday:2007he,Drummond:2007bm,Drummond:2008aq,Bern:2008ap,Alday:2009yn}.

For non-MHV  superamplitudes, no duality with Wilson loops or their generalizations is known at present, so it is {\it a priori} not clear how to determine the dual conformal anomaly.
We recall however that this anomaly is due to the infrared divergences of the scattering amplitudes. The latter have a universal, helicity independent form in $\mathcal{N}=4$ SYM thus suggesting that the dual conformal anomaly might also be universal  for MHV, NMHV, N${}^2$MHV, ... superamplitudes. This property can be formulated
in a compact form by introducing the so-called ratio function. In the NMHV case it is defined by
\begin{align}\label{mhvfac}
\mathcal{A}_n^{\rm NMHV} = \mathcal{A}_n^{\rm  MHV} \left[R_n^{\rm NMHV}+ O(\epsilon)\right]\,,
\end{align}
where $\mathcal{A}_n^{\rm  MHV}$ and $\mathcal{A}_n^{\rm NMHV}$ stand for the complete (all-loop) superamplitudes and $\epsilon$ is the parameter of dimensional regularization. The ratio function $R_n^{\rm NMHV}$
defined in this way is infrared finite and, therefore, one would expect that the symmetries, broken by the divergent factor $\mathcal{A}_n^{\rm  MHV}$, could be (partially) restored in $R_n^{\rm NMHV}$.  In particular,  if the dual conformal
anomaly is universal, then $R_n^{\rm NMHV}$ should be {\it dual conformally
invariant}. Indeed, it has been shown in \cite{Drummond:2008vq}  that the ratio function is given, to lowest order in the 't Hooft coupling $a=g^2 N$, by the following expression (see Eq.~\re{NMHV-678} below)
\begin{align}\label{R-loop}
 R_n^{\rm NMHV} = \sum_{ s,t=1}^n w_{st} R_{1st} \left[1+ a V_{1st}(x)  + O( a^2) \right]  +\text{cyclic}\,,
\end{align}
where the sum runs over the {\it linearly independent} superinvariants $R_{1st}$ and  $w_{st}$ are ($a-$independent)  rational numbers. {The terms needed to make $R_n^{\rm NMHV}$ invariant under cyclic shifts of the labels of the $n$ particles are denoted by  `cyclic'.}
Most importantly, the scalar functions $V_{1st}(x)$, which encode the loop corrections  in \p{R-loop}, are {\it dual conformally
invariant}. Thus, dual conformal symmetry is a general property of the ratio function~\cite{Drummond:2008vq,Drummond:2008bq,Brandhuber:2009xz,Elvang:2009ya}.

{We recall that the superinvariants $R_{1st}$, and thus the NMHV tree superamplitude \p{NMHV-tree0} have a larger, dual {\it super}conformal
symmetry. An obvious question is whether the dual conformal symmetry of $V_{1st}$ can also be promoted to dual superconformal symmetry. The latter is obtained by adding Poincar\'e supersymmetry to dual conformal symmetry (the rest follows from the $\cN=4$ superconformal algebra).  In this paper we  show that the dual Poincar\'e $\bar Q-$supersymmetry of the ratio function (which is also equivalent to ordinary special conformal supersymmetry, see \cite{Drummond:2008vq} and Eq.~\p{dualbarq} below)  is broken at one loop. We trace this one-loop anomaly back to the so-called holomorphic anomaly \cite{Cachazo:2004by,Bena:2004xu,Cachazo:2004dr} of the tree superamplitudes. Strictly speaking, the tree superamplitudes like the MHV one \p{MHVsua}, are invariant under $\bar Q-$supersymmetry only up to contact terms, due to the collinear pole singularities when $\vev{i\, i+1} \to 0$.  When loops are made out of trees via unitarity, such singularities are integrated over and induce an anomalous behavior under dual Poincar\'e (and ordinary special conformal) supersymmetry.%
\footnote{Very recently, the role of the collinear singularities and the associated holomorphic anomaly has been studied in \cite{Bargheer:2009qu}. The authors argue that one can deform the generators by including the holomorphic anomaly and maintaining the superconformal algebra. Requiring exact invariance under the deformed symmetry leads to recursive relations between tree amplitudes with different numbers of particles. However, it is not clear whether this approach can be efficiently pursued at loop level.} We illustrate this effect by an explicit calculation of the one-loop $\bar Q-$anomaly of the multi-particle discontinuity of the
$n=6$ NMHV superamplitude. At the same time, we show that the holomorphic anomaly does not affect the dual conformal symmetry of this amplitude.}

\subsection{Organization of the paper}

Section \ref{csmnmta} is devoted to proving ordinary superconformal symmetry of the tree superamplitudes. We first show this for the MHV superamplitude \p{MHVsua}. Compared to the original treatment in \cite{Witten:2003nn}, we exhibit a new feature, the equivalence of $\bar s-$conformal supersymmetry with the condition for twistor space collinearity from \cite{Witten:2003nn}. We argue that the combination of ordinary and dual superconformal symmetry fixes the form
of the MHV superamplitude \p{MHVsua}, up to a normalization constant. We then show that {\it each term} in the NMHV superamplitude \p{NMHV-tree0} is separately invariant under both ordinary and dual superconformal symmetries. The $\bar s-$conformal supersymmetry is equivalent to the condition that the tree NMHV amplitude is supported on three intersecting lines in twistor space.

In  Section \ref{singsec} we study the analytic properties of the NMHV superamplitudes in the various singular limits: multi-particle and two-particle (or collinear) physical singularities and the unphysical spurious singularities. We first recall the factorization property of tree amplitudes at multi-particle poles and the associated notion of discontinuity. We then examine the physical and spurious singularities of the NMHV superinvariants $R_{rst}$. Afterwards we show how the $n=6,7,8$ NMHV superamplitudes can be uniquely reconstructed, starting from an arbitrary linear combination of $n-$point superinvariants and imposing the conditions for correct singular behavior. In doing this, the absence of spurious singularities turns out to be the most powerful condition, fixing all relative coefficients of the superinvariants. The multi-particle factorization property is only needed to determine the overall normalization. We then repeat the analysis for the $n=6,7$ one-loop superamplitudes and show that the absence of spurious singularities implies strong restrictions on the dual conformal loop corrections to each superinvariant. At the end of the section, as an alternative, we discuss the role of the collinear (two-particle) singularities to  fix the unique form of the superamplitude, this time recursively.

In  Section \ref{dsaha} we investigate the role of the holomorphic anomaly as the source of the breakdown of dual Poincar\'e $\bar Q-$supersymmetry (or ordinary $\bar s-$conformal supersymmetry). Instead of discussing the amplitude itself, where the effect of the holomorphic anomaly is mixed up with  infrared divergences, we consider the multi-particle cut of the one-loop MHV and NMHV superamplitudes. This quantity is infrared finite and, as a consequence, it inherits the dual conformal symmetry of the constituent tree superamplitudes. However, dual supersymmetry is broken and we compute the corresponding anomaly.

Section \ref{concu} contains concluding remarks. Some technical details are presented in the  appendices.

\section{Conformal supersymmetry of MHV and NMHV tree superamplitudes}\label{csmnmta}

In this section, we discuss the constraints imposed on the tree MHV and NMHV superamplitudes by the superconformal symmetry in $\mathcal{N}=4$ SYM theory.
To this end it is sufficient to consider the action of the odd generators of Poincar\'e ($q,\bar q$) and conformal ($s, \bar s$) supersymmetry \cite{Witten:2003nn},
\begin{align}
 &  q_\a^A = \sum_{i=1}^n \la_{i\a} \eta_i^A\,, && \hspace*{-20mm} \bar q_A^{\da} = \sum_{i=1}^n \bl_{i}^{\da} \frac{\pa}{\pa\eta_i^A}\,, \nn
 \\
  &  s_A^\a = \sum_{i=1}^n  \frac{\pa^2}{\pa \la_{i\a}\pa\eta_i^A}\,, &&\hspace*{-20mm} \bar s_{\da}^A = \sum_{i=1}^n  \eta_i^A\frac{\pa}{\pa \bl_{i}^{\da}}\,,
   \label{gencsu}
\end{align}
the rest follows from the superconformal algebra $su(2,2|4)$ (see Eq.~\re{comm-rel}).
For example, computing the anticommutator $ \{{s}_{\a A},\overline{{s}}_{\da}^{ B} \} = \delta_A^B {k}_{\a \da}$ we can obtain the well-known expression for the generator of special conformal transformations \cite{Witten:2003nn},
\begin{align}\label{k}
{k}_{\a \da} = \sum_{i=1}^n \frac{\partial^2}{\partial \la_i^\a \partial\bl_i^\da}\,.
\end{align}
It is second-order with respect to the  spinor variables, which reflects the fact that
the conformal symmetry acts non-locally in the momentum representation.
Comparing \re{k} with \re{gencsu}, we note that the generators \re{gencsu}  are at most first-order in the spinor derivatives. Therefore, their action on the superamplitudes (after Fourier transforming the $\eta$ dependence) is linear and local. This is why we will concentrate on the verification of the invariance of the superamplitudes under $\bar s$ and $s$ supersymmetry, the action of $q$ and $\bar q$ being quite obvious.

{The realization of dual superconformal symmetry in the dual superspace \p{super-mom} and the proof that the MHV superamplitudes, as well as each term inside the NMHV superamplitudes  are covariant, has been presented in detail in \cite{Drummond:2008vq}. Here we only recall the partial overlap between the two superconformal algebras, namely, the following generators of ordinary (denoted by lower case letters) and dual (upper case) symmetries coincide:
\begin{eqnarray}
   \bar q_A^{\da} &\equiv& \bar S_A^{\da}\,,\qqqquad  \bar s^A_{\da} \equiv \bar Q^A_{\da}
  \label{dualbarq}\,.
\end{eqnarray}
Also, the dual Poincar\'e supersymmetry $Q_{\a A}=\sum_i \partial/\pa {\theta_i^{\a A}}$ is a trivial consequence of the change of variables \p{super-mom}. Thus, having proven $\bar s \equiv \bar Q$ symmetry, the statement about dual superconformal symmetry of the amplitudes is essentially reduced to their dual conformal $K-$covariance.
}

\subsection{MHV tree superamplitude}

As was shown
in Ref.~\cite{Witten:2003nn}, the tree-level MHV superamplitude is invariant
under the full native $\cN=4$ superconformal algebra. In this section we take a slightly different route. We generalize \re{MHVsua} by introducing the possible dependence on the bosonic variables $\la,\bl$ through an arbitrary function,
\begin{equation}\label{MHVsuamodi}
    {\cA}_n^{\rm MHV;0} = \delta^{(4)}(\sum_{i=1}^n \la_i\bl_i)\ \delta^{(8)}(\sum_{i=1}^n \la_i \eta_{i})\ f(\la, \bl)\,.
\end{equation}
Then we ask the question to what extent ordinary and dual superconformal symmetries restrict the function $f(\la,\bl)$. Notice that the expression in the right-hand side of
\re{MHVsuamodi} has to be a homogenous polynomial of degree 8 in the $\eta$'s and, therefore, the function $f(\la,\bl)$ is $\eta-$indepedent. In addition,
$f(\la, \bl)$ carries the helicity weights of the scattered (super)particles, i.e. it accounts for the scaling behavior ${\cA}_n^{\rm MHV;0} \to \e^{i\sum_k\chi_k}{\cA}_n^{\rm MHV;0}$
 under $\la_k\to \e^{-i\chi_k/2}\la_k$, $\bl_k \to \e^{i\chi_k/2}\bl_k$ and $\eta_k\to \e^{i\chi_k/2}\eta_k$.
This suggests to use the following ansatz,
\begin{align}\label{varphi}
f(\la, \bl) = \frac{\varphi(\la,\bl)}{\vev{12}\vev{23}\ldots \vev{n1}}\,,
\end{align}
with $\varphi(\la,\bl)$ being a helicity neutral function. There are two ways to construct such functions, by allowing dependence only through the helicity-free momenta \p{p=ll} or through purely holomorphic combinations like $\vev{ij}\vev{kl}/\vev{ik}\vev{jl}$. Below we argue that the former are ruled out by the conformal supersymmetry $\bar s$, and the latter by dual conformal symmetry. Thus, the MHV superamplitude can be fixed in the form \re{MHVsua} (up to normalization) by symmetries alone.

So, let us demand that all the generators in \p{gencsu} (and hence all generators of $su(2,2|4)$) annihilate ${\cA}_n^{\rm MHV;0}$ in \p{MHVsuamodi}.
Obviously, the two Poincar\'e supersymmetries $q$ and $\bar q$ defined in \p{gencsu}  do so, due to the fermionic and bosonic delta functions in \p{MHVsuamodi}. In the following subsections we discuss the action of the special conformal supersymmetry generators $\bar s$ and $s$.

\subsubsection{Conformal supersymmetry $\bar s$ and collinearity in twistor space}\label{cscts}

When applied to \p{MHVsuamodi}, the conformal supersymmetry generator $\bar s$, Eq.~\p{gencsu}, acts on the dotted spinor variables $\bl$. They are present both in the
function $f(\la, \bl)$ and in the argument of the momentum conservation delta function $\delta^{(4)}(\sum_{i=1}^n \la_i\bl_i)$. The variation of the latter is suppressed by the fermionic delta in \p{MHVsuamodi}.

Thus, acting on ${\cA}_n^{\rm MHV;0}$, the generator $\bar s$ goes through the delta functions in \p{MHVsuamodi} and directly hits $f(\la, \bl)$, giving
\begin{equation}\label{giving}
     \bar s^A_{\da} f(\la, \bl) = \sum_{i=1}^n \eta_i^A \frac{\pa}{\pa\bl_i^\da} f(\la, \bl) = 0\,.
\end{equation}
We should not require each term in this variation to vanish, since not all of the $\eta$'s are linearly independent. Indeed, the Grassmann delta function in \p{MHVsuamodi} imposes a linear relation, $\sum_{i=1}^n \la^\a_i \eta_i^A=0$. Projecting this relation with, e.g., $\lan{n}$ and $\lan{1}$, we can solve it for
\begin{equation}\label{slfo}
    \eta_1 = \frac{1}{\vev{1n}} \sum_{i=2}^{n-1}\vev{ni} \eta_{i}\,, \qquad \eta_n = \frac{1}{\vev{n1}} \sum_{i=2}^{n-1}\vev{1i} \eta_{i}\,.
\end{equation}
Then the relation \p{giving} reduces to
\begin{equation}\label{redvato}
    \bar s^A_\da f(\la, \bl) = \frac{1}{\vev{1n}} \sum_{2}^{n-1} \eta_i^A \left(\vev{1n} \frac{\pa}{\pa\bl_i^\da} + \vev{ni} \frac{\pa}{\pa\bl_1^\da} + \vev{i1} \frac{\pa}{\pa\bl_1^\da} \right) f(\la, \bl) =0 \,.
\end{equation}
Since all the $\eta$'s in this relation are linearly independent, we must impose the constraints
\begin{equation}\label{coll}
    F_{1,i,n} f(\la, \bl) \equiv \left(\vev{1n} \frac{\pa}{\pa\bl_i} + \vev{ni} \frac{\pa}{\pa\bl_1} + \vev{i1} \frac{\pa}{\pa\bl_n} \right)f(\la, \bl) = 0\,, \qquad
(2\le i \le n-1)
\,.
\end{equation}
We recognize that the operator $F_{1,i,n}$ coincides with the well-known operator of collinearity in twistor space \cite{Witten:2003nn}. Thus, the conformal supersymmetry $\bar s$ of the MHV superamplitude \p{MHVsuamodi} implies that
${\cA}_n^{\rm MHV;0}$ has to satisfy a collinearity condition, meaning that ${\cA}_n^{\rm MHV;0}$ has support on a single line in twistor space defined by the points $(1,n)$.

This is a very strong condition on the $\bl-$dependence of the function $f(\la,\bl)$. The  supersymmetry $\bar s$ does not constrain the $\la-$dependence, so we could in principle imagine an additional dependence on $\la$ through purely holomorphic combinations without helicity like $\vev{ij}\vev{kl}/\vev{ik}\vev{jl}$. However, such combinations are not compatible with dual conformal symmetry (see Appendix \ref{B}). This implies that the dependence
of the function $\varphi$ introduced in \re{varphi} on $\la$ and $\bl$ must come through the helicity-free momenta \re{p=ll}, that is $\varphi=\varphi(p_1,\ldots,p_n)$. In Appendix \ref{A1} we  show that \p{coll} yields  $\varphi={\rm const}$.

We conclude that the requirements of simultaneous invariance of the MHV tree superamplitude, Eqs.~\re{MHVsuamodi} and \re{varphi}, under ordinary and dual superconformal symmetry fixes its form up to an overall constant factor.

We wish to make an important comment on the discussion in this subsection.
The denominator in \re{varphi},  although naively holomorphic (a function of $\la$ but not of $\bl$) is singular {for $\vev{i\, i+1}\to 0$ and}, therefore, it is annihilated by $\bar s$ up to contact terms. This phenomenon is closely related to the so-called ``holomorphic anomaly" of amplitudes in twistor space  \cite{Cachazo:2004by}.
Of course, the above symmetry argument does not take into account the holomorphic anomaly of the amplitude.
This anomaly becomes important when tree amplitudes are used to form loops, via the unitarity cut technique \cite{Cachazo:2004by,Bena:2004xu,Cachazo:2004dr}. As we show in Section \ref{singsec}, it makes the conformal supersymmetry $\bar s$ (or the dual Poincar\'e supersymmetry $\bar Q$) anomalous at loop level.

\subsubsection{Grassmann Fourier transform}

We still have one more supersymmetry condition to verify, $s^\a_A \cA^{\rm MHV;0} =0$. This is not so simple, since the generator $s^\a_A$ in \p{gencsu} is given by a second-order differential operator. Although it is possible to show that $s^\a_A  {\cA}_n^{\rm MHV;0} = 0$ directly (see \cite{Witten:2003nn}), here we prefer to use another approach, more suitable for generalization to the NMHV case. We first Fourier transform  the amplitude with respect to the odd variables $\eta$, which renders the generator $s^\a_A$ first-order and makes the check much easier.

The Grassmann Fourier transform of an $\cN=4$ superamplitude is defined by the $n-$fold Grassmann integral
\begin{equation}\label{gftsu}
    \tilde {\cA}_n(\bet) = \int \prod_{i=1}^n d^4\eta_i \ \e^{\sum_{i=1}^n \bet_{iA} \eta^A_i} \cA_n(\eta)\,.
\end{equation}
As explained in \cite{Drummond:2008vq}, \cite{ArkaniHamed:2008gz}, the superamplitude $\tilde {\cA}_n(\bet)$ is equivalent to the PCT conjugate of $\cA_n(\eta)$ (hence the use of complex conjugate odd variables $\bet_A = (\eta^A)^*$). Thus, for the $n-$particle MHV superamplitude $\cA^{\rm MHV}$ containing, e.g., MHV gluon amplitudes with only two negative-helicity gluons, its transformed version $\tilde {\cA}^{\rm MHV}_n$ contains only two gluons of positive helicity.

Let us perform the Fourier transform of the MHV superamplitude \p{MHVsua} by taking
into account that
the fermionic delta function in \p{MHVsua} can be factorized into two four-dimensional ones, e.g. (cf. \p{slfo}),
\begin{equation}\label{impocon}
    \delta^{(8)}(\sum_{1}^n \la_i \eta_{i}) = \vev{1n}^4 \delta^{(4)}\left( \eta_1 - \frac{1}{\vev{1n}} \sum_{2}^{n-1}\vev{ni} \eta_{i} \right) \delta^{(4)}\left( \eta_n - \frac{1}{\vev{n1}} \sum_{2}^{n-1}\vev{1i} \eta_{i} \right)\,.
\end{equation}
This relation can be used to do the Fourier integrals with respect to $\eta_1$ and $\eta_n$. The remaining $(n-2)$ integrals take the form
\begin{align}
  \int \prod_{1}^n d^4\eta_i \ \e^{\sum_{1}^n \bet_{i} \eta_i} \delta^{(8)}(\sum_{1}^n \la_i \eta_{i})\nn
&=  \vev{1n}^4 \int \prod_{2}^{n-1} d^4\eta_i \ \exp\left\{\sum_{2}^{n-1} \left(\bet_i+ \frac{\vev{ni}}{\vev{1n}} \bet_1 + \frac{\vev{1i}}{\vev{n1}} \bet_n \right)\eta_i\right\}\nn \\
&= \vev{1n}^4 \prod_{2}^{n-1}\delta^{(4)}\left(\bet_i+ \frac{\vev{ni}}{\vev{1n}} \bet_1 + \frac{\vev{1i}}{\vev{n1}} \bet_n \right)\,.
\end{align}
Collecting the bosonic factors from \p{MHVsua}, we obtain the Grassmann Fourier transform of the tree MHV superamplitude:
\begin{equation}\label{41}
    \tilde {\cA}_n^{\rm MHV;0} = i \vev{1n}^{4(3-n)}\prod_{i=1}^n \vev{i\, i+1}^{-1}\  \delta^{(4)}(\sum_{i=1}^n \la_i\bl_i)\prod_{i=2}^{n-1}\  \delta^{(4)}\lr{\vev{1n} \bar \eta_{i}+\vev{ni} \bar \eta_{1}+\vev{i1} \bar \eta_{n} }\,.
\end{equation}
Note that this expression is a homogenous polynomial in $\bet$'s of degree $4(n-2)$.

\subsubsection{Conformal supersymmetry $s$}\label{QMHV}

Let us proceed to showing that \p{41} is invariant under the action of the supersymmetry generator $s_{\a A}$ defined in Eq.~\p{gencsu}. After the Fourier transform \re{gftsu} it becomes a first-order differential operator,
\begin{equation}\label{42}
     s_{\a A} = \sum_{i=1}^n \frac{\pa}{\pa \la^\a_i} \bet_{iA}\,.
\end{equation}
We have put the $\bet$'s to the right of the $\la$ derivatives on purpose. When acting on the amplitude \p{41}, we will use the Grassmann delta functions  in it to express $\bet_i$ (with $i=2 \ldots n-1$) in terms of $\bet_{1}$ and $\bet_n$. This introduces some $\la$ dependence, so we will have to push the $\la$ derivatives in \p{42} through it. The result is
\begin{equation}\label{43}
    s_{\a A} \tilde {\cA}_n^{\rm MHV;0} = \left[\frac{\bet_{1 A}}{\vev{1n}} \left(- \sum_{i=1}^n \vev{ni} \frac{\pa}{\pa \la^\a_i} +(n-2) \frac{\la_{n\a}}{\vev{1n}}   \right) -(1 \leftrightarrow n) \right] \tilde {\cA}_n^{\rm MHV;0}\,.
\end{equation}
Now, we take into account that the amplitude is annihilated by the (chiral) Lorentz generator $m^\a_\b = \sum_{i=1}^n \lr{ \la^\a_i  \pa_{i\b} - \frac12 \delta^\a_\b   \la^\gamma_i \pa_{i\gamma}}$, and obtain from \p{43}
\begin{equation}\label{44}
    s_{\a A} \tilde {\cA}_n^{\rm MHV;0} = \frac{\bet_{1 A}\la_{n\a} - \bet_{n A}\la_{1\a}}{2\vev{1n}} \left( \sum_{i=1}^n \la^\beta_i \frac{\pa}{\pa \la^\beta_i} +2(n-2)  \right)  \tilde {\cA}_n^{\rm MHV;0}\,.
\end{equation}
The operator in the parentheses counts the degree of homogeneity of
$\tilde {\cA}_n^{\rm MHV;0}$
in the $\la$'s. In Eq.~\re{41}, each $\vev{ij}$ gives 2, and the momentum conservation delta function gives $(-4)$, so the total of $4-2n$ cancels against the constant in the parentheses in \p{44}, leading to
\be
s_{\a A} \tilde {\cA}_n^{\rm MHV;0} =0\,.
\ee

We remark that the proof of this result can be simplified with the help of the
Poincar\'e supersymmetry $q_{i\a}^A  = \sum \la_{i\a} \pa/\pa\bet_{iA}$ (after the Fourier transform, see \p{gencsu}). This symmetry has eight fermionic parameters which can be used to `gauge away' eight components of the $\bet$'s, e.g., $\bet_1^A=\bet_n^A=0$. After this the
$(n-2)$ Grassmann delta functions in
 \p{41} imply that all the remaining $\bet$'s are zero, so the generator \p{42} vanishes when acting on the amplitude. Note that this gauge fixing is legitimate because the anticommutator $\{q,s\}$ (see \p{comm-rel})  vanishes when applied to the amplitude.

\subsection{NMHV tree superamplitude}

Let us now extend the analysis to the $n$-particle NMHV tree superamplitudes. The explicit expression for $\cA_n^{\rm NMHV;0}$ was found in \cite{Drummond:2008vq}:
\begin{align}\label{NMHV-tree'}
\mathcal{A}_n^{\rm NMHV;0} & = \mathcal{A}_n^{\rm MHV;0}      R_{n}^{\rm NMHV;0}  = i \frac{\delta^{(4)}(\sum_1^n \la_i\bl_i)\delta^{(8)}(\sum_1^n \la_i \eta_i)}{\vev{12}\ldots \vev{n-1\, n}\vev{n1}} R_{n} ^{\rm NMHV;0}\,,
\end{align}
where the tree-level ratio function admits two equivalent representations,
\begin{align}\label{NMHV-tree}
R_n^{\rm NMHV;0} & =   \sum_{4\le s+1< t\le n }   R_{1st}  =  \frac1n\sum_{r,s,t \in \mathcal{S}_n}  R_{rst}\,.
\end{align}
Here, in the second sum the indices $r,s,t=1,\ldots, n$ satisfy the conditions
\begin{align}\label{Sn}
\mathcal{S}_n: \qquad s-r \ge 2\ \text{(mod $n$)} ,\qquad t-s \ge 2\ \text{(mod $n$)} ,\qquad  r-t \ge 1\ \text{(mod $n$)} \,.
\end{align}
Each term in the sum \p{NMHV-tree} represents a {\it dual superconformal invariant} \begin{align} \label{R}
R_{rst} = R_{rts} = \frac{\vev{s-1 s}\vev{t-1 t} \delta^{(4)}(\Xi_{rst})}{x_{st}^2
\vev{r|x_{rs}x_{st} |t-1}\vev{r|x_{rs}x_{st} |t}\vev{r|x_{rt}x_{ts}
|s-1}\vev{r|x_{rt}x_{ts} |s}}\,,
\end{align}
where the standard   notation is used for contractions of spinors, $\vev{i j} = \la_i^\a  \la_{j\a}$ and $\vev{i|x_{ij}x_{jk}|k} = \la_i^\a (x_{ij})_{\a\da}(x_{jk})^{\da\b}\la_{k,\b}$; also, $x_{jk}=x_j-x_k$ and
$p_i^{\da\a}= x_{i,i+1} ^{\da\a} =\la_i^\a \tilde \la_i^{\da}$ is the  $i-$th particle momentum. The dependence of the superinvariants \re{R} on the particle supermomenta $\la_{i,\a} \eta_i^A = (\theta_i-\theta_{i+1})^{A}_\a$ comes through the Grassmann delta function $\delta^{(4)}(\Xi_{rst})$ with argument
\begin{align}\notag
\Xi_{rst}^A = \Xi_{rts}^A
&= \sum_{r+1}^{t-1} \vev{r|x_{rs}x_{st}|i}\eta^A_i + \sum_{r+1}^{s-1}\vev{r|x_{rt}x_{ts}|i}\eta^A_i
\\ \label{Xi}
&= \vev{r|x_{rs}x_{st}|\theta_t^A}+\vev{r|x_{rt}x_{ts}|\theta_s^A} +
x_{st}^2 \vev{r\,\theta_r^A}
\,.
\end{align}

It is convenient to use a diagrammatic representation for $R_{rst}$ as a box diagram shown in  Fig.~\ref{fig-R}.\footnote{The  superinvariants $R_{rst}$, and hence their diagrammatic representation,  are in one-to-one correspondence with the three-mass box coefficients from Ref.~\cite{Bern:2004bt}. } It has $n$ external legs ordered clockwise. One of the vertices is `massless', i.e. it has only one external leg with index $r$ attached to it, while the opposite vertex is always `massive', i.e. it has at least two external legs with indices $s,\ldots,t-1$ attached. It is easy to see that $\Xi_{rst}$, Eq.~\p{Xi}, vanishes if the restrictions \re{Sn} on the values of the labels $r,s,t$ are not fulfilled.

\begin{figure}[h]
\psfrag{R=}[cc][cc]{$R_{rst}~~=$}
\psfrag{r}[cc][cc]{$r$}\psfrag{r1}[cc][cc]{$r-1$}\psfrag{r2}[cc][cc]{$r+1$}\psfrag{t}[cc][cc]{$t$}\psfrag{s}[cc][cc]{$s$}\psfrag{s1}[cc][cc]{$s-1$}\psfrag{t1}[cc][cc]{$t-1$}\psfrag{r3}[cc][cc]{$r-2$}
\psfrag{t14}[cc][cc]{ }\psfrag{t11}[cc][cc]{$s-1$}\psfrag{t12}[cc][cc]{$s$}\psfrag{t13}[cc][cc]{$r-1$}
\psfrag{t20}[cc][cc]{$t$}\psfrag{t24}[cc][cc]{ }\psfrag{t21}[cc][cc]{$r$}\psfrag{t22}[cc][cc]{$r+1$}\psfrag{t23}[cc][cc]{$t-1$}
\centerline{\includegraphics[width=70mm]{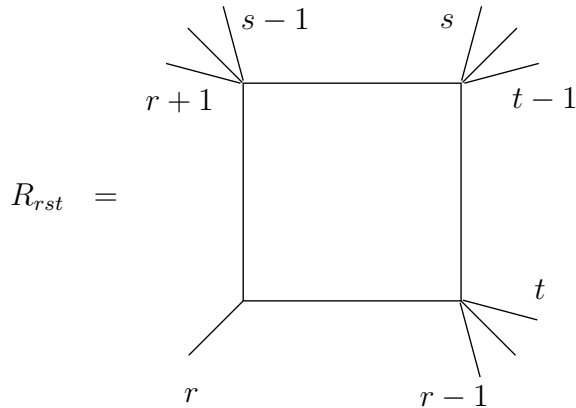}}
\caption{\small Diagrammatic representation of the dual superconformal invariant $R_{rst}$}
\label{fig-R}
\end{figure}

The equivalence between the two representations in \re{NMHV-tree} follows from the identity between the superinvariants \cite{Drummond:2008vq,Drummond:2008bq}
\begin{align}\label{R=R}
 \sum_{s,t \in \mathcal{S}_{n'}}  R_{1st} =  \sum_{s,t \in \mathcal{S}_{n'}}  R_{n'st}\,,\qquad (n'=5,\ldots,n)\,,
\end{align}
in which the indices satisfy the same conditions \re{Sn} with $r = 1$ in the left-hand side sum and $r = n'$ in the right-hand side sum. This identity holds without the help of the (super)momentum conservation delta functions $\delta^{(4)}(\sum_1^n \la_i\bl_i)\delta^{(8)}(\sum_1^n \la_i \eta_i)$.
It was found in \cite{Drummond:2008bq}  as a self-consistency condition for the scattering amplitudes in $\mathcal{N}=4$ SYM theory within the generalized unitarity cut method.
Another identity satisfied by the superinvariants is \cite{Drummond:2008vq,Drummond:2008bq}
\begin{align}\label{anothid}
R_{rs\,r-1} =  R_{r-2\, rs}
\end{align}
(for $n=5$ it is equivalent to \p{R=R}, but for $n>5$ it is independent). 

Using \re{R=R} and \p{anothid},  we can identify the set of linearly independent superinvariants~\cite{Drummond:2008bq,Elvang:2009ya} and express the NMHV superamplitude \re{NMHV-tree'} as a linear combination of those $R$'s.
For $n=6$ and $n=7$ the corresponding expressions are
\begin{align}\notag
R_6^{\rm NMHV;0} &=  \frac12 R_{146}+\text{cyclic} \,,
 \\[3mm] \label{NMHV-67}
R_7^{\rm NMHV;0} &=  \frac17 R_{146}+\frac27 R_{147}+ \frac37 R_{157} + \text{cyclic} \,.
\end{align}
Here `cyclic' stands for the terms obtained by cyclic shifts of the indices, $i\mapsto i+1$. They are needed for the cyclic symmetry of the superamplitude.

The properties of this amplitude under dual superconformal symmetry have been discussed in detail in \cite{Drummond:2008vq}. There it has been shown that {\it each term} \p{R} in the sum \p{NMHV-tree} is separately invariant under the full dual superconformal symmetry $SU(2,2|4)$.%
\footnote{The dual conformal symmetry of \p{R} and \p{Xi} under the inversion rules \p{conin} is manifest \cite{Drummond:2008vq}. Then, it is sufficient to show invariance under $\bar s \equiv \bar Q$ (see \p{dualbarq}) to prove the full dual superconformal symmetry of each term in the amplitude.} Certainly, the NMHV tree superamplitude, like  any divergence-free tree amplitude,  should also be invariant under the ordinary superconformal symmetry, Eqs.~\p{gencsu} and  \p{comm-rel}. What is not obvious, however, is that {\it each term} \p{R} in the sum \p{NMHV-tree} is separately invariant, as we show in the following subsections.

\subsubsection{Conformal supersymmetry $\bar s$ and line structure in
twistor space}\label{cslsts}

As before, checking the two Poincar\'e supersymmetries, $q
\mathcal{A}_n^{\rm NMHV;0}=\bar q\mathcal{A}_n^{\rm NMHV;0}=0$ requires
no particular effort. The conformal supersymmetry with generator $\bar
s$ is less trivial to verify. In \cite{Drummond:2008vq} a proof of $\bar
s\mathcal{A}_n^{\rm NMHV;0}=0$ was given, based on the equivalence
between $\bar s$ and the  dual Poincar\'e supersymmetry generator $\bar
Q \equiv \bar s$, see \p{dualbarq}. Here we present an alternative proof
which exhibits the twistor line structure of the NMHV superamplitude.

To this end, it is convenient to rewrite the tree NMHV superamplitude
\re{NMHV-tree'}  as
\begin{align}\nn
&\mathcal{A}_n^{\rm NMHV;0} =\sum_{4\le s+1<t\le n}  {\cal A}_{1st}\,,
\\
\label{rewrasfo1}
& {\cal A}_{1st} =   \delta^{(4)}(\sum_1^n \ran{i} [i|)\
\delta^{(8)}(\sum_1^n \ran{i}\eta_i)\ \delta^{(4)}\left(\sum_2^{s-1}
\vev{1i}\eta_i+ \sum_{s}^{t-1}\vev{1|x_{1s} x_{st}^{-1}
|i}\eta_i\right)\ f_{1st}(\la,\bl)\,,
\end{align}
where ${\cal A}_{1st}=\mathcal{A}_n^{\rm MHV;0} R_{1st}$
and the function  $f_{1st}(\la,\bl)$ carries the dependence on the on-shell
momenta
of the particles,
\begin{align} \label{somebosfa}
f_{1st} & =   \Bigg[x^2_{st} \prod_1^{s-2} \vev{i\, i+1}  \prod_s^{t-2}
\vev{j\, j+1}  \prod_t^{n} \vev{k\, k+1}
\\
  &\times    \vev{1|x_{1t}x^{-1}_{ts}|s-1} \vev{1|x_{1t}x^{-1}_{ts}|s}
\vev{1|x_{1s}x^{-1}_{st}|t-1} \vev{1|x_{1s}x^{-1}_{st}|t}
\Bigg]^{-1}.  \nn
\end{align}
Here $x_{ij} = \sum_{k=i}^{j-1} p_k = \sum_{k=i}^{j-1} \ran{k} [k|$ and
$x^{-1} \equiv x/x^2$.

As our first step, we use the Poincar\'e supersymmetry $\bar q$,
Eq.~\re{gencsu}, to gauge away a particular linear combination of $\eta$'s,
\begin{equation}\label{bqgqu}
    \mbox{$\bar q-$gauge:} \qqqquad \sum_s^{t-1}\ran{i}\eta_i^A = 0\,.
\end{equation}
As before, the gauge is legitimate because $\{\bar q, \bar s\}=0$ on the
amplitude.
The advantage of this gauge is the significant simplification of the
argument of the
Grassmann delta function in \re{rewrasfo1},
\begin{align} \label{rewrasfo}
& {\cal A}_{1st} =   \delta^{(4)}(\sum_1^n \ran{i} [i|)\
\delta^{(8)}(\sum_1^n \ran{i}\eta_i)\ \delta^{(4)}\left(\sum_2^{s-1}
\vev{1i}\eta_i\right)\ f_{1st}(\la,\bl)\,.
\end{align}
In addition to the gauge condition \p{bqgqu}, the Grassmann delta
functions in \p{rewrasfo}   impose further linear relations among the
odd variables $\eta_i$:
\begin{equation}\label{furlinrel}
    \sum_1^n \ran{i}\eta_i^A = 0\,, \qqqquad \sum_2^{s-1}
\vev{1i}\eta_i^A = 0\,.
\end{equation}
We can solve these constrains for five of the $\eta$'s. For example,
projecting Eq.~\p{bqgqu} onto the spinors $\lan{t-1}$ and $\lan{s}$, we
obtain $\eta_s$ and $\eta_{t-1}$, respectively.  Projecting the first of
Eqs.~\p{furlinrel} onto $\lan{s-1}$ and $\lan{1}$, we solve for $\eta_1$
and $\eta_t$, respectively. Finally, from the second of
Eqs.~\p{furlinrel} we find $\eta_{s-1}$:
\begin{align} \nn
  \eta_{s-1} &=  \frac{1}{\vev{s-1\, 1}} \sum_2^{s-2} \vev{1 i}\eta_i\,, &&
   \hspace*{-20mm}
  \eta_s  =  \frac{1}{\vev{s\, t-1}} \sum_{s+1}^{t-2} \vev{t-1\,
i}\eta_i \,, \\
  \eta_{t-1} &=  \frac{1}{\vev{t-1\, s}} \sum_{s+1}^{t-2} \vev{s i}\eta_i\,,
  &&  \hspace*{-20mm}
  \eta_t =  \frac{1}{\vev{t1}} \sum_{t+1}^{n} \vev{1 i}\eta_i\,, \nn
  \\  \label{expr5et}
  \eta_1 &
  =  \frac{1}{\vev{1\, s-1}}\sum_2^{s-2} \vev{s-1\, i}\eta_i -
\frac{1}{\vev{t1}} \sum_{t+1}^{n} \vev{t i}\eta_i \,.
\end{align}
Let us now apply the generator $\bar s$ given by \p{42} to the partial
amplitude ${\cal A}_{1st}$ defined in \re{rewrasfo}. As in the MHV
case,  the generator $\bar s$ goes through all the delta functions in
\p{rewrasfo} and hits $f_{1st}(\la,\bl)$. Taking into account the
relations \p{expr5et}, we find
\begin{equation}\label{linestru}
    \bar s f_{1st}  = \left(\frac{1}{\vev{1\, s-1}}\sum_2^{s-2}  \eta_i
F_{1,i,s-1} +  \frac{1}{\vev{s\, t-1}} \sum_{s+1}^{t-2} \eta_i
F_{s,i,t-1} + \frac{1}{\vev{t1}} \sum_{t+1}^{n}  \eta_i
F_{t,i,1}\right)f_{1st} \,,
\end{equation}
where the collinearity operators $F_{i,j,k}$ are given by \p{coll} after
appropriate identification of the indices. Since the $\eta$'s appearing
in \p{linestru} are linearly independent, we have to impose three sets
of collinearity conditions on the bosonic factors $f_{1st}(\la,\bl)$:
\begin{eqnarray}
  F_{1,i,s-1} f_{1st}(\la,\bl) &=& 0\,, \qquad (i=2,\ldots,s-2)\,, \nn
\\[2mm]
  F_{s,i,t-1} f_{1st}(\la,\bl) &=& 0\,, \qquad (i=s+1,\ldots,t-2)\,, \label{3colli}
\\[2mm]
  F_{t,i,1} f_{1st}(\la,\bl) &=& 0\,, \qquad (i=t+1,\ldots,n) \, . \nn
\end{eqnarray}

{Let us verify that the bosonic factor $f_{1st}$, Eq.~\p{somebosfa}, satisfies the  constraints \re{3colli}. Ignoring the possible
holomorphic anomalies, we see that the $\bl-$dependence in
\p{somebosfa}  comes only through dual coordinates (i.e., sums of
consecutive momenta) $x_{u\, v+1} = \sum_{u}^{v}p_i = \sum_{u}^{v}
\ran{i} [i|$. Such strings of $\bl$'s are annihilated by $F_{a,i,b}$ if
$u\leq a < b \leq v$, or, in a trivial way, if the intervals $[u,v]$ and $[a,b]$ do not overlap. This is indeed the case of the three strings $x_{1s},
x_{1t}, x_{st}$ in \p{somebosfa}.%
\footnote{The operators $F_{1,i,s-1}$ and $F_{t,i,1}$ have point 1 in common, so, for example, $F_{t,i,1}\, x_{1s} \neq 0$. However, in \p{somebosfa} $x_{1s}$ always appears with the projection $\bra{1}x_{1s} = \bra{1}x_{2s}$, hence $F_{t,i,1}\bra{1}x_{1s} = 0$. }}
The twistor space interpretation of this fact is that the NHMV
superamplitude is supported on three lines passing through the points
$(1, s-1)$, $(s, t-1)$ and $(t,1)$, respectively. In addition, these
lines lie in a single plane, which can be tested by applying the
operator of coplanarity $K_{i,j,k,l}$ \cite{Witten:2003nn}. In the case
of gluon NMHV tree amplitudes these properties were established in
Refs.~\cite{Bern:2004ky}, \cite{Bern:2004bt}. The coplanarity of the
NMHV tree superamplitudes is a direct consequence of the dual
supersymmetry $\bar Q$ (equivalent to $\bar s$) and of a simpler,
fermionic analog of coplanarity \cite{Drummond:2008vq}.

{At this point we could ask the question discussed at the end
of Sect.~\ref{cscts} in the context of the MHV tree superamplitude: To
what extent do the symmetries fix the form of the individual terms in
the sum \p{rewrasfo1}, as given in  \p{somebosfa}? Let us try to multiply $f_{1st}$ by an arbitrary function,
\begin{equation}\label{modfbyf}
    f_{1st} \ \to \ \varphi(\la,\bl) f_{1st}\,.
\end{equation}
We wish this function to satisfy the collinearity constraints \p{3colli} and, in addition, to be a dual conformal invariant and to have vanishing helicity, in order not to modify the properties of $f_{1st}$. In Appendix \ref{A1} we show that the only solution to these conditions is $\varphi = {\rm const}$.

}

\subsubsection{Grassmann Fourier transform and conformal supersymmetry $s$}

We now turn to the conformal supersymmetry with generator $s$
defined in \re{gencsu}. As in the MHV case,  in order to render the generator first-order,
we shall first Fourier transform $\mathcal{A}_{1st}$, Eq.~\re{rewrasfo1}, with respect to the
$\eta$'s. In doing this we cannot use  the gauge condition \re{bqgqu}, so we shall work with the general expression \re{gencsu}.

The Fourier transform \re{gftsu} is carried out in several steps. Firstly, we use $\delta^{(8)}(\sum_1^n \ran{i}\eta_i)$ inside \re{rewrasfo1} and apply the identity \re{impocon} to do the Fourier transform with respect to $\eta_1$ and $\eta_n$. Next, we gauge away $\bet_1=\bet_n=0$ with the help of the supersymmetry $q = \sum \la_i \pa/\pa\bet_i$. We then use the  second Grassmann delta function in the right-hand side of
\re{gencsu}
to perform the Fourier transform with respect to $\eta_2$. Doing the remaining Fourier integrals, we finally obtain
\begin{equation}\label{46}
\tilde{\cal A}_{1st}  = (\vev{n1}\vev{12})^4 f_{1st}\delta^{(4)}(\sum_1^n \ran{i} [i|)\ \prod_3^{s-1} \delta^{(4)}(\bet_i- {\vev{\check{2}i}}\bet_2) \
    \prod_s^{t-1} \delta^{(4)}(\bet_i - \vev{\hat{2} i}\bet_2) \
    \prod_t^{n-1} \delta^{(4)}(\bet_i)\,,
\end{equation}
where the notation was introduced for the composite spinors
\begin{equation}\label{47}
\bra{\check{2}}_\a = \frac{\bra{1}_\a}{\vev{12}}\,, \qqqquad \bra{\hat{2}}_\a = - \frac{(\lan{1}x_{1s}x^{-1}_{st})_\a}{\vev{12}}\,.
\end{equation}
Now, we want to apply the supersymmetry generator \p{42} to this amplitude. As explained in subsection \ref{QMHV}, we can use the Grassmann delta functions in \p{46} to express the various $\bet$'s in terms of $\bet_2$ or to set them to zero. We then push the derivatives $\pa/\pa\la$ through the resulting bosonic factors and find
\begin{equation}\label{48}
    s_{\a A}\tilde{\cal A}_{1st}  = \bet_{2\, A} \left[\Delta_\a -(s-3) \la_{\check{2}\a} - (t-s-2) \la_{\hat{2} \a} \right]\tilde{\cal A}_{1st} \,,
\end{equation}
where $\Delta_\a = \sum_2^{s-1} \vev{\check{2}i}\pa_{\a\, i} + \sum_s^{t-1} \vev{\hat{2}i}\pa_{\a\, i}$. Replacing $\tilde{\cal A}_{1st}$ by its explicit expression \re{46}
and going through some algebra we finally obtain (see Appendix \ref{A2} for details)
\begin{equation}\label{52}
   s_{\a A} \tilde{\cal A}_{1st} = 0\,.
\end{equation}
Thus, the partial NMHV amplitude ${\cal A}_{1st}$ is invariant under the superconformal
$\mathcal{N}=4$ symmetry.

\section{Analyticity constraints on the NMHV superamplitude}\label{singsec}

A characteristic feature of tree ratio function  \re{NMHV-tree} is that all superinvariants enter the sum with unit coefficients. The question arises where {does} this property come from. It can be
neither conventional, nor dual superconformal symmetry, since any linear combination of $R$'s will respect both symmetries simultaneously. Therefore, the values of
the coefficients of the superinvariants in \re{NMHV-tree} and \re{NMHV-67}
should follow from some new requirements, in addition to all  known symmetries.
We notice that $R_{rst}$, Eq.~\re{R}, considered as a function
of the particle momenta, has singularities due to the vanishing of the various factors in the denominator in \re{R}. As we will show in a moment, some of these singularities are physical, that is they correspond to the expected analytic properties of the superamplitude, while other singularities
are spurious and, therefore, should cancel against each other in the sum of superinvariants
in \re{NMHV-tree} and \re{NMHV-67}. This suggests that the coefficients in \re{NMHV-tree} and \re{NMHV-67} could be fixed by matching the analytic
properties of the linear combinations of superinvariants with those of the NMHV superamplitude. This is indeed the case, as we explain in this section.

\subsection{Factorization and discontinuities of superamplitudes}

{A generic $n-$particle tree-level superamplitude is a {meromorphic} function of the Mandelstam kinematic invariants with poles corresponding to the vanishing of some of these invariants~\cite{Mangano:1990by,Dixon:1996wi}. Color-ordered amplitudes can only have poles when multi-particle invariant masses of the contiguous type, $s_{i\ldots j-1} = (p_i+\ldots+p_{j-2} + p_{j-1})^2 = x^2_{ij}$, vanish.

A unique feature of the MHV superamplitude is that $ \mathcal{A}_n^{\rm MHV;0}$ only has poles  corresponding
to the vanishing of two-particle invariant masses $s_{i,i+1}=(p_i+p_{i+1})^2= x^2_{i,i+2} = \vev{i\ i+1}[i+1 \ i]$. Such singularities are called collinear, since $(p_i+p_{i+1})^2=0$ implies $p_i \sim p_{i+1}$ or $\la_i \sim \la_{i+1}\,$. At the same time, a tree-level non-MHV superamplitude has both two-particle and multi-particle poles.
The former describe the collinear limits of the non-MHV superamplitude and have the same
properties as for the MHV superamplitude. For multi-particle poles,  that is for $s_{i\ldots j-1} \to 0$ with $j-i \geq 3$, the tree-level non-MHV superamplitudes admit the multi-particle factorization (with  $P_\ell=p_i+\ldots+p_{j-1}$)~\cite{Mangano:1990by,Dixon:1996wi}
\begin{align}\label{fact}
\mathcal{A}^{\rm tree}(1,\ldots,n)\stackrel{P_\ell^2\to 0}{\sim} \int d^4\eta_\ell\, \mathcal{A}^{\rm tree}(i,\ldots,j-1,-\ell)
\,\frac{i}{P_\ell^2}\, \mathcal{A}^{\rm tree}(\ell,j,\ldots,i-1)\,,
\end{align}
where  $\ell$ and $(-\ell)$ denote on-shell states with (super)momenta $(P_\ell,\eta_\ell)$ and $(-P_\ell,\eta_\ell)$, respectively, and
the integration over $\eta_\ell$ accounts for the exchanges of particles of all possible helicities
~\cite{Georgiou:2004by,Brandhuber:2004yw,Bianchi:2008pu}.   For NMHV superamplitudes both tree superamplitudes in the right-hand side of \re{fact} are of the MHV type. }

Another way to represent the factorization property \re{fact} is in terms of discontinuities \cite{Eden}. The amplitudes are real functions \footnote{In the spinor helicity formalism the amplitude is a complex function of the complex spinor variables $\lambda, \bl$, but this does not affect the unitarity argument.} of the real kinematic invariants $s_{i...j-1}$,  which are well defined in the kinematic region where all $s_{i...j-1}<0$. The singularities {(poles or branch cuts)} occur when some of the $s_{i...j-1}$ change sign. The behavior near a singularity is described by the discontinuity defined as
\begin{equation}\label{defdisco}
    {\rm Disc}  f(s) = f(s+i0)-f(s-i0) = 2 i \Im f(s+i0)\,.
\end{equation}
For a pole singularity at $s_{i\ldots j-1} \equiv P_\ell^2 =0$ this formula reads
\begin{equation}\label{discproo}
    {\rm Disc}\lr{\frac{1}{P_\ell^2}} = 2i \Im \lr{\frac{1}{P_\ell^2+i0}} =-2i\pi \delta(s_{i\ldots j-1})\,.
\end{equation}
Further, standard unitarity arguments \cite{Eden,Mangano:1990by,Dixon:1996wi} relate the imaginary part, i.e. the discontinuity, of the amplitude to the product of two subamplitudes, obtained by cutting through the singular `propagator' $1/P_\ell^2$.

Applied to the NMHV tree superamplitude, the unitarity relation reads
\begin{align}\label{NMHV-disc'}
{\rm Disc}_{s_{i\ldots j-1}} \mathcal{A}_n^{\rm NMHV;0} = 2\pi \delta(s_{i\ldots j-1})
\int d^4 \eta_\ell\, \mathcal{A}^{\rm MHV;0}( i,\ldots, j-1, -\ell)   \mathcal{A}^{\rm MHV;0}(\ell,j,\ldots, i-1)\,.
\end{align}
We see that Eq.~\p{NMHV-disc'} is equivalent to the singular part of the factorization relation \p{fact}.
Let us replace the MHV superamplitudes in the right-hand side of \re{NMHV-disc'} by their explicit expressions \p{MHVsua}. The latter involve spinor variables at the exchange legs $(-\ell)$ and $\ell$, defined by writing the on-shell exchanged momentum  $P_\ell=p_i+\ldots+p_{j-1}$ (with $P_\ell^2=0$), as a product of two commuting spinors,
\be\label{P-tau}
P_\ell  = \ket{\ell}[\tilde\ell|\,.
\ee
Then we perform the
$\eta_\ell-$integration in \p{NMHV-disc'} with the help of the identity
\begin{align}
\int d^4 \eta_\ell\, \delta^{(8)}\bigg({\sum_s^{t-1} \eta_i \ket{i} +\eta_\ell \ket{\ell}}\bigg)
\delta^{(8)}\bigg({\sum_t^{s-1} \eta_i \ket{i} -\eta_\ell \ket{\ell}}\bigg)
= \delta^{(8)}\bigg({\sum_1^{n} \eta_i \ket{i} } \bigg)\delta^{(4)}\bigg({\sum_s^{t-1} \eta_i \vev{i \ell}}\bigg)\,,
\end{align}
and obtain (for $ j-i \geq 3$)
\begin{align}\label{NMHV-disc}
{\rm Disc}_{s_{i\ldots j-1}} \mathcal{A}_n^{\rm NMHV;0} = 2\pi i \,\delta(s_{i\ldots j-1})
 \frac{\vev{i-1i}\vev{j-1 j}\ \delta^{(4)}(\vev{\ell\, \q_{ij}})}{\vev{\ell i-1}\vev{\ell i}\vev{\ell j-1}\vev{\ell j}} \mathcal{A}_n^{\rm MHV;0}\,.
\end{align}

Let us now substitute the NMHV superamplitude in \re{NMHV-disc} by its expression \re{NMHV-tree'} in terms of the ratio function times the tree-level MHV
superamplitude. The latter has only two-particle poles (collinear singularities), which are the same as those of the NMHV superamplitude. Therefore,  the ratio function $R_n^{\rm NMHV;0}$ can only have multi-particle poles
at $s_{i\ldots j-1}=x_{ij}^2=0$ for $j-i \geq 3$. Moreover, it has to satisfy the following relation
\begin{align}\label{disc-R}
 {\rm Disc}_{x_{ij}^2} R_n^{\rm NMHV;0} = 2\pi i \,\delta(x_{ij}^2)
 \frac{\vev{i-1i}\vev{j-1 j}\ \delta^{(4)}(\vev{\ell\, \q_{ij}})}{\vev{\ell i-1}\vev{\ell i}\vev{\ell j-1}\vev{\ell j}}\,.
\end{align}
We recall that the ratio function $R_n^{\rm NMHV;0}$ is given by a linear combination of the superinvariants defined in \re{R}. Therefore,
knowing the analytic properties of the $R$'s, we can apply the relation \re{disc-R} to restrict the values of their
coefficients.

\subsection{Analytic properties of the superinvariants}

The singularities of the superinvariant $R_{rst}$  originate from the vanishing of one of the factors in the denominator in  \re{R}.  The first factor there produces a pole at $x_{st}^2=0$, corresponding to the vanishing of a multi-particle invariant mass $(p_s+\ldots+p_{t-1})^2=0$ (for $t-s \geq 3$). \footnote{The case $t-s=2$ is exceptional, see subsection \ref{phypo}.} Since the NMHV scattering amplitude does have such poles, we
shall refer to $x_{st}^2=0$ as to `physical poles'. In terms of the diagrams shown in Fig.~\ref{fig-R-cut} (a), the physical poles correspond to the vanishing of the invariant mass of the particles  with labels $s,\ldots,t-1$ attached to one of the vertices.

The remaining four factors in the denominator of \re{R}  produce poles at  $\vev{i|x_{ik}x_{kj}|j}=0$ for various values of the  indices $i, j, k$ that can be read off from \re{R}.   In Minkowski space-time this is equivalent to the condition   $|\vev{i|x_{ik}x_{kj}|j}|^2 = \vev{i|x_{ik}x_{kj}|j} [i|x_{ik} x_{kj}|j ] =0$, leading to the following relation
\footnote{In Minkowski signature $(+---)$ the spinor projections $[i|x_{ik} x_{kj}|j ]$ and $\vev{i|x_{ik} x_{kj}|j}$ are complex conjugate to each other while
in split signature $(++--)$ they are real and independent from each other.  }
\begin{align}\label{cubic}
-x_{ij}^2 x_{i+1,k}^2 x_{j+1,k}^2 + x_{i+1,j}^2 x_{ik}^2 x_{j+1,k}^2 - x_{i+1,j+1}^2 x_{ik}^2 x_{jk}^2 + x_{i,j+1}^2 x_{i+1,k}^2 x_{jk}^2 =0\,.
\end{align}
Since $x_{ij}^2=s_{i\ldots j-1}$, for generic values of the indices the pole at $\vev{i|x_{ik}x_{kj}|j}=0$ corresponds to a cubic relation among  multi-particle invariant masses. The scattering amplitudes do not
have such singularities and we shall refer to  these poles as to `spurious poles'~\cite{Bern:2004bt,Bern:2004ky}.

Notice, however, that there exist special values of the indices for which three of the four terms in the left-hand side of \re{cubic} vanish in virtue of $x_{i,i+1}^2=0$, so  the relation \re{cubic} yields the vanishing of a multi-particle invariant mass. This occurs  for $s=r+2$ and $t=r-1$, which are the boundary values of the indices (see \re{Sn}), so we shall refer to the corresponding superinvariants as to `boundary' ones.
In these cases more then one vertex in the diagrams in Fig.~\ref{fig-R-cut} (b) and (c) become massless, i.e. they have only one leg attached to them.  Indeed, for $s=r+2$ we have
\begin{align}\label{physpo}
\vev{r|x_{rt} x_{ts}|s-1} =  \vev{r|x_{rt} x_{t,r+1}|r+1} = -\vev{r\, r+1} x_{t,r+1}^2 
\end{align}
and, therefore, the corresponding superinvariant $R_{r,r+2,t}$ has an additional physical pole at $x_{t,r+1}^2=s_{r+1\ldots t-1}=0$. Similarly, $R_{r,s,t-1}$ has an additional physical pole at $x_{rs}^2=s_{r\ldots s-1}=0$. {These properties of the boundary superinvariants are in a one-to-one correspondence with similar properties of the three-mass-box coefficients
for the one-loop NMHV gluon amplitudes discussed in \cite{Bern:2004bt,Bern:2004ky}.}

\subsubsection{Physical poles}\label{phypo}

{For generic values of its indices, $R_{rst}$ in \re{R} has  a physical pole at $x_{rs}^2=0$.  In close
analogy with \re{NMHV-disc'}, let us compute the discontinuity of $R_{rst}$ in $x_{rs}^2$.}
Using the definition \p{discproo} in the form
\begin{align}
{\rm Disc}_{x_{st}^2} \frac1{x_{st}^2} = -2\pi i \delta(x_{st}^2)\,,
\end{align}
we can evaluate the remaining factors in \re{R} for $x_{st}^2=0$. Namely,  we write
the $2\times 2$ matrix $x_{st}^{\a \da}$ as a product of two commuting spinors,
\begin{equation}\label{remartha}
    x_{st} = \ran{\ell} [\tilde\ell| \,,
\end{equation}
and apply the identity $\bra{r} x_{rs} x_{st} =  \lan{r}x_{rt}|\tilde\ell] \lan{\ell}$ to simply $\Xi_{rst}$,
Eq.~\re{Xi},  as
\begin{align}
\Xi_{rst} = -\lan{r}x_{rt}|\tilde\ell]\vev{\ell\, \q_{st}}\,.
\end{align}
In this way, we can express the discontinuity of $R_{rst}$ as
\begin{align}\label{disc-R2}
{\rm Disc}_{x_{st}^2} R_{rst} = 2\pi i\delta\lr{x_{st}^2} \frac{\vev{s-1s}\vev{t-1 t}\ \delta^{(4)}(\vev{\t\, \q_{st}})}{\vev{\t s-1}\vev{\t s}\vev{\t t-1}\vev{\t t}}\,.
\end{align}
We remark that the right-hand side of Eq.~\p{disc-R2} is dual conformal.

We point out that the relation \re{disc-R2} holds for $t-s \geq 3$. If $t-s=2$, then the limit $x^2_{s, s+2} = s_{s, s+1} \to 0 $ corresponds to a collinear singularity.\footnote{For a more extensive discussion of the collinear singularities see Section \ref{collsingu}.} The MHV factor in \p{NMHV-tree'} already has the required collinear singular behavior and, therefore, the invariant $R_{r,s,s+2}$ is expected to be regular in this limit. Indeed, when the momenta $p_s$ and $p_{s+1}$ are aligned along the same
light-like direction, we have $p_s \sim p_{s+1} \sim \ket{\ell} [\tilde\ell|$, i.e. $\ket{s} \sim \ket{s+1}\sim \ket{\ell}$. In this limit the singularity of the denominator of $R_{r,s,s+2}$ is $x_{s,s+2}^2 \vev{\ell s}  \vev{\ell s+1} = O(x_{s,s+2}^4)$ but it is cancelled by the Grassmann delta function in the numerator, $\delta^{(4)}(\vev{\t\, \q_{st}})= O(x_{s,s+2}^4)$, so that $R_{r,s,s+2}$ remains finite, or
\begin{equation}\label{nodics2p}
    {\rm Disc}_{x^2_{s,s+2}} R_{r,s,s+2} = 0 \,.
\end{equation}

\begin{figure}[t]
\vspace*{5mm}
\psfrag{r}[cc][cc]{$r$}\psfrag{r1}[cc][cc]{$r-1$}\psfrag{r2}[cc][cc]{$r+1$}\psfrag{t}[cc][cc]{$t$}
\psfrag{s}[cc][cc]{$s$}\psfrag{s1}[cc][cc]{$s-1$}\psfrag{t1}[cc][cc]{$t-1$}
\psfrag{r3}[cc][cc]{$r-2$}\psfrag{r4}[cc][cc]{$r+2\ $}
\psfrag{t14}[cc][cc]{ }\psfrag{t11}[cc][cc]{$s-1$}\psfrag{t12}[cc][cc]{$s$}\psfrag{t13}[cc][cc]{$r-1$}
\psfrag{t20}[cc][cc]{$t$}\psfrag{t24}[cc][cc]{ }\psfrag{t21}[cc][cc]{$r$}\psfrag{t22}[cc][cc]{$r+1$}\psfrag{t23}[cc][cc]{$t-1$}
\includegraphics[width=170mm]{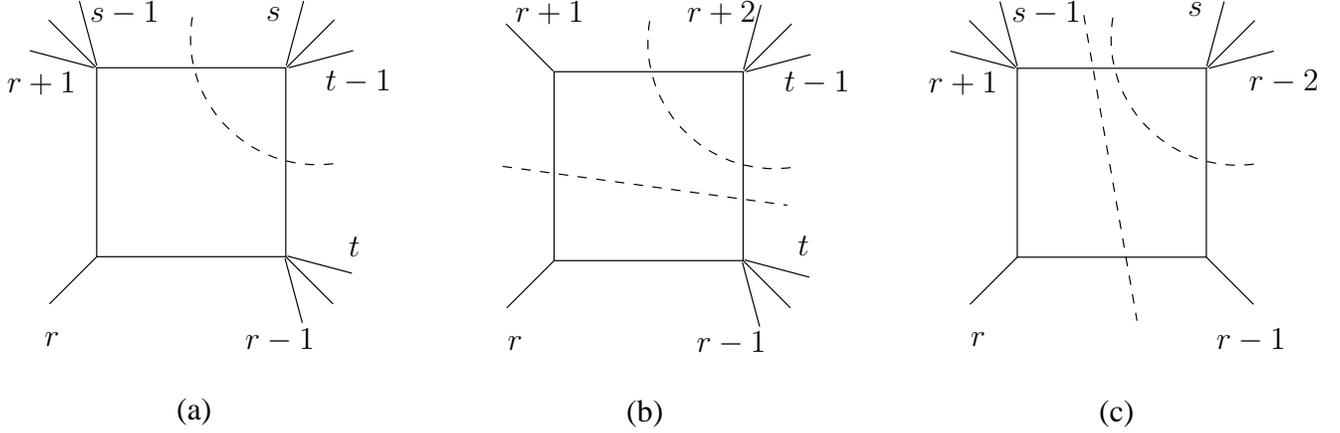}
\caption{\small Three different configurations of superinvariants with cuts (dashed lines) corresponding to physical poles: {generic case (a) and `boundary' terms with $s=r+2$ (b) and $t=r-1$ (c)}.}
\label{fig-R-cut}
\end{figure}

{Let us now consider the `boundary' invariants $R_{rst}$ with $r=s-2$ or $r=t+1$. We recall that they   have additional  physical (i.e., multi-particle) poles, besides the generic ones at $x_{st}^2=0$. }  In particular, for $r=s-2$, this pole is located at
$x_{t,r+1}^2=0$ (see Eq.~\re{physpo}).  As in \re{remartha}, for $x_{t,r+1}^2=0$ we define the spinor variables
\begin{equation}\label{remartha''}
    x_{t,r+1} = \ran{\ell} [\tilde\ell|
\end{equation}
and simplify  $R_{r,r+2,t} $ with the help of the identities
\begin{align} \notag
 & x_{r+2,t}^2 = (x_{t,r+1}+p_{r+1})^2 =  \vev{r+1\ \t}[\tt\ r+1]\,,
\\
 & \Xi_{r,r+2,t} =  \vev{r+1\ r+2} [\tt\ r+1] \vev{\t\ \q_{r+1,t}}\,.
\end{align}
In this way, we obtain the discontinuity of $R_{r,r+2,t} $ in the form
\begin{align}\label{bounter}
{\rm Disc}_{x^2_{t,r+1}}R_{r,r+2,t} = 2\pi i\delta(x_{t,r+1}^2)\frac{\vev{r\, r+1}\vev{t-1\, t}\delta^{(8)}(\vev{\t\ \q_{r+1,t}})}{\vev{\t  r}\vev{\t \ r+1}\vev{\t\ t-1}\vev{\t t}}\,.
\end{align}
The second case with $r=t+1$ can be treated similarly.
Moreover, taking into account the identity \p{anothid},
we find that the discontinuity of $R_{r,s,t-1}$ in $x_{rs}^2$ can be derived from the discontinuity
of $ R_{r-2,r,s}$ in $x_{rs}^2$  (see Eq.~\re{disc-R2})
\begin{align}\label{disc-Rc}
{\rm Disc}_{x^2_{rs}} R_{r,s,r-1} = 
 2\pi i\delta\lr{x_{rs}^2} \frac{\vev{s-1s}\vev{r-1\, r}\ \delta^{(4)}(\vev{\t\, \q_{rs}})}{\vev{\t\, s-1}\vev{\t s}\vev{\t\, r-1}
\vev{\t r}}\,.
\end{align}
The diagrammatic representation of the discontinuities of $R_{rst}$ in the three different cases,
Eqs.~\re{disc-R2}, \re{bounter} and \re{disc-Rc},   is shown in Fig.~\ref{fig-R-cut}.

We observe the striking similarity between the expressions for the discontinuity of the ratio function \re{disc-R}
and of the invariants $R_{rst}$, Eqs.~\re{disc-R2}, \re{bounter} and \re{disc-Rc}. In fact,
these expressions coincide after the appropriate identification of the indices. Then, it is easy to verify that
the known form  \re{NMHV-tree} of the tree-level NMHV amplitude is in agreement with the unitary condition \re{disc-R}. We may now invert the logic
and ask ourselves if the condition \re{disc-R} is sufficiently strong to fix all the coefficients in \re{NMHV-tree}. We can immediately see that this is not the case. Indeed, the sum in \re{NMHV-tree} involves, among others, superinvariants  $R_{rst}$  with $t=s+2$. As follows from \re{nodics2p}, such invariants do not have physical poles (except for $r=s-2$ and $r=t+1$) and, therefore,
they can not be detected from the unitarity condition. As we show in the next subsections,  the really powerful
condition on the coefficients of the superinvariants follows from the cancellation of spurious poles.

\subsubsection{Spurious poles}\label{sect-spur}

The spurious poles of $R_{rst}$  are generated by the four factors in the denominator
in \re{R} involving the brackets $\vev{ \ldots }$.  Let us start with the kinematical configuration
\begin{align}\label{spur}
\vev{r|x_{rt}x_{ts} |s} = 0
\end{align}
and compute the residue of   $R_{rst}$ at this pole. We  resolve this condition  as %
\footnote{Since the expression for $R_{rst}$ is invariant under rescaling of $\bra{r}$, we can choose
the normalization factor of $\bra{r}$  at our convenience.}
\begin{align}\label{p-spur}
 \bra{r} = \bra{s}x_{st}x_{tr}
\end{align}
and  substitute it into the expression \re{Xi} for $\Xi_{rst}$ to get
\begin{align}\notag
\Xi_{rst} &=  \bra{s} x_{st} x_{tr} x_{rs}x_{st}\ket{\theta_t}+x_{st}^{2}
x_{tr}^{2}\vev{s\theta_s} + x_{st}^{2} \vev{s|x_{st}x_{tr}|\theta_r}
\\[3mm] \label{Xi-new}
&= x_{st}^{2}  \bra{s} x_{sr}x_{rt}\ket{\theta_t}+ x_{st}^{2}
x_{tr}^{2}\vev{s\theta_s} + x_{st}^{2}  \vev{s|x_{st}x_{tr}|\theta_r} \,,
\end{align}
where we have applied  the identity
$x_{st} x_{tr} x_{rs}x_{st} =  x_{st}( x_{ts}+x_{sr}) x_{rs}x_{st} = x_{st}^2 x_{sr}x_{rt}$.
Comparing the second relation in \re{Xi-new} with \re{Xi}, we  observe that, in the kinematical configuration corresponding to the spurious pole \p{spur}, $\Xi_{rst}$ reduces to $\Xi_{str}$
\begin{align}
\Xi_{rst}  =
x_{st}^{2}\,  \Xi_{str}\,.
\end{align}
Then, we substitute \re{p-spur} into the remaining nonsingular factors in the denominator
of $R_{rst}$ and  obtain, after some algebra,
\begin{align}\label{R-residue}
R_{rst} = -\frac1{\vev{r|x_{rt}x_{ts} |s}}
\lr{\frac{ x_{tr}^{-2}\vev{t-1\, t}\delta^{(4)}(\Xi_{str})}{\vev{s|x_{sr}x_{rt}|t-1}
\vev{s|x_{sr}x_{rt}|t}} } +\ldots\ ,
\end{align}
where we have displayed the spurious pole contribution and the ellipsis denotes regular terms.
We note that the right-hand side of this relation resembles the expression for
$R_{str}$,
\begin{align}\label{R-another}
R_{str} = \frac{\vev{t-1\, r}\vev{t-1\, t}
\delta^{(4)}(\Xi_{str})}{x_{rt}^2\vev{s|x_{st}x_{tr} |r-1}\vev{s|x_{st}x_{tr}
|r}\vev{s|x_{sr}x_{rt} |t-1}\vev{s|x_{sr}x_{rt} |t}}\,.
\end{align}
Indeed, it has the same spurious pole \re{spur} and we use \re{p-spur} to substitute
$\vev{s|x_{st}x_{tr} |r-1} = \vev{r\, r-1}$ in the denominator of \re{R-another}. We then
verify that the residue of $R_{str}$ at the spurious pole \re{spur} is the same as in \re{R-residue},
up to a  sign. Therefore, the spurious pole cancels in the sum
\begin{align}\label{spur-canc}
R_{rst} + R_{str} = 0\times \frac{\delta^{(4)}(\Xi_{str})}{\vev{r|x_{rt}x_{ts}
|s}}+ \ldots\,,
\end{align}
where the ellipsis denotes terms regular at $\vev{r|x_{rt}x_{ts}
|s}=0$.
Notice that the two invariants in the left-hand side of \re{spur-canc} have cyclicly shifted indices.

Similarly, we find that the spurious pole at $\vev{r|x_{rs}x_{st}|t}=0$ cancels in the sum
\begin{align}\label{r1}
{R_{rst} + R_{trs} =  0\times
\frac{\delta^{(4)}(\Xi_{trs})}{\vev{r|x_{rs}x_{st} |t}}} + \ldots\ .
\end{align}
For the spurious poles at
$\vev{r|x_{rt}x_{ts}|s-1}=0$ and $\vev{r|x_{rs}x_{st}|t-1}=0$ we get, respectively,
\begin{align}\notag
R_{rst} - R_{s-1\, tr} &= 0\times
\frac{\delta^{(4)}(\Xi_{s-1\, tr})}{\vev{r|x_{rt}x_{ts}|s-1}}+\ldots\,,
\\ \label{r2}
R_{rst} - R_{t-1\, rs} &= 0\times
\frac{\delta^{(4)}(\Xi_{t-1\, rs})}{\vev{r|x_{rs}x_{st}|t-1}} + \ldots\ .
\end{align}
Compared to \re{spur-canc}, the additional minus sign in the first relation
in \p{r2} is due to the antisymmetry of  $R_{rst}$ under the exchange of the indices $s$
and $s-1$.
Substituting $s\to s-1$ and $t\to t-1$  in \re{spur-canc} and
\re{r1}, respectively, and combining them with \re{r2} we obtain other (equivalent)
relations
\begin{align}\notag
R_{rst}+ R_{rs\, t-1} & = 0\times
\frac{\delta^{(4)}(\Xi_{t-1\, rs})}{\vev{r|x_{rs}x_{st}|t-1}}+\ldots
\\ \label{rr2}
R_{rst}+ R_{rs-1\, t} & = 0\times
\frac{\delta^{(4)}(\Xi_{s-1\, tr})}{\vev{r|x_{rt}x_{ts}|s-1}}+\ldots\ .
\end{align}
As follows from \re{spur-canc}, \re{r1} and \re{r2}, all spurious poles of $R_{rst}$
cancel in  the following
linear combination of invariants,
\begin{align}\label{sum-R}
R_{rst} + \lr{ R_{str} + R_{trs} - R_{s-1\, tr}- R_{t-1\, rs}}\,.
\end{align}
This shows that in order to verify the cancellation of
spurious poles in the ratio function \re{R}, it is sufficient to check that each $R_{rst}$ appears in it accompanied by the superinvariants inside
the parentheses in \re{sum-R}.

So far we have assumed that the indices $r,s,t$ of  the superinvariants $R_{rst}$ take generic values such that all the superinvariants in \re{sum-R} are well defined. However, when we go close to the boundary in  \re{Sn}, some of the $R$'s in  \re{sum-R} are not well defined, i.e. their indices do not
verify the condition \re{Sn}. This happens, in particular, to the boundary superinvariants
mentioned above. For example, for $t=r-1$ three out of the four $R$'s inside the brackets
in \re{sum-R} are not well defined. The reason for this is that the factors
producing the spurious poles of $R_{r,s,r-1}$ become `shorter' for $t=r-1$, e.g.,
\begin{align}
\vev{r|x_{r,r-1}x_{r-1,s}|s} = \vev{r-1\, r} [r-1|x_{r-1,s}\ket{s}
\end{align}
and the pole is located at $ [r-1|x_{r-1,s}\ket{s} =0$. The cancellation of such
poles of the boundary $R$'s requires special care and we shall return to this
question in the next subsection (see Eqs.~\re{R+R} and \re{R-R-R} below).

\subsection{Reconstructing the NMHV superamplitude from physical and spurious singularities}

In this subsection we illustrate how the requirement of cancellation of spurious poles supplemented with the unitarity
condition (presence of physical singularities {only}) allows us to determine all the coefficients in the ratio function \p{NMHV-tree'}. We are not going to present a general proof valid for all $n$, but we consider three examples, $n=6$, $n=7$ and $n=8$. This is sufficient to understand the general mechanisms through which the coefficients are fixed.

\subsubsection{$n=6$ NMHV}\label{sect:n6}

Let us start with the simplest NMHV superamplitude, $n=6$.%
\footnote{The case $n=5$ looks like an NMHV superinvariant, which however corresponds to the `googly' (Grassmann Fourier transformed) $\overline{\rm MHV}$ five-particle amplitude \cite{Drummond:2008vq}.} Solving the linear relations \p{R=R} and \p{anothid} among the  superinvariants, we find that the independent
ones are $R_{146}$ and its cyclic images $R_{1+i,\, 4+i,\, 6+i}$ with $i=1,\ldots,5\ ({\rm mod}\ 6)$. Thus, the most general expression for the ratio function $R_6^{\rm NMHV;0}$, consistent
with all symmetries of the superamplitude, is
\begin{align} \nn
R_6^{\rm NMHV;0} &= \alpha R_{146} + \text{cyclic}
\\[2mm] \label{R6-c}
&= \alpha \lr{R_{146} +R_{251}+R_{362}+R_{413}+R_{524}+R_{635}} \,.
\end{align}
We remark that in this, and only in this special case, the NMHV superamplitude is effectively determined by its symmetries up to the overall constant factor $\alpha$.

For $n=6$ the general expression \re{R} for  $R_{146}$ simplifies considerably because the various strings $\vev{\ldots}$ in the numerator and denominator factorize. After a few obvious steps, we find
\begin{align}  \label{R6-ini}
R_{146} &= - \frac{\vev{34}\vev{45}\vev{56}\vev{61}}{x_{14}^2
[45][56]\bra{1}x_{14}|4][6|x_{63}\ket{3}} \delta^{(4)}(\eta_4
[56]+\eta_5[64]+\eta_6[45])\,.
\end{align}
We observe that $R_{146}$ has a physical pole at $x_{14}^2=0$ (see \p{physpo}), but it is regular at $x_{46}^2=0$, see \p{nodics2p}. In addition, it has
two spurious poles at $[4|x_{41}\ket{1} =0$ and $[6|x_{63}\ket{3} = 0$.

We start with the physical pole and examine the discontinuity of $R_{146}$ in
$x_{14}^2=(p_1+p_2+p_3)^2=(p_4+p_5+p_6)^2$.
Following \re{remartha}, we solve the condition  $x_{14}^2=0$ by introducing the spinors $x_{14} = \ket{\ell} [\tilde\ell|$ and apply
the identity
\be
\vev{\ell\theta_{14}}  = -\sum_{i=4}^6 \vev{\ell i} \eta_i = -\frac{\vev{\ell 4}}{[56]}
(\eta_4
[56]+\eta_5[64]+\eta_6[45])\,,
\ee
together with $\vev{\ell 4}/[56]=\vev{\ell 5}/[64]=\vev{\ell 6}/[45]$, to get from
\re{R6-ini}
 \begin{align}  \label{R6-disc}
{\rm Disc}_{x_{14}^2}\, R_{146} &= 2\pi i \delta(x_{14}^2) \frac{\vev{34}\vev{61}\delta^{(4)}(\vev{\ell\theta_{14}})}{
 \vev{\ell 3}\vev{\ell 4}\vev{\ell 6}\vev{\ell 1}}  \,,
\end{align}
in agreement with \re{disc-Rc} for $r=1$ and $s=4$.

Now, let us impose the unitarity condition \re{disc-R} on the ratio function \re{R6-c}.
We have
\begin{align} \label{c-eq}
{\rm Disc}_{x_{14}^2}  R_6^{\rm NMHV;0}  = 2 \alpha \, {\rm Disc}_{x_{14}^2} R_{146}\,,
\end{align}
where the additional factor 2 is due to the invariance of
the right-hand side of \re{R6-disc}  under cyclic shifts of the indices $i\mapsto i+3$.  Substituting \re{R6-disc}
into \re{c-eq} and matching the resulting expression  with \re{disc-R}, we obtain
\be\label{sol6}
\alpha=\frac12\,,
\ee
in agreement with the known result \re{NMHV-67}. Thus, the unitarity condition combined with
the symmetries of the $n=6$ NMHV superamplitude allowed us to reconstruct the tree-level expression
for the ratio function.

Since we have already fixed the unique form of the amplitude, we know that it must be  free from spurious poles. Still, let us show this explicitly.
We start with the spurious pole at
\be
\bra{1}x_{14}|4] = \bra{1}(p_1+p_2+p_3)|4] = \bra{1}(p_2+p_3)|4] = 0\,.
\ee
This pole corresponds to the kinematical configuration { (a `planar singularity' in the terminology of \cite{Bern:2004bt})}
\begin{align} \notag
p_2+p_3 = c_1 p_1 + c_4 p_4\,,
\end{align}
with $c_1$ and $c_4$ arbitrary. Unlike the physical poles,
it does not correspond to vanishing invariant masses and implies instead that
the sum of two momenta lies in the plane defined by two other momenta. Working out $|\bra{1}x_{14}|4]|^2=0$ in terms of dual coordinates, we find
\be\label{quadric}
x_{14}^2 x_{25}^2 - x_{15}^2 x_{24}^2 = 0\,,
\ee
which is a quadratic relation between two- and three-particle invariant masses.  \footnote{This is a special case of the more general cubic relation \p{cubic}.}

To understand the spurious singularities of $R_{146}$, we use the
identity \p{anothid} for $n=6$,
\begin{align}\label{iden}
R_{146} = R_{351} = \mathbb{P}^2 R_{135} = \mathbb{P}^4 R_{136}\,,
\end{align}
where $\mathbb{P}$ generates a cyclic shift of the indices $i\mapsto i+1$ with the periodicity condition $i+6\equiv i$. Let us now apply the relation \re{spur-canc} for $r=1,s=3, t=5$:
\begin{align} \label{R+R}
R_{135} + R_{351} = 0\times \frac{\delta^{(4)}(\Xi_{351})}{\vev{1|x_{15}x_{53}
|3}}+ \ldots = 0\times \frac{\delta^{(4)}(\Xi_{351})}{\bra{1}x_{14}|4]\vev{43}
}+ \ldots\ .
\end{align}
As follows from our analysis in Sect.~\ref{sect-spur}, both $R_{135}$ and $R_{351}$ have spurious poles at $\bra{1}x_{14}|4]=0$, but they cancel in the sum of the two invariants. Making use of the identity \re{iden}, we can rewrite the left-hand side of \re{R+R} as
\be\label{combin}
 R_{135}+ R_{351}
=R_{135}+R_{146}=R_{146}+ \mathbb{P}^4 R_{146}\,.
\ee
Thus, the sum $R_{146}+ \mathbb{P}^4 R_{146} =R_{146}+R_{524}$ does not contain the spurious pole at $ \bra{1}x_{14}|4] = 0 $, but it still contains the spurious pole at $[6|x_{63}\ket{3}=0$ coming from $R_{146}$, and
another spurious pole at
$[2|x_{25}\ket{5}=0$ coming from $\mathbb{P}^4 R_{146}=R_{524}$. Applying \re{spur-canc} and \re{r1}
for $r=1,s=3, t=5$, we find that all spurious poles cancel at once in the linear
combination
\be\label{aux1}
R_{135}+ R_{351} + R_{513} = R_{146}+ \mathbb{P}^2
R_{146} + \mathbb{P}^4 R_{146} = R_{146} + R_{136} + R_{135}\,.
\ee
This is just the $n=6$ NMHV ratio function \re{NMHV-tree} but we remark that the requirement of vanishing spurious singularities does not fix the overall factor $\alpha$ in \p{R6-c}.

We conclude that $R_6^{\rm NMHV;0}$ is free from spurious singularities while its physical poles
satisfy the unitarity conditions \re{disc-R}.

\subsubsection{$n=7$ NMHV}\label{sect:n7}

The case $n=7$ is the first example of the generic situation where we use  the spurious singularity condition in order to fix the relative coefficients of the superinvariants in the ratio function, while the unitarity condition is needed to determine the overall normalization.

In this case, taking into account the linear relations among the superinvariants \re{R=R} and \re{anothid}, we identify the basis of linearly independent invariants as $R_{146}$, $R_{147}$ and $R_{157}$ and their cyclic counterparts. Then, the general form of the cyclicly invaraint $R_7^{\rm NMHV;0}$ involves three arbitrary coefficients,
\begin{align}\label{R7-ini}
R_7^{\rm NMHV;0} =  \alpha R_{146} + \beta R_{147} + \gamma R_{157} + \text{cyclic}\,.
\end{align}
The explicit expressions for the superinvariants entering \p{R7-ini} follow from \re{R}:
\begin{align} \notag
R_{146} & = - \frac{\vev{34}\vev{56}\delta^{(4)}(\Xi_{146})}{x_{46}^2 \vev{45}^2}
\lr{\bra{1} x_{14} |4] \bra{1} x_{15}|5]
\vev{1|x_{16}x_{63}|3}\vev{6|x_{64}x_{41}|1} }^{-1}\,,
\\ \notag
R_{147} &= \frac{\vev{34}\vev{67}\delta^{(4)}(\Xi_{147})}{x_{47}^2 x_{14}^2
\vev{71}^3} \lr{\bra{4} x_{47} |7] \bra{3} x_{37}|7] \vev{6|x_{64}x_{41}|1}
}^{-1}\,,
\\ \label{R7-R}
R_{157} &= \frac{\vev{45}\vev{67}\delta^{(4)}(\Xi_{157})}{x_{57}^2 x_{51}^2
\vev{71}^3\vev{56}^2 [67]} \lr{\bra{4} x_{47} |7] \bra{1} x_{15}|5] }^{-1}\ .
\end{align}
We notice that all three invariants involve both physical poles corresponding to $x_{ij}^2=0$ and
spurious poles generated by the factors $\langle \ldots ]$ and $\vev{ \ldots
}$.

Let us first examine the physical poles and impose the unitarity condition \re{disc-R}.
We notice that $R_{146}$ does not have physical poles, as an example of the exception in Eq.~\p{nodics2p}. Thus,  $R_{146}$ does not contribute to the discontinuity
of the ratio function in \re{disc-R} and, therefore, the unitarity condition \re{disc-R}
cannot be used to determine the value of the constant $\alpha$. To obtain a condition on the other two coefficients $\beta$ and $\gamma$, we remark that since the ratio function \re{R7-ini} is cyclicly invariant, it suffices to examine its discontinuity with respect to any three-particle invariant mass, say $x_{14}^2$. It follows from
\re{R7-R} that ${\rm Disc}_{x_{14}^2} R_7^{\rm NMHV;0}$ receives a nonzero contribution from
three terms, $R_{147}$, $\mathbb{P}^4R_{147}=R_{514}$ and $\mathbb{P}^3R_{157}=R_{413}\,$. The discontinuities of each of these terms can be read off from \re{disc-R2}, \re{bounter} and \re{disc-Rc}. Then, the unitarity condition \re{disc-R} implies
\be\label{cond0}
2 \beta + \gamma = 1\,.
\ee
Thus, unlike the case $n=6$, unitarity alone does not determine all the coefficients in \re{R7-ini}.

Additional relations among the coefficients follow from the spurious poles cancellation in $R_7^{\rm NMHV;0}$.
The examination of \re{R7-R} reveals that, up to cyclic shifts of the indices, the superinvariants
have three types of spurious poles: $\vev{6|x_{64}x_{41}|1}=0$, $\bra{1} x_{14} |4]=0$ and $\bra{1} x_{15}|5]=0$.

Let us start with the first spurious pole, which, according to \re{R7-R}, is present in  $R_{146}$ and $R_{147}$. It follows from the first relation in \re{rr2} that for $r=1, s=4, t=7$ the spurious pole at $\vev{6|x_{64}x_{41}|1}=0$ cancels in the sum $R_{146}+R_{147}$, hence the residues of $R_{146}$ and $R_{147}$ at this pole have opposite signs.
Then, examining the cyclicly invariant combinations of superinvariants, we obtain
\footnote{To simplify the formulae, here we do not display the $\delta^{(4)}(\Xi)$ fatcors in the right-hand side.}
\begin{align} \notag
R_{146} + \text{cyclic} & \sim  \frac{2}{\vev{6|x_{64}x_{41}|1}} +  \ldots\,,
\\
R_{147} + \text{cyclic} & \sim - \frac{1}{\vev{6|x_{64}x_{41}|1}} +  \ldots\,,
\end{align}
where the ellipses denote regular terms. The additional factor 2 in the first relation is due to the fact that $\mathbb{P}^5R_{146}$ also has the same spurious pole. Then, requiring the ratio
function \re{R7-ini} to be regular at $ \vev{6|x_{64}x_{41}|1} = 0$, we obtain
\begin{align}\label{cond1}
2\alpha - \beta  = 0 \,.
\end{align}
Next, we examine the spurious pole at $\bra{1} x_{14} |4]=0$.  We find that the residues of the
superinvariants at this pole coincide up to signs,
\begin{align} \notag
& R_{146} + \text{cyclic} \sim  \frac{1}{\bra{1} x_{14} |4]} +  \ldots\,,
 \\ \notag
& R_{147} + \text{cyclic} \sim  \frac{1}{\bra{1} x_{14} |4]} +  \ldots\,,
 \\ \label{R-R-R}
& R_{157} + \text{cyclic} \sim  -\frac{1}{\bra{1} x_{14} |4]} +  \ldots\ .
\end{align}
Requiring the ratio function \re{R7-ini} to be regular at $\bra{1} x_{14} |4]=0$, we derive
\begin{align}\label{cond2}
 \alpha + \beta - \gamma = 0\,.
\end{align}
One can verify that the same condition ensures regularity at $\bra{1} x_{15}|5]=0 $.~\footnote{This should not be surprising since  the ratio function should be invariant under relabeling of the indices of the external legs $i\mapsto n-i+1$. The same transformation maps the spurious poles at $\bra{1} x_{14}|4]=0$ and $\bra{1} x_{15}|5]=0$ into each other (modulo cyclic shift of indices).}

Combining \re{cond1} and \re{cond2}, we find that the condition of compensation of spurious poles fixes the relative values of the coefficients, $\beta=2\alpha$ and $\gamma=3\alpha$, but not their absolute values. The latter is determined from the unitarity condition \re{cond0}, so we finally obtain
\begin{align}\label{sol7}
\alpha = \frac17\,,\qquad \beta = \frac27\,, \qquad  \gamma = \frac37\,.
\end{align}
Substitution of these values into \re{R7-R} yields the known expression \re{NMHV-67} for the $n=7$
ratio function.

\subsubsection{General $n$}\label{sect:n7'}

It is straightforward to extend the above consideration to arbitrary $n$. Our analysis for $n=6$ and $n=7$ suggests that the requirement of cancellation of spurious poles is sufficient
to fix all relative coefficients while the unitarity condition is  needed to fix the overall normalization. To test this conjecture, we consider the case $n=8$.

The general form of the $n=8$ ratio function is
\begin{align}\label{R8}
R_8^{\rm NMHV;0} =  \alpha R_{147} + \beta R_{148} + \gamma R_{157} +
\delta R_{158} + \varepsilon R_{168} +
\text{cyclic}\,.
\end{align}
Examining the expressions for the linearly independent superinvariants in \p{R8}, we find that the ratio function \p{R8} has two series of physical poles
at $x_{15}^2=0$ and $x_{14}^2=0$ (and their cyclic images)  and six series of spurious poles specified below.  Going through the same analysis as for $n=7$, we obtain that
the unitarity condition  \re{disc-R} for the discontinuity of $R_8^{\rm NMHV;0}$  in
$x_{15}^2$ and $x_{14}^2$ leads to the following
relations
\begin{align}\label{sys1}
2(\beta + \delta)=1 \,,\qquad \alpha + \beta+\delta+\varepsilon=1\,.
\end{align}
Further, the cancellation of the spurious poles in $\vev{1|x_{14}x_{47}|7}$, $\vev{1|x_{15}x_{57}|7}$, $\vev{1|x_{17}x_{74}|4}$, $\bra{1} x_{13}|3]$, $\bra{1}x_{15}|5]$ and $\bra{1} x_{16}|6]$ results in six relations between the coefficients,
\begin{align}\label{sys2}
\alpha - \beta = \alpha +\gamma-\delta = 2 \alpha-\gamma = \delta-\varepsilon = \beta+\gamma-\delta = \beta+\gamma-\varepsilon=0\,.
\end{align}
Comparing the two systems of equations, Eqs.~\re{sys1} and \re{sys2}, we notice that the former is undetermined while the latter is overdetermined. Still, the two
systems are consistent and have a unique solution.
In agreement with our expectations, the set \re{sys2} allows us to determine the relative coefficients
while \re{sys1} fixes the overall normalization as follows,
\begin{align}\label{sol8}
\alpha= \beta=\frac18\,,\qquad \gamma=\frac28\,,\qquad \delta=\varepsilon=\frac38\,.
\end{align}
Substituting this result into \re{R8} we obtain the expression for the tree ratio function
which can be shown to be equivalent to \re{NMHV-tree} for $n=8$ in virtue of the identities \re{R=R} and \re{anothid}.

\subsection{Spurious poles at loop level}

So far our discussion has been restricted to tree level. The loop corrections to the scattering amplitude suffer from infrared divergences which break the (dual) conformal symmetry of the tree amplitudes. As in Eq.~\re{NMHV-tree'}, we introduce the all-loop ratio function $R_n^{\rm NMHV}$, Eq.~\re{mhvfac}.
The new ratio function has a
perturbative expansion in powers of the `t Hooft coupling  $a = g^2N$,
\be
R_n^{\rm NMHV} = R_n^{\rm NMHV;0}+ a R_n^{\rm NMHV;1}+O(a^2)\,,
\ee
with the tree-level expression $R_n^{\rm NMHV;0}$ given by \re{NMHV-tree} and \re{NMHV-67}.
{The infrared divergences of the superamplitudes $\mathcal{A}_n^{\rm MHV}$ and $\mathcal{A}_n^{\rm NMHV}$ have the same form and, therefore, the ratio function $R_n^{\rm NMHV}$ is infrared finite. Moreover, it has been shown  in \cite{Drummond:2008vq,Drummond:2008bq}
that the one-loop ratio function has the form \re{R-loop} and it
is given by a sum over the {linearly independent} superinvariants multiplied by the scalar functions $V_{1st}(x)$ and weighted with rational coefficients $w_{st}$.}

{Since the coefficients $w_{st}$ in  \re{R-loop} do not depend on the coupling constant,} they can be determined from the analyticity conditions on the tree-level superamplitudes, as we  have shown
in the previous subsection for $n=6$ and $n=7$ (see Eq.~\re{NMHV-67}). In these two cases, the loop corrections to the ratio function have the form~\cite{Drummond:2008vq,Drummond:2008bq}
\begin{align}\notag
R_6^{\rm NMHV;1} &=  \frac12 R_{146} V_{146}+\text{cyclic} \,,
 \\[3mm] \label{NMHV-678}
R_7^{\rm NMHV;1} &=  \frac17 R_{146}V_{146} +\frac27 R_{147}V_{147} + \frac37 R_{157}V_{157} + \text{cyclic} \,.
\end{align}
{For general $n$, the analogous expressions for the ratio function were recently found in  Refs.~\cite{Brandhuber:2009xz,Elvang:2009ya}.}
At one loop, the functions $V_{1st}$ are given by linear combinations of one-loop scalar box Feynman integrals, having some  remarkable properties. By construction, they are free
from infrared divergences and, most importantly, they depend on the dual conformal coordinates only
through conformally invariant  cross-ratios
\begin{align}\label{u}
u_{ijkl} = \frac{x_{il}^2 x_{jk}^2}{x_{ik}^2 x_{jl}^2}\,,\qqqquad u_{ijkl} = u_{klij} = 1/u_{ijlk}\,.
\end{align}
This shows that the functions $V_{1st}$ enjoy {\it exact} dual conformal symmetry, unlike the MHV factor in \p{mhvfac}, which satisfies an {\it anomalous} dual conformal Ward identity \cite{Drummond:2007au}. We recall
that $R_{1st}$ are full dual {\it super}conformal  invariants while
$V_{1st}$ are only dual conformal invariants. In  the next section we will explain
that the breakdown of dual supersymmetry by loop corrections has to do with the so-called holomorphic anomaly of the scattering amplitudes.

As we have seen in the previous subsection,  the superinvariants have spurious poles  coming
from the vanishing of the factors $\vev{i|x_{ik} x_{kj}|j}$  in the denominator of \re{R}. The NMHV superamplitude, or equivalently the ratio function $R_6^{\rm NMHV}$, should be free from spurious singularities at all perturbative orders. At tree level, this restricts the coefficients $w_{st}$ in \re{R-loop}.
At one-loop level, the requirement of cancellation of the spurious poles leads to conditions
on the scalar functions $V_{1st}(u_{ijkl})$. In this subsection we formulate these conditions for $n=6$ and $n=7$ and we show that the actual combinations of box integrals appearing in the superamplitude verify them.

{We should mention that the analogous verification of the absence of spurious poles in the one-loop seven-gluon amplitudes has been carried out in \cite{Bern:2004ky}. The advantage of dealing with superamplitudes is that, as we show here, the calculation is quite simple, while in \cite{Bern:2004ky} the check had to be done numerically.}

We recall that in Minkowski space-time the spurious pole at $\vev{i|x_{ik} x_{kj}|j}=0$ yields the cubic relation \re{cubic} among the distances $x^2_{ij}$ in dual space. The condition $\vev{i|x_{ik} x_{kj}|j}=0$, and hence its corollary \re{cubic} are dual conformally covariant. So,  \re{cubic} can be recast as a relation among dual conformal cross-ratios \re{u} as follows,
\begin{align}\label{u-cubic}
u_{i,i+1,k,j+1} + u_{j,j+1,k,i+1} -  {u_{i,i+1,k,j}u_{j,j+1,k,i}}{u_{i,i+1,j,j+1}} = 1\,.
\end{align}
In general, the cross-ratios $u_{ijkl}$ depend on four dual points and for
$n$ particles one can construct $O(n^4)$ different $u$'s. However, only  $n(n-5)/2$ of them are independent (taking into account the light-like separation between adjacent points, see \cite{Drummond:2007au}).
We can choose the following basis of independent $u$'s:
\footnote{We would like to thank David Kosower for a useful discussion of this point.}
\be\label{u-set}
\{ u_{1245}\,,\ u_{1256}\,, \ldots , u_{12,k,1+k}\,,\  u_{12,1+k,2+k}\}\,,
\ee
with $k=[n/2]$ being the integer part of $n/2$. Notice that $u_{1234}=0$ in virtue of the relation $x_{i,i+1}^2=0$ and that all cross-ratios of the form $u_{i,i+1,j,j+1}$ can be obtained from those in \re{u-set} by cyclic shifts of the indices. The remaining
cross-ratios can be obtained as products or ratios of various $u_{i,i+1,j,j+1}$.

We have seen in the previous subsection that some of the
superinvariants entering \re{NMHV-678} have simple poles at the spurious kinematical configuration
$\vev{i|x_{ik} x_{kj}|j}=0$. Computing  and setting to zero the residue of the ratio function \re{R-loop} at this pole, we obtain linear relations among the functions $V(u)$,
evaluated on the hypersurface defined by \re{u-cubic}. To illustrate this procedure, let us consider the two special cases $n=6$ and $n=7$.

\subsubsection{$n=6$ NMHV}

For $n=6$ the set \re{u-set} reduces to  $u_{1245}$ and we identify the three independent cross-ratios as
\be\label{ui}
u_1=u_{1245}\,,\qquad u_2=u_{2356} = \mathbb{P}\, u_{1245}\,,\qquad u_3=u_{3461}= \mathbb{P}^2 u_{1245}\,.
\ee
The ratio function $R_6^{\rm NMHV}$ is given by \re{NMHV-678} and it involves the invariant
$R_{146}$ and the scalar function $V_{146}(u_1,u_2,u_3)$.
As was shown in Sect.~\ref{sect:n6},  $R_{146}$ has a spurious pole at $\bra{1}x_{14} |4] =0$. According to \re{quadric}, this pole corresponds to the kinematical configuration
\be\label{u=1}
u_{1245} = 1\,,
\ee
with $u_{2356}$ and $u_{3461}$ being arbitrary. We recall that the spurious poles
cancel in the sum $R_{146} +R_{135}=R_{146} + \mathbb{P}^4 R_{146}$ (see
Eq.~\re{combin}). Then, the condition for the ratio function \re{NMHV-678} to have zero residue at this pole is
$V_{146} - \mathbb{P}^4 V_{146} = 0$ for $u_{1245}=1$. Using the identity $\mathbb{P} V_{146}(u_1,u_2,u_3) = V_{146}(\mathbb{P} u_1,\mathbb{P} u_2,\mathbb{P} u_3) = V_{146}(u_2,u_3,u_1)$, we can rewrite the same condition as
\begin{align}\label{zero}
V_{146}(1,u_2,u_3) - V_{146}(u_3,1,u_2) =0\,,
\end{align}
with $u_2$ and $u_3$ arbitrary. {Notice that this relation should hold for
the scalar function $V_{146}(u_1,u_2,u_3)$ to all loops.}

Let us verify the relation \re{zero} using the one-loop expression for $V_{146}$ found in~\cite{Drummond:2008vq,Drummond:2008bq}
\begin{align}\label{V-def}
V_{146}(u_1,u_2,u_3) = -\ln u_1 \ln u_3 + \frac12 \sum_{k=1}^3 \left[ \ln u_k \ln u_{k+1} + {\rm Li_2}(1-u_k) \right] -\frac{\pi^2}6\,,
\end{align}
with $u_i$ defined in \re{ui}. It is easy to see that
\begin{align}
V_{146}(u_1,u_2,u_3) - V_{146}(u_3,u_1,u_2) = \ln u_{1} \ln\lr{ \frac{ u_{2}}{ u_{3}}}
\end{align}
and it indeed vanishes for $u_{1}=1$ in agreement with \re{zero}.

We recall that $R_{146}$ has another spurious pole at $[6| x_{63}\ket{3}=0$ which cancels against a similar pole of $R_{136}=\mathbb{P} ^2 R_{146}$. Repeating the analysis, we obtain that the cancellation of  this spurious pole leads to a condition equivalent to \re{zero}.

\subsubsection{$n=7$ NMHV}

For $n=7$ we have seven independent cross-ratios,  $u_{1245}$ and its six cyclicly shifted images. The ratio function $R_7^{\rm NMHV}$ is given by \re{NMHV-678} where the relative coefficients in front of the superinvariants are fixed by the condition of spurious pole cancellation at tree level. The spurious poles are at $\vev{1|x_{14} x_{46} |6}=0$,  $\bra{1} x_{14} |4]=0$ and $\bra{1} x_{15} |5]=0$.
According to \re{u-cubic}, the first kinematical configuration implies the relation
$
u_{1246} +u_{2467}  = 1
$.
Since the spurious pole $\vev{1|x_{14} x_{46} |6}=0$ cancels in the sum $R_{146}+R_{147}$, it will also cancel in the ratio function \re{NMHV-678} provided that
\begin{align}\label{n=7-rel}
V_{146} - V_{147} = 0\, \quad \text{for  $u_{1246} +u_{2467}  = 1$}\,.
\end{align}
Similarly to the case $n=6$, the spurious pole at $\bra{1} x_{14} |4]=0$ corresponds to $u_{1245}=1$. Repeating the argument of Sect.~\ref{sect:n7}, we find that this pole cancels in the ratio function provided that
\begin{align}\label{equiv}
V_{146} + 2\,\mathbb{P}^4  V_{147} - 3\, \mathbb{P}^4 V_{157} = 0 \,,\quad \text{for $u_{1245}=1$}\,,
\end{align}
where the operator $\mathbb{P}$ cyclicly shifts the indices of the conformal ratios inside the $V-$functions, $\mathbb{P}  V_{147} (u_{1245},\ldots) = V_{147} (u_{2356},\ldots)$. Finally, just as for the tree-level ratio function, the cancellation of the spurious pole at
$\bra{1} x_{15} |5]=0$ leads to a relation  equivalent to \re{equiv}.

Now, let us verify the relations \re{n=7-rel} and \re{equiv} using the expressions for $V_{146}$,
$V_{147}$ and $V_{157}$ found in \cite{Drummond:2008bq}:
\begin{align}\label{V7}
V_{147} = V_{146} + \frac74 V_{\rm II}\,,\qquad V_{157} = V_{146} + \frac76 V_{\rm I}
\end{align}
with
\begin{align} \notag
  V_{146} & = \frac12\left[{\rm Li_2}\lr{1-u_{1245}}-{\rm Li_2}\lr{1-u_{1246}}-\ln u_{1245}\ln u_{3467}
\right] + \text{cyclic}\,,
  \\[3mm] \notag
V_{\rm I\phantom{I}} &  ={\rm Li_2}\lr{1-u_{1246}}+{\rm Li_2}\lr{1-u_{2467}} +{\rm
Li_2}\lr{1-u_{2754}}-{\rm Li_2}\lr{1-u_{1245}}
\\ \notag & \qqquad
- \ln u_{1256} \ln u_{4571} +\ln u_{2467}\ln u_{1256} +\ln u_{2467} \ln u_{2754} -\frac{\pi^2}{6}\,,
\\ \label{VVII}
V_{\rm II} &  ={\rm Li_2}\lr{1-u_{1246}}+{\rm Li_2}\lr{1-u_{2467}} + \ln u_{1246}\ln
u_{2467}-\frac{\pi^2}{6}\,,
\end{align}
where ${\rm Li}_2$ is the dilogarithm function.
Substituting these relations into \re{n=7-rel} and \re{equiv} and taking into account that
$\mathbb{P}\, V_{146} = V_{146}$, we obtain
\begin{align} \notag
V_{\rm II}& = 0\,, \quad \text{for  $u_{1246} +u_{2467}  = 1$}\,,
\\[3mm] \label{n7-spur-canc}
V_{\rm I} - V_{\rm II}  &= 0\,, \quad \text{for $u_{4571}=\mathbb{P}^3 u_{1245}=1$}\,.
\end{align}
Indeed, replacing $u_{2467} = 1-u_{1246}$ in the expression for $V_{\rm II}$, Eq.~\re{VVII}, we find that  $V_{\rm II} = 0$ for arbitrary $u_{1246}$ in virtue of a well-known dilogarithm identity. Then, we use \re{VVII} to evaluate
\begin{align} \label{VI-VII}
 V_{\rm I} - V_{\rm II} &  = {\rm Li}_2 (1-u_{2754}) - {\rm Li}_2 (1-u_{1245}) - \ln u_{1256} \ln u_{4571}
  + \ln u_{2467}  \ln\frac{u_{1256} u_{2754}}{
    u_{1246} }\,.
\end{align}
The cross-ratios \re{u} satisfy the identity
$u_{2754} =  u_{4571} u_{4512} = u_{4571}  u_{1246}/u_{1256}$, which is used to simplify \re{VI-VII} as
\begin{align}
 V_{\rm I} - V_{\rm II} &  = {\rm Li}_2 (1-u_{1245} u_{4571}) - {\rm Li}_2 (1-u_{1245}) - \ln u_{4571}\ln(u_{1256}
  u_{2467}) \,.
\end{align}
Obviously, it vanishes for $u_{4571} =1$, in perfect agreement with \re{n7-spur-canc}.

In conclusion, in this subsection we have shown that the absence of spurious singularities imposes non-trivial restrictions on the loop corrections to the ratio function.

\subsection{Collinear singularities}\label{collsingu}


In the previous subsections, our consideration was based on the analysis of two types of singularities of the ratio function: The requirement that the spurious singularities cancel out in the tree-level ratio function fixes the relative coefficients in the expression for  $R_n^{\rm NMHV}$ as a linear combination of independent superinvariants whereas the additional unitarity condition (i.e., the correct behavior at multi-particle singularities) fixes the overall normalization. This completely determines the unique form of the tree-level NMHV superamplitude. As a consequence, it is guaranteed to have the expected collinear singular behavior \cite{Berends:1988zn,Mangano:1990by,Bern:1994zx,Dixon:1996wi}.  In this subsection we turn our attention to the collinear (two-particle) singularities and show that the
same values of the coefficients can be determined by imposing the collinear limit condition, without reference to the spurious or multi-particle singularities.

In the collinear limit, the on-shell momenta of two color-adjacent particles become
aligned $p_i \to z P_\ell$ and $p_{i+1} \to (1-z) P_\ell$ with  $P_\ell^2=0$ and the longitudinal momentum fraction $0 < z <1$. In this limit, generic color-ordered scattering
amplitudes have the following behavior at tree level~\cite{Mangano:1990by,Bern:1994zx,Dixon:1996wi}
\begin{align}\label{gluon-col}
A^{\rm tree}_n(\ldots, i^{\lambda_i},(i+1)^{\lambda_{i+1}},\ldots)
\stackrel{i\| i+1}{\to} \sum_{\lambda=\pm} {\rm Split}^{(0)}_{-\lambda}(z; i^{\lambda_i},(i+1)^{\lambda_{i+1}}) A^{\rm tree}_{n-1}(\ldots, \ell^{\lambda}, \ldots)\,,
\end{align}
where $i^{\lambda_i}$ denotes the $i-$th particle with helicity $\lambda_i$
and on-shell momentum $p_i$ and the sum runs over the helicities of the intermediate particle $ \ell^{\lambda}$ with momentum $P_\ell=p_i+p_{i+1}$. The splitting
amplitudes are universal functions depending on the quantum numbers
of two collinear particles only. For example, the six-gluon helicity-split NMHV amplitude
behaves in the collinear limit $6\|1$ as~\cite{Bern:1994cg}
\begin{align}\nn
A^{\rm NMHV;0}_6(1^+,2^+,3^+,4^-,5^-,6^-) & \stackrel{6\| 1}{\to}    {\rm Split}^{(0)}_-(6^-,1^+) \times A^{\rm MHV;0}_5(2^+,3^+,4^-,5^-,\ell^+)
\\[2mm] \label{6-col}
& + \  {\rm Split}^{(0)}_+(6^-,1^+) \times A^{\rm NMHV;0}_5(2^+,3^+,4^-,5^-,\ell^-)\,,
\end{align}
where $A^{\rm NMHV;0}_5$ coincides with the `googly' $\rm \overline{MHV}$ amplitude.
Notice that in the collinear limit the NMHV amplitude is expressed in terms of both
MHV and NMHV amplitudes. The same takes place for generic non-MHV amplitudes.

We recall that the amplitudes  $A_n(\ldots, i^{\lambda_i},(i+1)^{\lambda_{i+1}},\ldots)$ appear in the expansion of the $n-$particle {\it super}amplitude $\mathcal{A}_n(\ldots, i,(i+1),\ldots)$. It is straightforward to translate the relation \re{gluon-col} into the collinear limit of the
superamplitude. In the on-shell superspace approach, the limit $i\| i+1$
corresponds to~\cite{Drummond:2008cr}
\begin{align}\notag
& \lambda_i \to \sqrt{z}\, \lambda_\ell\,, &&  \tilde\lambda_i \to \sqrt{z}\, \tilde\lambda_\ell\,,
&&  \eta_i \to \sqrt{z}\, \eta_\ell\,,
\\[2mm] \label{col-def}
&  \lambda_{i+1} \to \sqrt{1-z} \lambda_\ell\,,&&  \tilde\lambda_{i+1} \to \sqrt{1-z}\, \tilde\lambda_\ell
 \,,&& \eta_{i+1} \to  \sqrt{1-z} \, \eta_\ell\,,
\end{align}
so that the exchanged  particle $\ell$ carries the total momentum ($P_\ell$) and supercharge ($ \lambda_\ell\, \eta_\ell$)
of the two particles,
 $p_i \to z P_\ell$, $p_{i+1} \to (1-z) P_\ell$ and $\lambda_i \eta_i + \lambda_{i+1}\eta_{i+1} \to \lambda_\ell\, \eta_\ell$. It should not be surprising that the collinear behavior
of color-order amplitudes \re{gluon-col} simplifies enormously when expressed in terms of superamplitudes. Indeed, in $\mathcal{N}=4$ SYM, all {tree-level} scattering amplitudes  (MHV, NMHV, ...) have the same universal collinear behavior~\cite{Drummond:2008cr}
\begin{align}\label{col-limit}
\mathcal{A}^{\rm tree}_n(\ldots,i,i+1,\ldots) \stackrel{i \| i+1}{\to} \frac{1}{\sqrt{z(1-z)}\vev{i\,i+1}}
\mathcal{A}^{\rm tree}_{n-1}( \ldots,\ell,\ldots) \,.
\end{align}
The superamplitude in the right-hand side of \p{col-limit} has one external particle less and it is obtained by replacing two collinear particles $i$ and $i+1$ by a single on-shell
state $\ell$  carrying the supermomentum $(P_\ell,\eta_\ell)$. Notice that the two
superamplitudes in \re{col-limit} are of the same type (MHV, NMHV, ...) which seems
to contradict to \re{6-col}. As we explain in Appendix \ref{apD}, the two relations, Eqs.~\re{6-col} and \re{col-limit}, are perfectly consistent with each other. Namely, all terms of  the `wrong' type (like the MHV contribution in the right-hand side of \re{6-col}) cancel out when summed up within the superamplitude.

The collinear factorization \re{gluon-col} also holds for color-ordered $n-$particle
amplitudes at loop level~\cite{Kosower:1999xi,Kosower:1999rx,Bern:1999ry,Anastasiou:2003kj,Bern:2004cz}. The corresponding all-loop  $\mathcal{N}=4$ splitting amplitudes are known to be proportional to the tree-level ones with the proportionality factor $r(s_{i,i+1},z,1/\epsilon)$
independent of the helicity $\lambda$ of the exchanged particle. Extending this result to superamplitudes, we find that their collinear limit is described by \re{col-limit}
with the only difference that the right-hand side of \re{col-limit} is multiplied by the universal all-loop function $r(s_{i,i+1},z,1/\epsilon)$. In other words, all superamplitudes (MHV, NMHV, $\ldots$) have the same collinear behavior. When translated into
properties of the ratio function, Eq.~\re{mhvfac}, this means that, firstly, the ratio function $R_n$ should be finite in the collinear limit and, most importantly, it should reduce to
the $(n-1)-$particle ratio function,
\begin{align}\label{reduction}
R_n(\ldots,i,i+1,\ldots) \stackrel{i \| i+1}{\to} R_{n-1}(\ldots,\ell,\ldots)\,.
\end{align}
We would like to emphasize that this relation should hold to all loops and for all
(NMHV, N${}^2$MHV,...) ratio functions. As we will show below, it imposes
strong constraints on the form of the ratio function.

Applying \re{reduction}, we can use the recurrence in $n$ to relate $R_n$ to the ratio function with the smallest possible number of particles for which it exists.
In the NMHV case, this number is $n=5$~\footnote{In general, for N${}^p$MHV the minimal value is $n=3+2p$.} and the ratio function \re{NMHV-tree} reads~\cite{Drummond:2008bq}
\begin{align}\label{R-exact}
R^{\rm NMHV}_5 = R_{135} \,.
\end{align}
Notice that this relation is exact to all loops. The reason for this is that the $n=5$ NMHV superamplitude is related to the $n=5$ MHV superamplitude through {PCT conjugation} and a Grassmann Fourier transform. However, the loop corrections to both superamplitudes are described by the same scalar function of dual coordinates, which drops out in the ratio function. Then,
it follows from \re{R-exact} and \re{reduction} that for $n=6$ the loop corrections to $R^{\rm NMHV}_6$ should vanish in the collinear limit and so on.

Let us verify that the relation \re{reduction} holds for one-loop NMHV ratio function $R_n^{\rm NMHV;1}$. In Sect.~\ref{phypo} we have shown that the superinvariants do not have singularities
in the two particle-invariant masses $x_{i,i+2}^2$ and, therefore, any linear combination of them will be regular in the collinear limit \re{col-def}. In addition, some of the superinvariants
vanish in this limit. To show this, let us make use of the cyclic symmetry of the $n-$particle ratio function and consider the collinear limit \re{col-def} with $i=n-1$, or, equivalently, $n-1\| n$. Examining the general expression for $R_{rst}$, Eq.~\re{R}, we
observe that for $s=n$ or for $t=n$ it is proportional to $\vev{n\, n-1}$ and, therefore,
it vanishes in the collinear limit
\be\label{zero1}
R_{rnt} = R_{rsn}=0\,,
\ee
with the indices $s$ and $t$ satisfying \re{Sn}. Similarly, for $r=n$ we
use the invariance of $R_{nst}$ under rescalings of $\lambda_n$ to replace
$\lambda_n\to  \lambda_\ell$,
\be\label{zero2}
R_{nst} = R_{\ell st}\,,\qquad R_{ns,n-1} = 0\,,
\ee
where $2\le s \le t+2\le n$ and the second relation follows from the vanishing of $\Xi_{ns n-1}$, Eq.~\re{Xi}. The remaining invariants $R_{rst}$ coincide, in the limit \re{col-def}, with the corresponding superinvariants for the $(n-1)-$particle
amplitude, with the only difference that the index $(n-1)$ is now associated with the line $\ell$. Applying \re{zero1} and \re{zero2},
it is straightforward to verify~\cite{Drummond:2008cr} that $R_n^{\rm NMHV;0}$ \re{NMHV-tree} indeed satisfies the relation \re{reduction},
\begin{align}\label{Rn}
R_n^{\rm NMHV;0} = \sum_{s,t\in \mathcal{S}_n} R_{nst}  =   \sum_{2\le s\le t-2\le n-2} R_{nst}  +\sum_{s=2}^{n-3}  R_{ns n-1} \stackrel{n-1\| n}{\to}  \sum_{s,t\in \mathcal{S}_{n-1}} R_{\ell st}  \,.
\end{align}
Here, in the first relation, we used the cyclic symmetry of the ratio function \re{NMHV-tree} to choose the first index of the superinvariant to be $n$.

Let us now return to the central question formulated at the beginning of this subsection: What is the general solution to \re{reduction} and how unique is the tree-level expression  \re{Rn} for the ratio function?
To answer this question, we have to expand $R_n^{\rm NMHV;0}$ over the complete set of independent superinvariants,
and then impose the condition \re{reduction}. We would like to emphasize that the number of independent
$R$'s is bigger than the number of terms $R_{nst}$ in \re{Rn}. For instance, for $n=6$ the general
expression for $R_6^{\rm NMHV;0}$, Eq.~\re{NMHV-67}, involves 6 terms,  against only 3 terms in \re{Rn}.
The two representations for the tree-level ratio function, Eqs.~\re{NMHV-tree} and \re{NMHV-67}, are equivalent, but starting from one loop
each independent superinvariant gets dressed by its own scalar function $V(u)$, Eqs.~\re{NMHV-678} and \re{R-loop}, and the simplicity
of the tree-level expression \re{NMHV-tree} is lost.

Let us substitute the general expression for the $n=6$ ratio function into \re{reduction} and examine the collinear
limit $5\| 6$. Making use of the identities \re{zero1} and \re{zero2}, we get
\begin{align}\label{R6-col}
R_6^{\rm NMHV;0} \stackrel{5\| 6}{\to }&\  \alpha \lr{R_{413}+R_{524}} = \alpha \lr{\mathbb{P}^3 R_{135}+\mathbb{P}^4  R_{135}} = 2\alpha \,R_5^{\rm NMHV;0} \,,
\end{align}
where in the right-hand side the index $5$ of the superinvariants $R_{524}$ and $R_{135}$ refers to the on-shell state $\ell$ and the cyclic invariance of the $n=5$ ratio function was used, $\mathbb{P}  R_{135}=R_{135}$. Imposing the condition \re{reduction}, we find $\alpha=1/2$, in accord with \re{sol6}. The generalization to
higher $n$ is straightforward. For $n=7$ we substitute \re{R7-ini} into \re{reduction} and go through the same steps to get, in the collinear limit $6\| 7$,
\begin{align}\notag
R_7^{\rm NMHV;0} & \stackrel{6\| 7}{\to }  \alpha \lr{R_{146}+R_{513}+R_{624}+R_{635}}
\\[2mm] \nn & \qquad
+ \beta \lr{R_{251} + R_{362}+R_{514}+R_{625}} + \gamma \lr{R_{413}+R_{524}+R_{635}}
\\[2mm] \label{R7-col}
& = \bigg[{(\alpha+\beta) +2\beta\, \mathbb{P} + (\alpha +\beta)\mathbb{P} ^2 + (\alpha +\gamma)\mathbb{P} ^3 + \gamma  \mathbb{P} ^4 + (\alpha +\gamma)\mathbb{P} ^5 }\bigg]R_{146}\,.
\end{align}
Here in the first relation we took into account \re{zero1} and \re{zero2} and in the second
one applied the identity \re{anothid}.
Comparing \re{R7-col} with the tree-level expression for the $n=6$ ratio function, Eqs.~\re{NMHV-tree} and \re{aux1},
\begin{align}
R_6^{\rm NMHV;0} = \lr{1+ \mathbb{P}^2 +\mathbb{P}^4}  R_{146} =\lr{\mathbb{P}+ \mathbb{P}^3 +\mathbb{P}^5}  R_{146} \,,
\end{align}
we find that the relation \re{reduction} is satisfied provided that the coefficients verify the relations
\begin{align}
 \alpha+3\beta-1= \alpha-2\beta+\gamma = \alpha+\beta-\gamma =0\,.
\end{align}
The solution to this linear system coincides with \re{sol7}. Repeating the same analysis for $n=8$, we also reproduce the correct expression for $R_8^{\rm NMHV;0}$, Eqs.~\re{R8} and \re{sol8}. Thus, we conclude that the relation \re{reduction} fixes the form of the tree level ratio function.

Let us now see how the collinear limit condition \re{reduction} works at loop level, by considering the
simplest example of the $n=6$ NMHV ratio function.
As we already explained, the relation \re{reduction} implies that the loop corrections
to $R_6^{\rm NMHV}$ should vanish in the collinear limit \re{col-def}. The one-loop
expression for the $n=6$ NMHV ratio function is given by \re{NMHV-678}. Substituting \re{NMHV-678} into \re{reduction}, we find that
the only difference compared to the tree-level calculation \re{R6-col}
is that each superinvariant  in the right-hand side of \re{R6-col} should be multiplied by the corresponding scalar function $V(u)$.
In this way, we obtain from  \re{R6-col}
\begin{align}\label{VV}
R_6^{\rm NMHV;1} \stackrel{5\| 6}{\to }&\  \alpha \lr{R_{413} V_{413}+R_{524} V_{524}} 
=   \alpha R_{135} \big[{V_{146}(u_1,u_2,u_3)+ V_{146}(u_2,u_3,u_1)}\big] \,,
\end{align}
where the scalar functions $V_{413}=\mathbb{P}^3 V_{146}$ and  $V_{524} = \mathbb{P}^4  V_{146}$ can be obtained from $V_{146}(u_1,u_2,u_3)$, Eq.~\re{V-def},
by a cyclic shift of its arguments, $V_{413}=V_{146}(u_1,u_2,u_3)$ and
$V_{524} =V_{146}(u_2,u_3,u_1)$.
The cross-ratios $u_i$, Eq.~\re{ui}, scale in the collinear limit $5\| 6$ (or equivalently $x_{51}^2\to 0$) as~\cite{Drummond:2007bm}
 \begin{align}\label{lim}
u_1 \to 0\,,\qquad u_2\to u \,,\qquad u_3\to 1-u\,.
\end{align}
{Then, the relation \re{reduction} implies that the right-hand side of \re{VV} should vanish in the limit \re{lim}
\be\label{col-zero}
\lim_{u_1\to 0}\big[ V_{146}(u_1,u,1-u)+ V_{146}(u,1-u,u_1)\big] = 0\,.
\ee
Notice that this relation should hold for the function $V_{146}(u_1,u_2,u_3)$ to all loops.}

To one loop, we apply \re{V-def}
and evaluate the sum of the two functions $V$ in the right-hand side of \re{VV} as
\begin{align}
 \ln u\ln (1-u) + {\rm Li}_2(u) + {\rm Li}_2(1-u) -\frac{\pi^2}6 = 0\,,
\end{align}
so that $R_6^{\rm NMHV;1}$ vanishes in the collinear limit, as expected. For $n=7$ it becomes a tedious exercise to verify that one-loop ratio function $R_7^{\rm NMHV;1}$, Eqs.~\re{NMHV-678} and \re{V7}, reduces $6\| 7$ to $R_6^{\rm NMHV;1}$ in the collinear limit, in agreement with \re{reduction}.

The fact that the one-loop NMHV ratio function, Eq.~\re{NMHV-678},
has the correct structure of collinear singularities should not be surprising.  The one-loop expressions for the functions $V$ were first obtained in \cite{Drummond:2008vq} by matching
the general expression for the NMHV superamplitude with the known one-loop  $n-$particle NMHV gluon amplitudes~\cite{Bern:1994cg,Bern:2004ky,Bern:2004bt}, which do have correct analytic properties. They
were later reproduced in \cite{Drummond:2008bq} from generalized unitarity, where the
analytic properties of the amplitudes are built in from the start.

\subsection{Spurious poles cancellation versus collinear factorization}\label{spcvcf}

In this section, we have demonstrated that the tree-level expression for the NMHV ratio function
is uniquely fixed either by imposing the condition of spurious pole cancellation
supplemented with multi-particle factorization, or by requiring the amplitude to have the
correct behavior in the two-particle collinear limit. The two approaches are different but they lead to the same expression for $R_n^{\rm NMHV;0}$.
The important difference is that the fist approach only uses the properties of the superinvariants with a fixed number of particles $n$. At the same time, the second approach is recursive, to determine the $n-$particle ratio function one needs to
know the $(n-1)-$particle ratio function, etc.

A natural question is how powerful the two approaches are after including loop corrections to the ratio function.  We have shown that, for the NMHV ratio function, the two approaches lead to different constraints on the scalar functions $V(u)$, c.f. Eqs.~\re{zero} and \re{col-zero}.  It remains unclear which one is more restrictive but there is strong evidence that
the collinear constraints are not powerful enough. Indeed, the collinear factorization
was one of the main motivations for the BDS ansatz \cite{Bern:2005iz} for the all-loop MHV amplitudes. It was later shown in \cite{Drummond:2008aq,Bern:2008ap} that the BDS ansatz fails for the six-gluon two-loop MHV amplitude. The deviation of the  true amplitude from the BDS form is the `remainder function' which  vanishes in the
collinear limit and, therefore, can not be determined from the collinear factorization
condition alone.

\section{Dual supersymmetry and holomorphic anomaly}\label{dsaha}

As we already emphasized in the previous section, the ratio function $R_n^{\rm NMHV}$  is a dual superconformal invariant at tree level only. At loop level the perturbative
corrections to $R_n^{\rm NMHV}$ involve the scalar functions $V_{rst}$ which depend on
the dual space coordinates $x_i$ through the conformal cross-ratios \re{u}. The latter are dual conformal, but not {\it super}conformal invariants. \footnote{In fact, in a chiral  superspace $(x^{\a\da},\q^{A\, \a})$ there exists no supersymmetric invariant interval. As a consequence, it is impossible to construct regular supersymmetric invariants, other than the exceptional delta function ones that we find in the tree-level superamplitudes.}
This suggests that the loop corrections to  $R_n^{\rm NMHV}$ break the
dual Poincar\'e (or, equivalently, ordinary special conformal) supersymmetry with generator $\bar Q^A_{\da} \equiv \bar s^A_{\da}$ (see \p{dualbarq}).
This phenomenon was first pointed out in \cite{Drummond:2008vq}. In this section we
show that its origin  can be traced back to the
so-called holomorphic anomaly of the scattering amplitudes \cite{Cachazo:2004by,Bena:2004xu,Cachazo:2004dr}.

\subsection{Dual supersymmetry anomaly of the MHV superamplitude}

To illustrate the main idea of our approach, let us first consider the one-loop corrections to the  $n-$particle MHV superamplitude. At tree level,  $\mathcal{A}_n^{\rm MHV;0}$ is a (dual)superconformal covariant while at loop level this symmetry is broken by infrared divergences. In addition, $\mathcal{A}_n^{\rm MHV;0}$ only has poles in the two-particle invariant masses $s_{i,i+1}$, while at loop level it develops discontinuities (branch cuts) with respect to the multi-particle invariant masses $s_{i\ldots j}=(p_i+\ldots+p_{j})^2$ with $j-i\ge 3$. At one loop, the latter are given by the two-particle cut
\begin{align}\label{disc-1-loop}
{\rm Disc}_{s_{1\ldots j}} \mathcal{A}_n^{\rm MHV;1}  = \mathcal{A}^{\rm MHV;0}(-\ell_1,1,\ldots,j,-\ell_2)
\star \mathcal{A}^{\rm MHV;0}(\ell_2,j+1,\ldots,n,\ell_1)\,,
\end{align}
where we have used the same notations for the on-shell superstates $\ell_i$ and $(-\ell_i)$ as in \re{NMHV-disc} and  `$\star$' stands for the integration over the phase space of the two superstates
propagating through the cut with the following measure
\be\label{measure}
(2\pi)^4\int \frac{d^4 \ell_1}{(2\pi)^3} \delta_+(\ell_1^2)\, d^4 \eta_{\ell_1}
\int \frac{d^4 \ell_2}{(2\pi)^3} \delta_+(\ell_2^2)\, d^4 \eta_{\ell_2}\,,
\ee
where $\delta_+(\ell^2) = \theta(\ell_0)\delta(\ell^2)$.

\subsubsection{Dual conformal symmetry of the discontinuity}\label{dcsotds}

For the one-loop gluon MHV amplitude,  the discontinuity \re{disc-1-loop} was computed in \cite{Bena:2004xu}.
Since the loop corrections to MHV amplitudes have a universal form independent of the types of scattered particles, we can easily generalize the results of \cite{Bena:2004xu} to the MHV superamplitude as follows,
\begin{align}\label{disc-IR-finite}
& {\rm Disc}_{s_{1\ldots j}} \mathcal{A}_n^{\rm MHV;1}  = 2\pi i c_\Gamma \mathcal{A}_n^{\rm MHV;0}  \theta(s_{1\ldots j})
 \ln \bigg[ \frac{(s_{n...j-1}s_{1...j}-s_{1 ...j-1}s_{n...j})}{ (s_{1...j-1}s_{2...j}-s_{1 ...j}s_{2...j-1})}
  \frac{(s_{2...j+1}s_{1...j}-s_{1 ...j+1}s_{2...j})}{( s_{n...j}s_{1...j+1}-s_{1 ...j}s_{n...j+1})}
 \bigg],
\end{align}
with $c_\Gamma = 1/(4\pi)^2 + O(\epsilon)$. We notice that unlike $\mathcal{A}_n^{\rm MHV;1}$ itself, its discontinuity \re{disc-IR-finite} is infrared finite~\footnote{At the same time the
two-particle invariant mass discontinuity ${\rm Disc}_{s_{i,i+1}} \mathcal{A}_n^{\rm MHV;1}$ is infrared singular.} and, most importantly, it is dual conformally invariant. This property becomes manifest after one replaces $s_{ij}=x_{i,j+1}^2$ and rewrites \re{disc-IR-finite} in terms of the dual cross-ratios \re{u},
\begin{align}\label{agree}
& {\rm Disc}_{x_{1,j+1}^2} \mathcal{A}_n^{\rm MHV;1}  =  2\pi i c_\Gamma \mathcal{A}_n^{\rm MHV;0}\ln \bigg[ \frac{(1-u_{1,n,j,j+1})(1-u_{1,2,j+2,j+1})}{(1-u_{1,2,j,j+1})(1-u_{1,n,j+2,j+1})}\bigg]\,.
 \end{align}
{So, dual conformal symmetry is
restored in the infrared finite discontinuity of $\mathcal{A}_n^{\rm MHV;1}$, but dual supersymmetry is still broken due to the non-trivial dependence on the dual coordinates $x_i$. We may ask the question: Where does this difference in the two symmetries come from?

The restoration of dual conformal invariance in ${\rm Disc}_{s_{1\ldots j}} \mathcal{A}_n^{\rm MHV;1}$  can be explained in two ways.} Firstly, we may use the remarkable duality \cite{Alday:2007hr,Drummond:2007aua,Brandhuber:2007yx} between the MHV superamplitude $\mathcal{A}_n^{\rm MHV} $ and a light-like $n-$gon Wilson loop
$W_n$,
\be\label{W-dual}
\mathcal{A}_n^{\rm MHV} =\mathcal{A}_n^{\rm MHV;0} \, W_n\,,\qquad W_n = 1+ a W_{n}^{(1)} + O(a^2)  \,,
\ee
so that $\mathcal{A}_n^{\rm MHV;1} = \mathcal{A}_n^{\rm MHV;0}\,W_{n}^{(1)}$ and so on.
{From this relation and from the fact that $\mathcal{A}_n^{\rm MHV;0}$ is dual conformal, it follows that the dual conformal behavior of $\mathcal{A}_n^{\rm MHV}$ can be derived from the conformal properties of the Wilson loop $W_n$, regarded as an $\mathcal{N}=4$ SYM correlator in dual space.} The conformal invariance of $W_n$ is broken by
cusp ultraviolet singularities and, as a consequence, $W_n$ satisfies an anomalous conformal Ward identity \cite{Drummond:2007au}
\begin{align} \notag
K^{\da\a}\, \ln W_n &=  \sum^n_{i=1} (2x_i^{\da\a}x_i\cdot\pa_i - x_i^2 \pa_i^{\da\a}) \ln W_n
\\ \label{anom}
  &=  -\sum_{l\ge 1}a^l \lr{\frac{\Gamma_{\rm cusp}^{(l)}}{ l\ep
}+  {\Gamma^{(l)}} }\sum_{i=1}^n\ \lr{-x_{i-1,i+1}^2\mu^2}^{l\ep} \ x^{\da\a}_i  + O(\epsilon) \,.
\end{align}
Here $K^{\da\a}$ generates conformal boosts of the dual coordinates, and
$\Gamma_{\rm cusp}^{(l)}$ and $\Gamma^{(l)}$ denote the perturbative values
of the cusp and collinear anomalous dimensions, respectively. Let us now consider the discontinuity of both sides of Eq.~\re{anom} with respect to $x_{1,j+1}^2$ (with $3\le j\le n-2$). Since the right-hand side only involves two-particle invariant masses $x_{i-1,i+1}^2=(p_{i-1}+p_i)^2$, we find that
\be\label{K-norm}
{\rm Disc}_{x_{1,j+1}^2} (K^{\da\a}\, \ln W_n) = K^{\da\a}( {\rm Disc}_{x_{1,j+1}^2}  \ln W_n) =0\,,
\ee
thus implying that the discontinuity of
$ \ln W_n=a  W_n^{(1)} + O(a^2)$ is a dual
conformal invariant. Since $ {\rm Disc}_{x_{1,j+1}^2} \mathcal{A}_n^{\rm MHV;1} = \mathcal{A}_n^{\rm MHV;0}\, {\rm Disc}_{x_{1,j+1}^2} W_{n}^{(1)}$,
we conclude that the ratio of the discontinuity of $\mathcal{A}_n^{\rm MHV;1}$ and the tree MHV superamplitude $\mathcal{A}_n^{\rm MHV;0}$ is also a
dual conformal invariant.

We would like to stress that \re{K-norm} only holds for multi-particle invariant mass discontinuities. The two-particle ones have a dual conformal anomaly  (see \p{anom}),
\be\label{K-norm1}
K^{\da\a}( {\rm Disc}_{x_{i-1,i+1}^2}  \ln W_n) =-2\pi i\, \Gamma_{\rm cusp}(a) x_i^{\da\a} \,,
\ee
controlled by the cusp anomalous dimension $\Gamma_{\rm cusp}(a)$ (see \cite{Basso:2009gh} and references therein).

{An alternative way to study the anomalies of the symmetries of  ${\rm Disc}_{s_{1\ldots j}} \mathcal{A}_n^{\rm MHV;1}$ is to directly act with the generators  on both sides of Eq.~\re{disc-1-loop}. The dual superconformal generators  are linear
differential operators, so we find the variation
\begin{align} \notag
\delta\lr{ {\rm Disc}_{s_{1\ldots j}} \mathcal{A}_n^{\rm MHV;1} } & = \delta \mathcal{A}^{\rm MHV;0}((-\ell_1),1,\ldots,j,(-\ell_2))
\star \mathcal{A}^{\rm MHV;0}(\ell_2,j+1,\ldots,n,\ell_1)
\\[2mm] \label{delta-A}
&+\mathcal{A}^{\rm MHV;0}((-\ell_1),1,\ldots,j,(-\ell_2))
\star\delta \mathcal{A} ^{\rm MHV;0}(\ell_2,j+1,\ldots,n,\ell_1)\,,
\end{align}
provided that the integration measure over the supermomenta \re{measure} is {covariant}
and its conformal weight compensates the conformal weights of the superamplitudes
corresponding to the cut legs.
In what follows we will apply this method to the two principal generators of the dual $su(2,2|4)$ algebra, $K^{\da\a}$  and $\bar Q^A_\da$, defined in \p{anom} and \p{dualbarq}, respectively.%
\footnote{The {covariance} of the measure \re{measure} under $K^{\da\a}$  and $\bar Q^A_\da$ has been proven in   \cite{Brandhuber:2008pf,Elvang:2009ya}.} In this way, we will confirm the absence of a dual conformal anomaly (see \re{agree} and \p{K-norm}), but we will reveal the anomaly of dual Poincar\'e supersymmetry.

Where can an anomaly come from?} The tree-level superamplitudes are invariant under dual superconformal transformations, so we would (naively) expect that $\delta \mathcal{A}^{\rm MHV;0}=0$ implies $\delta\lr{ {\rm Disc}_{s_{1\ldots j}} \mathcal{A}_n^{\rm MHV;1} }=0$. This property
is however affected by the so-called holomorphic anomaly \cite{Cachazo:2004by}. The MHV superamplitude \p{MHVsua} involves singular functions of the $\lambda$'s of the type $\vev{i\, i+1}^{-1}$. They are only  naively holomorphic, but  in reality
\begin{align}
\frac{\partial }{\partial \tilde\lambda^{\dot\alpha}} \frac{1}{\vev{\lambda\chi }}
=2\pi \tilde\chi_{\dot\alpha} \delta\lr{\vev{\lambda,\chi}}\delta({[\tilde\lambda,\tilde\chi]})\,.
\end{align}
Since the generators $K^{\da\a}$ and $\bar Q^A_\da$ involve derivatives with
respect to $\bl$, they act nontrivially on the denominator of the MHV superamplitude. A simple calculation shows that
\begin{align} \notag
K_{\alpha\dot\alpha} \frac1{\vev{i\, i+1}} &= \sum_{k} (x_{k+1})_{\alpha\dot\beta}
\tilde\lambda_{k\dot\alpha}\partial_{k}^{\dot\beta} \frac1{\vev{i\, i+1}}
\\ \notag
& = 2\pi \delta(\vev{i\, i+1})\delta([i i+1]) \lr{\tilde\lambda_{i\dot\alpha} {([i+1|x_{i+1})}_\alpha -\tilde\lambda_{i+1\dot\alpha} {([i|x_{i+2})}_\alpha}
\\ \notag
& = 2\pi \delta(\vev{i\, i+1})\delta([i i+1]) \lr{\tilde\lambda_{i\dot\alpha} {([i+1|x_{i+2})}_\alpha -\tilde\lambda_{i+1\dot\alpha} {([i|x_{i+2})}_\alpha}
\\[2mm]
& = 2\pi \delta(\vev{i\, i+1})\delta([i i+1]) [i i+1]  (x_{i+2})_{\alpha\dot\alpha} =0\,.
\end{align}
In a similar manner,
\begin{align} \notag
\bar Q^A_{\dot\alpha}\frac1{\vev{i\, i+1}} &= \sum_{k} \eta_{k}^A\partial_{k\dot\alpha} \frac1{\vev{i\, i+1}}
\\ \label{Q-anom}
& = 2\pi \delta(\vev{i\, i+1})\delta([i i+1]) \lr{\eta_{i}^A\tilde\lambda_{i+1\dot\alpha} -\eta_{i+1}^A\tilde\lambda_{i\dot\alpha}}\,.
\end{align}
We conclude that, unlike the supersymmetry generator $\bar Q$, the conformal one $K_{\alpha\dot\alpha}$ is not affected by the holomorphic anomaly and,
therefore, $K_{\alpha\dot\alpha}\lr{ {\rm Disc}_{s_{1\ldots j}} \mathcal{A}_n^{\rm MHV;1} }=0$
in agreement with \re{agree} and \p{K-norm}. {Another way to formulate this is that dual conformal symmetry is an {\it exact} symmetry of the MHV tree amplitude and it becomes anomalous at loop level (due to infrared singularities), while dual $\bar Q-$supersymmetry is anomalous already at tree level (see also \cite{Bargheer:2009qu} for an extensive discussion of this anomaly).}

We wish to point out that, unfortunately, the Ward identity method, so useful in the study of the dual conformal anomaly, cannot be applied to the dual $\bar Q-$supersymmetry. The reason is that at present we do not have a dual field theory model for {\it super}amplitudes. The Wilson loop dual to the (loop corrections to) MHV amplitudes is only a conformal, but not a {\it super}conformal correlator in dual space.

\subsubsection{Holomorphic anomaly and the breakdown of dual supersymmetry}

Let us now compute $\bar Q^{\dot\alpha,A}\lr{ {\rm Disc}_{s_{1\ldots j}} \mathcal{A}_n^{\rm MHV;1} }$
by using \re{delta-A} and by taking into account the holomorphic anomaly contribution \re{Q-anom}. In Sect.~\ref{cslsts} we have shown  that the action of the generator $\bar s \equiv \bar Q$, Eq.~\p{dualbarq},
on the MHV superamplitude can be expressed in terms of the collinearity operators $F$
defined in \re{coll}. The holomorphic anomaly in the action of the collinearity operator on the  one-loop MHV gluon amplitudes
was studied in \cite{Cachazo:2004by}.  For MHV superamplitudes the analysis goes along the same lines.
The anomalous contributions of the type \re{Q-anom} come from the denominator in the MHV superamplitudes,
\begin{align}\label{anomconbqu}
\bar Q^{A} _{\da}\mathcal{A}^{\rm MHV;0}((-\ell_1),1,\ldots,j,(-\ell_2)) \sim
\bar Q^{A} _{\da}\lr{\frac1{\vev{\ell_1 1} \vev{12}\ldots \vev{j \ell_2} \vev{\ell_2\ell_1}}}\,,
\end{align}
in the kinematical configurations where two neighboring momenta are aligned. Assuming that $s_{i,i+1}\neq 0$ for the external legs, we are left with three such configurations corresponding
to  $\delta(\vev{\ell_1 1})\delta([\ell_1 1])$,  $ \delta(\vev{\ell_2 j})\delta([\ell_2 j])$ and $ \delta(\vev{\ell_1 \ell_2})\delta([\ell_1 \ell_2])$. In addition, the integration over $\eta_{\ell_i}$ in \re{delta-A} produces the factor
\begin{align}\label{gr}
\int d^4\eta_{\ell_1} d^4\eta_{\ell_2} \delta^{(8)}(\sum_1^j \eta_k \lambda_k-\eta_{\ell_1} \lambda_{\ell_1}-\eta_{\ell_2} \lambda_{\ell_2}) = \vev{\ell_1 \ell_2}^4\,,
\end{align}
which suppresses  the third configuration corresponding
to the vanishing of the multi-particle invariant mass $(p_1+\ldots+p_j)^2$. Similarly,
$\bar Q^{A} _{\da}\mathcal{A}^{\rm MHV;0}(\ell_2,j+1,\ldots,n,\ell_1)$ is localized at two
kinematical configurations in which the momenta of the particles $\ell_2$ and $\ell_1$ are aligned with the external momenta of the particles $j+1$ and $n$, respectively.

Let us first consider the contribution coming from  $\delta(\vev{\ell_1 1})\delta([\ell_1 1])$.
The delta functions imply that the momenta $\ell_1$ and $p_1$ are collinear and
the on-shell condition  $\ell_2^2=(P-\ell_1)^2=0$ (with $P=\sum_1^j p_i = x_{1,j+1}$) leads to
\begin{align}
\ell_1 =  c \, p_1\,,\qquad c= \frac{P^2}{2(p_1 P)}= \frac{P^2}{\bra{1}P|1]}\,.
\end{align}
Following \cite{Bena:2004xu} we evaluate the first line in \re{delta-A} as
\begin{align}\label{rel-Q1}
2\pi i c_{\Gamma}\mathcal{A}_n^{\rm MHV;0}\bigg[
\frac{ P^2\vev{j j+1}[n 1]}{\bra{j} P|1]\bra{j+1} P|1][n\ell_1]}\lr{\eta_{1}^A\tilde\lambda_{\ell_1\dot\alpha} -\eta_{\ell_1}^A\tilde\lambda_{1\dot\alpha}} + \ldots\bigg]\,,
\end{align}
where the dots stand for the three remaining terms coming from  $\ell_1 \sim p_n$,   $\ell_2\sim p_j$ and   $\ell_2\sim p_{j+1}$. Then we use the Grassmann delta function in \re{gr} to replace  $\eta_{\ell_1} = -\sum_1^s \eta_i {\vev{i\ell_2}}/{\vev{\ell_1\ell_2}}$ and we obtain, after some algebra,
\begin{align}\label{rel-Q2}
P^2\frac{ [n 1]}{ [n\ell_1]}\lr{\eta_{1}^A\tilde\lambda_{\ell_1\dot\alpha} -\eta_{\ell_1}^A\tilde\lambda_{1\dot\alpha}} = \lr{P^2\eta_{1}^A -  \sum_1^j\eta_i^A \bra{i}P |1]}\tilde\lambda_{1\dot\alpha} = {\Xi_{2,j+1,1}^A}\tilde \lambda_{1\dot\alpha}/\vev{12}\,,
\end{align}
where  $\Xi_{2,j+1,1}$ is  defined in \re{Xi}. Putting together \re{rel-Q1} and \re{rel-Q2} and adding the contributions from the other three collinear configurations, we finally
get
\begin{align}  \notag
\bar Q^A_{\dot\alpha} \big( {\rm Disc}_{x_{1, j+1}^2}  &  \mathcal{A}_n^{\rm MHV;1}\big) = - 2\pi i c_\Gamma \mathcal{A}_n^{\rm MHV;0}
\\[2mm]  \notag
\times
\Bigg[ &
\frac{\tilde\lambda_{1\dot\alpha}\vev{j j+1}\, \Xi_{2,j+1,1}^A }{\vev{12}\bra{j}x_{j1}|1][1|x_{1,j+1}\ket{j+1}}
-  \frac{\tilde\lambda_{n\dot\alpha}\vev{j j+1}\,\Xi_{1,j+1,n}^A\vev{j j+1}}{\vev{n1}\bra{j}x_{jn}|n][n|x_{n,j+1}\ket{j+1}}
\\  \label{Qbar-MHV}
-&\frac{\tilde\lambda_{j\dot\alpha}\vev{n1}\, \Xi_{j+1,1,j}^A }{\vev{j j+1}\bra{n}x_{nj}|j][j|x_{j1}\ket{1}}
+\frac{\tilde\lambda_{j+1\dot\alpha}\vev{n1}\,\Xi_{j+2,1,j+1}^A}{\vev{j+1,j+2}\bra{n}x_{n,j+1}|j+1][j+1|x_{j+1,1}\ket{1}}
  \Bigg]\,.
\end{align}
Clearly, the four $\Xi$'s in \p{Qbar-MHV} are linearly independent, so the right-hand side cannot vanish.
It is not hard to verify that the same result is obtained by applying  $\bar Q^A_{\dot\alpha}$ directly to the known expression \re{agree} for the one-loop discontinuity.

{The relation \re{Qbar-MHV} gives the  $\bar Q-$anomaly of the discontinuity of the one-loop MHV superamplitude. We notice that although each term in the right-hand side of \re{Qbar-MHV} is dual conformally covariant, they transform with different dual conformal weights. Thus, the complete anomaly is not dual conformal. This is not surprising, since the generators $\bar Q$ and $K$ do not commute. Using the fact that $K^{\a\da}( {\rm Disc}_{x_{1, j+1}^2}  \mathcal{A}_n^{\rm MHV;1})=0$, from  \re{Qbar-MHV} we can also derive the anomaly in the dual superconformal generator $S=[K,\bar Q]$ (see the algebra \p{comm-rel}).}

In summary, in this subsection we have demonstrated that the  $\bar Q-$anomaly of the discontinuity
of the one-loop MHV superamplitude originates from the holomorphic anomaly of the tree-level amplitude. This implies
that the $\bar Q-$anomaly is also present in the MHV superamplitude itself, but in this
case the infrared divergences provide an additional contribution to the anomaly. In the next subsection, we extend our analysis to the NMHV superamplitude and in particular to the
$\bar Q-$anomaly of the ratio function.

\subsection{Dual supersymmetry anomaly of the NMHV superamplitude}\label{dsans}

As we have already pointed out, the NMHV superamplitude introduces a new object, the ratio function $R_6^{\rm NMHV}$, Eq.~\p{mhvfac}. It is dual superconformal at tree level, but only dual conformal at one-loop level. The mechanism of the breakdown of dual supersymmetry is similar to that in the MHV case. Here we present the results for the simplest $n=6$ NMHV superamplitude, the details of the calculation are given in Appendix \ref{C}.

As before, dual conformal symmetry is not affected by the holomorphic anomaly, but dual supersymmetry (or ordinary conformal supersymmetry) is. The computation of the $\bar Q-$anomaly of the discontinuity of the amplitude gives
%
\begin{align}  \notag
\bar Q^A_{\dot\alpha} \big( {\rm Disc}_{x_{14}^2} & \mathcal{A}_6^{\rm NMHV;1} \big) = 2\pi i c_{\Gamma} \mathcal{A}_6^{\rm MHV;0} R_{146}
\\ \label{Qb-NMHV}
&\times
 \lr{\eta_1 [23] + \eta_2[31]+\eta_3[12]}^A
\Bigg(
   \frac{\tilde\lambda_{1\dot\alpha}[6|x_{63}\ket{3} }{x_{14}^2 [61][12]}
+
 \frac{\tilde\lambda_{3\dot\alpha}[4|x_{41}\ket{1}}{x_{14}^2[23][34]}
\Bigg)
+ (i\to i+3)\,.
\end{align}
We have also directly verified that the known one-loop expression for the $n=6$ NMHV superamplitude indeed satisfies the relation \re{Qb-NMHV}.

It is instructive to compare \re{Qb-NMHV} with the similar relation for the one-loop $n=6$ MHV superamplitude \re{Qbar-MHV},
%
\begin{align} \notag
\bar Q^A_{\dot\alpha} \big( {\rm Disc}_{x_{14}^2} & \mathcal{A}_6^{\rm MHV;1} \big)  = 2\pi i  c_{\Gamma} \mathcal{A}_6^{\rm MHV;0}
\\ \label{Qb-MHV6}
&\times\lr{\eta_1[23]+\eta_2[31]+\eta_3[12]}^A
 \lr{\frac{\tilde\lambda_{1\dot\alpha}\vev{34}}{\bra{4}x_{41}|1][12]}+ \frac{\tilde\lambda_{3\dot\alpha}\vev{61}}{\bra{6}x_{63}|3][23]}}+ (i\to i+3)\,.
\end{align}
We observe that the two expressions have many features in common. In particular,
they involve  the same combinations of $\eta$'s, and the index $\da$ is carried by the spinors $\tilde\lambda_4$,
$\tilde\lambda_6$ and $\tilde\lambda_1$, $\tilde\lambda_3$,
which correspond to the external particles adjacent to the cut.

From these two results and from the definition \p{mhvfac}, we obtain the anomaly of the discontinuity of $R_6^{\rm NMHV;1}$ in the channel $x^2_{14}= (p_1+p_2+p_3)^2$:
%
\begin{align}\nn
\bar Q^A_{\dot\alpha} \big( {\rm Disc}_{x_{14}^2} & {R}_6^{\rm NMHV;1} \big)
=2\pi i c_{\Gamma}   \left[1+\xi(u)\right] R_{146}
\\ \label{power}
&\times
 \lr{\eta_1 [23] + \eta_2[31]+\eta_3[12]}^A
\Bigg(
   \frac{\tilde\lambda_{1\dot\alpha}[6|x_{63}\ket{3} }{x_{14}^2 [61][12]}
+
 \frac{\tilde\lambda_{3\dot\alpha}[4|x_{41}\ket{1}}{x_{14}^2[23][34]}
\Bigg)
+ (i\to i+3)
\,,
\end{align}
where
$
\xi
=[(u_{3461}+u_{1245}-1)/u_{2356}-1]^{-1}
$
is a dual conformal invariant. Comparing this relation with \re{Qb-NMHV} we notice
that the $\bar Q-$anomaly of the discontinuity of the $n=6$ NMHV superamplitude and of the ratio
function are in fact related to each other and only differ by a simple dual conformally invariant function. It would be interesting to work out the generalization for $n=7$, where the NMHV superamplitude involves several different (super)helicity structures.
Finally,  we replace ${R}_6^{\rm NMHV;1}$ in \re{power} by its expression \re{NMHV-678} and take into account $\bar Q-$invariance of $R_{146}$ to get
\begin{align}
\bar Q^A_{\dot\alpha} \big( {\rm Disc}_{x_{14}^2} & {R}_6^{\rm NMHV;1} \big)
= \frac12 R_{146} \bar Q^A_{\dot\alpha} \big( {\rm Disc}_{x_{14}^2}  {V}_{146} \big)
+ \text{cyclic}\,.
\end{align}
Matching this relation into \re{power} we can read off expression for
$ \bar Q^A_{\dot\alpha} \big( {\rm Disc}_{x_{14}^2}  {V}_{146} \big)$.
The fact that it is different from zero means that, in agreement with our
expectations, the scalar function
$V_{146}(u_1,u_2,u_3)$ does not respect dual $\bar Q-$supersymmetry.

\section{Discussion and concluding remarks}\label{concu}

In this paper we have studied the implications of the conventional and
dual superconformal symmetries for scattering amplitudes
in planar $\mathcal{N}=4$ SYM theory. At tree level, both symmetries are exact and, as a consequence, the ratio
functions
$R_n^{{\rm N}^p{\rm MHV}}$ should be expressed in terms of invariants of
these symmetries. For the NMHV superamplitude the invariants
$R_{rst}$ have a relatively simple form, but for the N${}^p$MHV superamplitudes (with $p=2, \ldots, n-4$)
their form becomes much more complicated.  The full expression of the ${\rm N}^p{\rm MHV}$ tree-level superamplitudes has been obtained by solving the supersymmetric version
of the BCFW recurrence relations \cite{Drummond:2008cr}. Each term in it is manifestly invariant under dual superconformal symmetry. They are also invariant under ordinary superconformal symmetry, which can be proven along the lines of Section \ref{csmnmta} (see \cite{toappear}).  One may wonder what is the most
general form of the $n-$point invariants of both symmetries and whether the set of invariants entering the tree-level
N${}^p$MHV  superamplitudes is complete. The answer will require  
a systematic classification of all invariants of the two symmetries. We should mention that a new type of dual superconformal invariant \footnote{The question if it is also an invariant of  ordinary superconformal symmetry  has not been investigated yet.} appears
for N${}^2$MHV and more complicated amplitudes at loop level \cite{Drummond:2008bq}. It
corresponds to the four-mass
scalar boxes which do not contribute to the tree amplitude. In this sense, the NMHV amplitudes  play a special role since the
one-loop
corrections do not induce new invariants that are not seen at tree level.

Even if we know the complete set of invariants, we are still left with the question what else is needed to fix the coefficients in the {\it unique} linear combination of these invariants, which corresponds to the amplitude. In this paper we argued that this additional information comes from
studying the analytic properties of the amplitude. We demonstrated that
two different approaches - one based on the cancellation of spurious poles,
supplemented with the multi-particle factorization property, and
another one relying on the correct collinear limit of the ratio function
\re{reduction},
lead to the same expression for the NMHV tree superamplitude. It would be
interesting to extend the analysis to the general case of N${}^p$MHV superamplitudes.

Turning on loop corrections to the scattering amplitudes, we find that both
conventional and dual superconformal symmetries are broken. The anomalous
contributions come from
two different sources, from the infrared divergences and from the holomorphic
anomaly of the tree amplitudes. Further progress in the understanding of the
all-loop $\cN=4$ scattering amplitudes is ultimately related to our ability to
control the anomalies of
both symmetries. For the conventional symmetry, the above mentioned anomalous
contributions
affect the conformal boosts ($k$) and special conformal supersymmetry ($s$ and $\bar
s$) generators, but their anomalies are not independent due to  $\{ s , \bar
s\} = k$.
For the dual symmetry, the anomaly affects the generators $K$, $\bar Q$ and
$S$, but the latter
is not independent due to $[K,\bar Q] = S$. Therefore, taking into account
that $\bar s = \bar Q$,
the list of independent anomalous generators for both symmetries includes
dual conformal
boosts $K$, dual supersymmetry $\bar Q$ and ordinary conformal supersymmetry $s$.
All of them
become anomalous because of infrared divergences.

We can circumvent the infrared divergences by considering the finite ratio
function defined in \re{mhvfac}. Quite remarkably, the dual conformal
anomaly cancels in the ratio of superamplitudes and the $K-$symmetry is
restored. However, this is not the case for the dual Poincar\'e  $\bar
Q-$supersymmetry. We demonstrated by an explicit one-loop calculation that
the NMHV ratio function is not invariant
under the action of $\bar Q$, and that the breakdown of dual $\bar
Q-$supersymmetry (but not of dual conformal symmetry) can be traced back
to the holomorphic anomaly of the tree amplitudes.
Since $\bar Q \equiv \bar s$, this also means that the ordinary special
conformal supersymmetry and, as a consequence, ordinary conformal symmetry
become anomalous. Such broken symmetries potentially lose their predictive
power, unless we can understand and control their anomalies.

An example where we have been able to make all-loop predictions, based on an
anomalous symmetry, is dual conformal symmetry. Through the  MHV
amplitude/Wilson loop duality we can trace the origin of this anomaly to
the cusp singularities of the light-like Wilson loop. These singularities
are localized at the cusp points, which makes the analysis of their
conformal behavior relatively easy. Moreover, the Wilson loop is a
correlator in dual space, which is calculated from a conformal Lagrangian.
The breaking of conformal symmetry is due to the ultraviolet regulator,
and the Lagrangian formalism provides us with the standard tools of the
Ward identities.  All of this resulted in a very simple all-loop dual
conformal anomalous Ward identity \cite{Drummond:2007au}.

One possible way to control the anomalies of scattering amplitudes would
be to extend
the above mentioned duality to non-MHV amplitudes and to identify the dual
object
describing the ratio function. Such an object should necessarily incorporate
the particle helicities and
it should be a function of the supercoordinates of the $n$ particles in the dual
superspace $Z_i=(x_i,\theta_i)$ (with $i=1,\ldots,n$).
In addition, as follows from our analysis, it should also have a number of
unusual features.
First of all, due to the $\bar Q-$anomaly, this object should
not be super-Poincar\'e
invariant at loop level. Secondly, the collinear limit of the ratio function
\re{reduction} suggests that,
viewed as a function of $Z_i$, the object dual to the $n-$particle ratio
function should
reduce to the $(n-1)-$particle one when three neighboring points lie on the
same line
in dual superspace, $Z_{i+1} = z Z_{i+2}+ (1-z) Z_i$.
It is natural to expect that such an object would be defined on an $n-$gon in the dual
superspace with vertices
located at the points $Z_i$ (with $i=1,\ldots,n$). We believe that solving this problem may provide the key to the complete understanding of the superamplitudes, with their anomalous symmetries.

\section*{Acknowledgments}

We would like to thank Nima Arkani-Hamed, Niklas Beisert, Zvi Bern, Lance Dixon, James Drummond, David Gross, David Kosower, Juan Maldacena, {Edward Witten} for interesting discussions. ES would like to acknowledge the hospitality of the Galileo Galilei Institute (Florence) and of the Kavli Institute for Theoretical Physics (Santa Barbara), where part of this work was done. This work was supported
in part by the French Agence Nationale de la Recherche under grant
ANR-06-BLAN-0142, by the CNRS/RFFI grant 09-02-00308 and by the National Science Foundation under Grant No. PHY05-51164.

\appendix


\section{Appendix: $\cN=4$ superconformal symmetry }\label{B}

Throughout the paper, we denote the generators of conventional $(q,\bar q, s, \bar s, \ldots)$ and dual superconformal symmetry $(Q,\bar Q, S, \bar S, \ldots)$ by lower-case and upper-case letters, respectively. Their explicit expressions can be found in \cite{Drummond:2008vq}. The nontrivial commutation relations for both symmetries are
\begin{align} \notag
& \{{q}_{\a }^A,\overline{{q}}_{\da B}\}  =  \delta_B^A {p}_{\a \da},
   &&  \{{s}_{\a A},\overline{{s}}_{\da}^B \} = \delta_A^B {k}_{\a \da}, &&
    [{k}_{\a \da},{p}^{\b \db}] = \delta_\a^\b \delta_\da^\db {d} +
{m}_{\a}{}^{\b}
 \delta_\da^\db + \overline{{m}}_{\da}{}^{\db} \delta_\a^\b,
\\[3mm] \notag
& {}[{p}_{\a \da},{s}^{\b}_A] = \delta_{\a}^{\b} \overline{{q}}_{\da A},
 &&  [{k}_{\a \da},{q}^{\b A}] = \delta_{\a}^{\b} \overline{{s}}_{\da A}, &&
  \{{q}^{\a A},{s}_{\b B}\} =  {m}^{\a}{}_{\b}
\delta_B^A + \delta^{\a}_{\b} {r}^{A}{}_{B} + \tfrac{1}{2}\delta^{\a}_{\b} \delta_B^A ({d}+{c}),
\\[3mm]  \notag
& {}[{p}_{\a \da},\overline{{s}}^{\db A}]  =  \delta^{\db}_{\da} {q}_{\a}^A,
&& [{k}_{\a \da}, \overline{{q}}^{\db}_A]  =  \delta_{\da}^{\db} {s}_{\a
A}, &&
\{\overline{{q}}^{\da}_{ A},\overline{{s}}_{\db}^{ B}\} = \overline{{m}}^{\da}{}_{\db} \delta_A^B  - \delta^{\da}_{\db} {r}^{B}{}_{A} + \tfrac{1}{2} \delta^{\da}_{\db}\delta_A^B
({d}-{c})\ .
\\
 \label{comm-rel}
 \end{align}
Here we find the generators of translations $p$, conformal boosts $k$, Lorentz $m, \bar m$, SU(4) rotations $r$, dilatation $d$ and central charge $c$.\footnote{The central charge of the superamplitudes vanishes, so we might as well consider the superalgebra $psu(2,2|4)$.}

As explained in \cite{Drummond:2008vq}, the conformal part of this algebra ($s$, $\bar s$, $k$) can be obtained form the Poincar\'e part ($q$, $\bar q$, $p$) by applying the discrete operation of conformal inversion $I$, namely, $k = IpI$, $s = I \bar q I$, $\bar s = I q I$. Thus, it is often more convenient to check dual conformal symmetry by just looking at the properties of the amplitude under inversion in dual superspace.
Here we list some of the inversion rules, needed in this paper (the details can be found in  \cite{Drummond:2008vq}):
\begin{align} \label{conin}
& I[x_{ij}] = x^{-1}_i x_{ij} x^{-1}_j  \,,\qquad
  I[\la_i] = \frac{x_i \ran{\la_i}}{\sqrt{x^2_i x^2_{i+1}}} = \frac{x_{i+1} \ran{\la_i}}{\sqrt{x^2_i x^2_{i+1}}} \,, \qquad I[\bl_i] = \frac{[\bl_i|x_{i}}{\sqrt{x^2_i x^2_{i+1}}} = \frac{[\bl_i|x_{i+1}}{\sqrt{x^2_i x^2_{i+1}}} \,,
\end{align}
%
where $(x^{-1})^{\a\da} \equiv x^{\a\da}/x^2$ and we have used the standard bra/ket notation for spinors. It is easy to see that the contraction of two adjacent spinors $\vev{i\, i+1}$ is covariant,
\begin{equation}
    I[\vev{i\, i+1}]  = \frac{\vev{i\, i+1}}{x^2_{i+1}\sqrt{x^2_i x^2_{i+2}}} \,,
\end{equation}
while any other contraction $\vev{ij}$ (with $j \neq i$) is not. This explains why the holomorphic helicity-free cross-ratio $\vev{ij}\vev{kl}/\vev{ik}\vev{jl}$ mentioned in Section \ref{cscts} cannot be dual conformal.
Further, a typical example of a dual conformally covariant expression is the following string:
\begin{equation}\label{inii}
    I[\bra{i} x_{ij} x_{jk} \ket{k}]  = \frac{\bra{i} x_{ij} x_{jk} \ket{k}}{x^2_{j}\sqrt{x^2_i x^2_{i+1} x^2_k x^2_{k+1}}} \,.
\end{equation}
Because of the two equivalent form of the transformations of $\la$ in \p{conin}, the above string remains covariant if we replace $\bra{i}\ \to \bra{i-1}$ or $\ket{k}\ \to \ket{k-1}$.

\section{Appendix: Solution of the collinearity conditions}\label{A1}

Consider the collinearity condition (a generalization of \p{coll})
\begin{equation}\label{coll'}
    F_{a,i,b} \varphi(\la, \bl) \equiv \left(\vev{ab} \frac{\pa}{\pa\bl_i} + \vev{bi} \frac{\pa}{\pa\bl_a} + \vev{ia} \frac{\pa}{\pa\bl_b} \right)\varphi(\la, \bl) = 0\,, \qquad (a+1\le i \le b-1)\,.
\end{equation}
Here $a<b$ are the end points of a line segment in twistor space, and $i=a+1, \ldots, b-1$ are the intermediate points. Without loss of generality, we can take $\varphi(\la_k, \bl_k) = \varphi(p_k)$, $k=1,\ldots,n$. Any additional  purely holomorphic dependence on $\la$ does not affect the collinearity condition (here we neglect the possible holomorphic anomalies).

For the sake of this argument, let us assume that we are in signature $(++--)$, where we can treat $\la$ and $\bl$ as independent variables. Then, the collinearity condition \p{coll'} can be interpreted as the invariance of $\varphi$ under simultaneous shifts of $\bl_i$, $\bl_a$ and $\bl_b$,
\begin{align}
\bl_i \to \bl_i + \tilde\epsilon_i  \vev{ab}\,,\qquad \bl_a \to \bl_a+\tilde\epsilon_i \vev{bi}\,,\qquad
\bl_b \to \bl_b +\tilde\epsilon_i \vev{ia}\, \qquad (a+1\le i \le b-1)
\end{align}
with arbitrary parameters $\tilde\epsilon_i$.
Let us choose $\tilde\epsilon_i = - \bl_i/\vev{ab}$, so that the shifted $\bl_i$ vanish. Then,
\begin{align}\label{sol-col}
\varphi(p_1,\ldots,p_a,\ldots,p_b,\ldots,p_n) = \varphi(p_1,\ldots,p_a',0,\ldots,0,p'_b,\ldots,p_n)\,,
\end{align}
where $p_i'=0$ for $ a+1\le i \le b-1$ and
\begin{align}\label{sol-col'}
p_a'= -\frac{\ket{a}\bra{b}x_{a\, b+1}}{\vev{ab}} \,,\qquad
p_b' = \frac{\ket{b}\bra{a}x_{a\, b+1}}{\vev{ab}} \,,\qquad x_{a\, b+1}=p_a+\ldots+p_b \,.
\end{align}
We verify that
$p_a'+p_b' =p_a+\ldots+p_b$, as it should be. In other words, the collinearity condition \p{coll'} means that the function $\varphi$ can depend on $\bl_i$ in the range $a \leq i \leq b$ only via the dual coordinates $ x_{a\, b+1}= x_a - x_{b+1}$.

In the special case of the collinearity condition \re{coll}, we set $a=1$, $b=n$ and we find $p_1+\ldots+p_n=0$, hence {$p_1'=p_n'=0$}. Thus, the only solution to  \p{coll} is $\varphi=\text{const}$.

Now, let us turn to the case considered in Section \ref{cslsts}, where we modified the known bosonic helicity structure $f_{1st}$ by a helicity-free and dual conformally invariant function $\varphi(\la,\bl)$, see \p{modfbyf}. As explained in Section \ref{cscts}, we can restrict it to be a function of the momenta, $\varphi(p_1, \ldots, p_n)$. This function, like $f_{1st}$ itself, should simultaneously satisfy the three collinearity conditions \p{3colli}. The particularity here is that two of the lines intersect at point 1. Repeating the steps above and taking care to shift $\bl_1$ along the lines $(1, s-1)$ and $(t,1)$ at the same time, we obtain the following general solution:
\begin{equation}\label{gensophinmhv}
    \varphi(p_1', 0, \ldots, 0, p_{s-1}', p_s', 0, \ldots, 0, p_{t-1}', p_t', 0, \ldots,0)\,.
\end{equation}
Here the five light-like non-vanishing momenta are defined by
\begin{align}\notag
&p_1'  = \frac{\ket{1}\bra{s-1}x_{s1}}{\vev{1\, s-1}} + \frac{\ket{1}\bra{t}x_{t1}}{\vev{t\, 1}}\\
&p_{s-1}' = \frac{\ket{s-1}\bra{1}x_{1s}}{\vev{1\, s-1}}\,, \qquad p_{s}' = \frac{\ket{s}\bra{t-1}x_{ts}}{\vev{s\, t-1}}\\
&p_{t-1}' = \frac{\ket{t-1}\bra{s}x_{st}}{\vev{s\, t-1}}\,, \qquad p_{t}' = \frac{\ket{t}\bra{1}x_{1t}}{\vev{t\, 1}}\ ,   \notag
\end{align}
and they satisfy the conservation condition
\begin{equation}\label{sumfive}
    p_1'+p_{s-1}'+p_{s}'+p_{t-1}'+p_{t}' = 0\,.
\end{equation}
From them we can form five independent Lorentz invariant dot products.%
\footnote{The six dot products of the four independent momenta satisfy a linear relation following from the light-likeness of the fifth vector in \p{sumfive}.} By inspecting the various combinations, we find five such dot products, which are dual conformal (up to irrelevant holomorphic factors depending only on $\la$):
\begin{align}\notag
 & p_1'\cdot p_{s-1}' \sim \bra{1}x_{1s}x_{st}\ket{t}\,, \qquad p_s'\cdot p_{s-1}' \sim \bra{1}x_{1s}x_{st}\ket{t-1}\\
 \notag &  p_t'\cdot p_{t-1}' \sim \bra{1}x_{1t}x_{ts}\ket{s} \,, \qquad  p_1'\cdot p_{t}' \sim \bra{1}x_{1t}x_{ts}\ket{s-1}\\
 &  p_s'\cdot p_{t-1}' \sim  x^2_{st} \,. \label{5var}
\end{align}
All of them appear in the bosonic function $f_{1st}$ \p{somebosfa} and give it the necessary dual conformal weights.

The additional factor $\varphi(p_k)$ in \p{modfbyf} must then be a dual conformally  invariant function of the five variables \p{5var}. However, the latter involve only five points in dual space, $1,s-1,s,t-1,t$, from which it is impossible to build a conformal cross-ratio. Thus, we conclude that $\varphi={\rm const}$.

\section{Appendix: Conformal $s-$supersymmetry of the NMHV tree superamplitude}\label{A2}

Let us show that the NMHV tree superamplitude satisfies the relation
\begin{equation}\label{app:48}
      \left[\Delta_\a -(s-3) \lambda_{\check{2}\a} - (t-s-2) \lambda_{\hat{2} \a} \right]\tilde{\cal A}_{1st} =0 \,,
\end{equation}
where $\tilde{\cal A}_{1st}$ is given by \re{46} and the operator $\Delta_\a$ has the form
\be
\Delta_\a = \sum_2^{s-1} \vev{\check{2}i}\frac{\pa}{\pa \lambda_{i}^\a} + \sum_s^{t-1} \vev{\hat{2}i}\frac{\pa}{\pa \lambda_{i}^\a}
\ee
with spinors ${\check{2}_\a}$ and ${\hat{2}_\a}$ defined in \re{47}.
The differential operator $\Delta_\a$ has the following properties:
\begin{eqnarray}
   && \Delta_\a \vev{\check{2}i} = 0\,, \qquad (1 \leq i \leq s-1) \nn\\[3mm]
   && \Delta_\a \vev{\hat{2}i} = 0\,, \qquad (s \leq i \leq t-1) \nn\\
  && \Delta_\a \sum_1^n \lambda_i^\b \bl_i^{\dot\b} = 0 \,,   \label{50}
\end{eqnarray}
which allow us to show that all the delta functions in \p{46} are annihilated by $\Delta_\a$ leading to
\begin{align}\label{cal-Delta}
 \Delta_\a \tilde{\cal A}_{1st} = \tilde{\cal A}_{1st} \Delta_\a \ln  \lr{\vev{n1}^4\vev{12}^4 f_{1st}}\,,
\end{align}
with $f_{1st}$ given by \re{somebosfa}.
Further properties of $\Delta_\a$ include
\begin{align}
\Delta_\a \ln \vev{k\, k+1} =  \left\{ \begin{array}{ccl}
                                          - \check{2}_\a  & \mbox{for} & 1 \leq k \leq s-2 \\
                                          - \hat{2}_\a  & \mbox{for} & s \leq k \leq t-2  \\
                                          0 & \mbox{for} & t \leq k \leq n
                                        \end{array}
     \right.
\end{align}
together with
\begin{align}\nn
& \Delta_\a \ln x^2_{st}=-\Delta_\a \ln \vev{\hat{2}\ t} =\Delta_\a  \ln \frac{\vev{1|x_{1t}x^{-1}_{ts}|s-1}}{\vev{12}} = - \hat{2}_\a
\\ \nn
&   \Delta_\a  \ln \frac{\vev{1|x_{1t}x^{-1}_{ts}|s}}{\vev{12}} = \check{2}_\a
\\[2mm]
&  \Delta_\a \ln \vev{\hat{2}\ t-1} =  0\,.
\end{align}
Applying these relations, we obtain from  \re{cal-Delta}
\begin{align}
\Delta_\a \ln  \lr{\vev{n1}^4\vev{12}^4 f_{1st}} =  (s-3) {\check{2}_\a} + (t-s-2) {\hat{2}_\a}\,.
\end{align}
Substituting this result into \re{cal-Delta}  yields  \re{app:48}.

We conclude by remarking that although the $q$-supersymmetry gauge $\bar\eta_n=\bar\eta_1=0$ simplifies the computation, one can verify \p{52} without fixing the gauge. In this case the operator $s_{\a A}$ has an additional term compared to \p{48},
\begin{align}
 s_{\a A} & = \left(\bet_{2 A} + \bet_{1 A} \frac{\vev{2n}}{\vev{n1}}+ \bet_{n A} \frac{\vev{12}}{\vev{n1}} \right)[\Delta_\a -(s-3) \check{2}_\a - (t-s-2) \hat{2}_\a  ]\nn
 \\
 &+  \frac{\bet_{n A}\lambda_{1\a} - \bet_{1 A}\lambda_{n\a}}{2\vev{n1}} \left( \sum_{i=1}^n \lambda^\a_i \frac{\pa}{\pa \lambda^\a_i} +2(n-3)  \right)\ .
\end{align}
The counting operator in the second line gives $4$ when applied to $\vev{n1}^4\vev{12}^4 f_{1st}$ and $(-4)$ when applied to the momentum conservation delta function, so \p{52} still holds.

\section{Appendix: Collinear limit of the $n=6$ NMHV superamplitude}\label{apD}

In this Appendix we show that the collinear limit of the six-gluon NMHV amplitude \re{6-col} follows from the analogous relation for the $n=6$ NMHV superamplitude \re{col-limit}.

The tree-level $n=6$ NMHV superamplitude has the form~\cite{Drummond:2008vq,Drummond:2008bq}
\begin{align}\label{app:6}
\mathcal{A}_6^{\rm NMHV;0} =\mathcal{A}_6^{\rm MHV;0}\lr{ R_{146}+R_{362}+R_{524} }\,.
\end{align}
Replacing the superinvariants by their explicit expressions \re{R}, we find after some
algebra
\begin{align} \nn
\mathcal{A}_6^{\rm NMHV;0} & = \delta^{(4)}({\sum_1^6 \lambda_i \tilde\lambda_i})
\delta^{(8)}({\sum_1^6 \eta_i  \lambda_i })\bigg[ \frac{\delta^{(4)}(\eta_4 [56]+\eta_5[64]+\eta_6[45])}{x_{14}^2 \vev{12}\vev{23}[45][56]\bra{1} x_{14}|4]\bra{3}x_{36}|6]}
\\ \label{3-terms}
& + \frac{\delta^{(4)}(\eta_6 [12]+\eta_1[26]+\eta_2[61])}{x_{36}^2 \vev{34}\vev{45}[61][12]\bra{3} x_{36}|6]\bra{5}x_{52}|2]}
+ \frac{\delta^{(4)}(\eta_2 [34]+\eta_3[42]+\eta_4[23])}{x_{52}^2 \vev{56}\vev{61}[23][34]\bra{5} x_{52}|2]\bra{1}x_{14}|4]} \bigg],
\end{align}
where the three terms inside the square brackets correspond to the three terms in \re{app:6}.
We expect that in the collinear limit this expression should reduce to the $n=5$ NMHV superamplitude~\cite{Drummond:2008vq,Drummond:2008bq}
\begin{align}\label{app:5}
\mathcal{A}_5^{\rm NMHV;0} & =  \delta^{(4)}({\sum_1^5 \lambda_i \tilde\lambda_i} )
\delta^{(8)}({\sum_1^5 \eta_i  \lambda_i }
)
 \frac{\delta^{(4)}(\eta_2[34]+\eta_3[42]+\eta_4[23])}{\vev{51}^4 [1 2][23][34][45][51]}\ .
\end{align}

Let us examine \re{3-terms} in the collinear limit $6\| 1$ defined in \re{col-def} for $i=6$. We observe that the first term in \re{3-terms} remains regular in this limit, while the two
other terms contain the vanishing factors $[61]$ and $\vev{61}$ in the denominator.
However, examining the argument of the Grassmann delta-function in the second term
in \re{3-terms} we find that it vanishes in the collinear limit
\begin{align}\label{app:zero}
\eta_6 [12]+\eta_1[26]+\eta_2[61] \stackrel{6\| 1}{\to} \sqrt{z\bar z} \,\eta_\ell [\ell 2] +
 \sqrt{z\bar z}\,\eta_\ell [2\ell] + \sqrt{z\bar z}\, \eta_2 [ \ell  \ell ] = 0,
\end{align}
with $\bar z= 1-z$. Therefore, the dominant contribution to $\mathcal{A}_6^{\rm NMHV;0}$ only comes from the last term in \re{3-terms}, leading to
\begin{align}\label{app:col}
\mathcal{A}_6^{\rm NMHV;0} & \stackrel{6\| 1}{\to}
 \frac1{\vev{61}\sqrt{z\bar z}} \delta^{(4)}({\sum \lambda_i \tilde\lambda_i}
 )
\delta^{(8)}({\sum \eta_i  \lambda_i }
)
 \frac{\delta^{(4)}(\eta_2[34]+\eta_3[42]+\eta_4[23])}{\vev{5\ell}^4 [\ell 2][23][34][45][5\ell]}\ ,
\end{align}
where the sums inside the delta functions run over $i=\ell,2,...,5$. Comparing this
relation with \re{app:5}, we conclude that
\begin{align}
\mathcal{A}^{\rm tree}_6(1,2,3,4,5,6) & \stackrel{6\| 1}{\to}  \frac1{\vev{61}\sqrt{z\bar z}}
\mathcal{A}^{\rm tree}_5(\ell,2,3,4,5)\,,
\end{align}
in agreement with \re{col-limit}.

Let us now apply \re{3-terms} to reproduce the collinear limit of the six-gluon helicity-split
amplitude \re{6-col}. Using \re{superstate}, we find that this
amplitude appears as a particular term in the expansion of the $n=6$ NMHV superamplitude,
\begin{align}\label{6-gluon-term}
\mathcal{A}_6^{\rm NMHV} = A_6(1^+2^+3^+4^-5^-6^-) (\eta_4)^4 (\eta_5)^4 (\eta_6)^4 + \ldots\,,
\end{align}
where $(\eta)^4\equiv \epsilon_{ABCD} \eta^A\eta^B\eta^C\eta^D/4!$ and the ellipsis denote
terms describing the remaining six-point NMHV amplitudes. The latter include six-gluon NMHV
color-ordered amplitudes with different ordering of negative and positive helicities as well as
six-particle NMHV amplitudes involving gluinos and scalars.

Let us now expand both sides of \re{3-terms} in powers of $\eta$'s and identify one particular
term displayed in the right-hand side of  \re{6-gluon-term}. The first term in the right-hand side
of \re{3-terms} is regular in the collinear limit and, therefore, can be discarded. Then, the term $\sim(\eta_4)^4 (\eta_5)^4 (\eta_6)^4$ can be easily extracted
from  \re{3-terms} by making use of the identities
\begin{align}\nn
\delta^{(8)}({\sum_1^6 \eta_i  \lambda_i }) \delta^{(4)}(\eta_6 [12]+\eta_1[26]+\eta_2[61])
& =\lr{[12]\vev{45}}^4 (\eta_4)^4 (\eta_5)^4 (\eta_6)^4 + \ldots\,,
\\ \label{app:rel}
\delta^{(8)}({\sum_1^6 \eta_i  \lambda_i }) \delta^{(4)}(\eta_2 [34]+\eta_3[42]+\eta_4[23])
& =\lr{[23]\vev{56}}^4 (\eta_4)^4 (\eta_5)^4 (\eta_6)^4 + \ldots\,,
\end{align}
where expressions in the right-hand side are homogenous polynomials in $\eta$'s of degree 12 and the ellipses denote remaining terms. Substituting \re{app:rel} into \re{3-terms} and making use of \re{6-gluon-term}, we find that the second and third term in the right-hand side of \re{3-terms}
provide the contribution to $A^{\rm NMHV;0}_6(1^+,2^+,3^+,4^-,5^-,6^-)$ which
coincides, respectively, with the first and second term in the right-hand side of \re{6-col}.

Let us now examine the first relation in \re{app:rel}  in the collinear limit $6\| 1$.
We recall  that, in virtue of \re{app:zero}, the Grassmann delta function in the left-hand side
vanishes in this limit and, as a consequence,
the second term in \re{3-terms} does not contribute to the singular behavior of the superamplitude \re{app:col}. It is easy to see however that the first term in the right-hand side of this relation is different from zero in the limit  $6\| 1$ and, therefore, it is compensated in the collinear limit
by the remaining terms shown by ellipses. More precisely,
\begin{align}
\delta^{(4)}(\eta_6 [12]+\eta_1[26]+\eta_2[61]) \stackrel{6\| 1}{\to}  [12]^4 (\eta_6)^4 + [12]^3[26] (\eta_6)^3 (\eta_1) + \ldots+ [26]^4 (\eta_1)^4 \to 0 \,,
\end{align}
where we applied \re{col-def} for $i=6$.  Here, in the second relation each term represents a
particular three-particle on-shell state which can be identified using \re{superstate}. For instance, the term $(\eta_6)^4$  describes a three-gluon state $G^-_6 G^+_1 G^+_2$,
the term $(\eta_6)^3\eta_1$  describes an antigluino-gluino-gluon state
${\bar\Gamma_6}{\Gamma_1} {G^+_2}$, and so on. Each of these states contributes to the
right-hand side of the first relation in \re{app:rel} but their sum vanishes.
Thus, the second term in \re{3-terms}
produces an MHV-type contribution to the collinear limit of the six-gluon NMHV
amplitude, but it cancels inside the  NMHV {\it super}amplitude after one takes into account
the contribution of the six-point NMHV amplitudes, in which the two collinear gluons $G^-_6G^+_1$
are replaced by a pair of two particles (gluons, gluinos, scalars) with the total helicity and
the SU(4) charge equal to zero.

\section{Appendix: $\bar Q-$anomaly of the $n=6$ NMHV superamplitude}\label{C}

The $n=6$ NMHV superamplitude has three-particle cuts and its discontinuity in, say, $s_{123}=(p_1+p_2+p_3)^2$ has the following form at one loop
\begin{align} \notag
{\rm Disc}_{s_{123}} \mathcal{A}_6^{\rm NMHV;1} & =  \mathcal{A}_5^{\rm NMHV;0}( -\ell_1 ,1,2,3, -\ell_2 )\star
\mathcal{A}_5^{\rm MHV;0}( \ell_2,4,5,6, \ell_1)
\\[2mm]
& \label{disc-NMHV0}
+ \mathcal{A}_5^{\rm MHV;0}(-\ell_1,1,2,3,-\ell_2)\star
\mathcal{A}_5^{\rm NMHV;0}(\ell_2,4,5,6,\ell_1)\,,
\end{align}
where we used the same notations as in \re{disc-1-loop}.
Here the second term can be obtained from the first one through shift of indices $i\mapsto i+3$.
Explicit expressions for 5-particle superamplitudes entering \re{disc-NMHV0} are
\begin{align} \notag
 & \mathcal{A}_5^{\rm NMHV;0}(-\ell_1,1,2,3,-\ell_2) = \delta^{(4)}(\sum \la_i \bl_i)\delta^{(8)} (\sum \la_i \eta_i)\frac{\delta^{(4)}(\eta_1[23]+\eta_2[31]+\eta_3[12])}{\vev{\ell_1\ell_2}^4 [\ell_1 \ell_2] [\ell_2 3][32][21][1\ell_1]}\,,
 \\ \label{A5}
 & \mathcal{A}_5^{\rm MHV;0}(\ell_2,4,5,6,\ell_1) =\delta^{(4)}(\sum \la_i \bl_i) \delta^{(8)} (\sum \la_i \eta_i)\frac1{\vev{\ell_24}\vev{45}\vev{56}\vev{6\ell_1}\vev{\ell_1\ell_2}}\,,
\end{align}
where the sum in the argument of delta functions runs over all external particles and
the conventions are used
$\lambda_{-\ell_i} = -\lambda_{\ell_i}$ and $\tilde\lambda_{-\ell_i} = \tilde\lambda_{\ell_i}$.

The tree-level superamplitudes respect dual superconformal symmetry but for $\mathcal{A}_6^{\rm NMHV} $ this symmetry is broken at loop level by infrared divergences.
Invoking the same arguments as for one-loop MHV superamplitude, we can use \re{disc-NMHV0} to argue that ${\rm Disc}_{s_{123}} \mathcal{A}_6^{\rm NMHV;1}$ is free from dual conformal $K-$anomaly. Going to all loops and to an arbitrary number of external particles, we write $\mathcal{A}_n^{\rm NMHV}$ in a close analogy with \re{W-dual} as
\be
\mathcal{A}_n^{\rm NMHV} =\mathcal{A}_n^{\rm MHV;0}\, W_n^{\rm NMHV}\,.
\ee
In distinction with \re{W-dual} the function $W_n^{\rm NMHV}$ is not related to
light-like Wilson loop $W_n$. Nevertheless,  $W_n^{\rm NMHV}$ satisfies the same Ward identities as $W_n$, Eq.~\re{anom} and \re{K-norm}.  As a consequence, ${\rm Disc}_{x_{1,j+1}^2}  \ln W^{\rm NMHV}_n$ is invariant under dual conformal transformations.

Let us now turn to dual supersymmetry and
apply the operator $\bar Q^A_\da=\sum_1^6 \eta_i^A \partial_{i,\da}$ to both sides of \re{disc-NMHV0}
\begin{align}\label{Qbar-A6}
\bar Q^A_\da ({\rm Disc}_{s_{123}} \mathcal{A}_6^{\rm NMHV;1}) & =  \bar Q^A_\da \mathcal{A}_5^{\rm NMHV;0}(-\ell_1,1,2,3,-\ell_2)\star
\mathcal{A}_5^{\rm MHV;0}(\ell_2,4,5,6,\ell_1)
\\[2mm] \notag
&
+ \mathcal{A}_5^{\rm NMHV;0}(-\ell_1,1,2,3,-\ell_2)\star
\bar Q^A_\da \mathcal{A}_5^{\rm MHV;0}(\ell_2,4,5,6,\ell_1) + (i\to i+3)\,.
\end{align}
As before, the right-hand side of this relation is different from zero only due to holomorphic
anomaly. It is generated by  angular brackets $\vev{\ldots}$ in the denominator
of tree-level superamplitudes. The superamplitude $\mathcal{A}_5^{\rm NMHV;0}(-\ell_1,1,2,3,-\ell_2)$, Eq.~\re{A5},  has only one such factor
$\vev{\ell_1 \ell_2}$   and it provides the contribution to $\bar Q^A_\da \mathcal{A}_5^{\rm NMHV;0}$ localized at $\ell_1 \sim \ell_2$
or equivalenltly $(\ell_1+\ell_2)^2=s_{123}=0$. Similar to MHV case, this contribution is
suppressed by the additional factor $\vev{\ell_1\ell_2}^4$ coming from integration over
$\eta_{\ell_i}$, Eq.~\re{gr}. As a result, the first term in the right-hand side of \re{Qbar-A6} does not contribute to the anomaly. The MHV superamplitude $A_5^{\rm MHV;0}(-\ell_2,4,5,6,-\ell_1)$, Eq.~\re{A5}, has two factors in the denominator,
$\vev{\ell_2 4}$ and $\vev{6\ell_1}$, whose contribution is localized at the kinematical
configurations with $\ell_2\sim p_4$ and $\ell_1\sim p_6$, respectively.

We can further simplify the calculation by writing down $n=5$ NMHV superamplitude entering the second term in \re{Qbar-A6}
in the factorized form
\begin{align}\notag
\mathcal{A}_5^{\rm NMHV;0}(-\ell_1,1,2,3,-\ell_2)& =\mathcal{A}_5^{\rm MHV;0}(-\ell_1,1,2,3,-\ell_2)
\\[2mm]
&\times
\mathcal{M}(\ell_1,\ell_2)\delta^{(4)}(\eta_1[23]+\eta_2[31]+\eta_3[12])
\end{align}
with
\begin{align}
\mathcal{M}(\ell_1,\ell_2) &= \frac{ \vev{\ell_11}\vev{12}\vev{23}\vev{3\ell_2}}{\vev{\ell_1\ell_2}^3[\ell_1 \ell_2] [\ell_2 3][32][21][1\ell_1]}
=\frac1{P^4}\frac{\vev{12}\vev{23}}{[12][23]}\frac{\vev{1|\ell_1\ell_2 |3}}{[1|\ell_1\ell_2 |3]}
\end{align}
and $P=\ell_1+\ell_2=p_1+p_2+p_3$.
Substituting this relation into \re{Qbar-A6} and taking into account that  $\mathcal{M}(\ell_1,\ell_2)$ is free from the holomorphic anomaly, we find that evaluation of \re{Qbar-A6} is analogous to that for MHV superamplitude \re{delta-A} with the only difference that
contribution of each kinematical configuration like \re{rel-Q1} is multiplied by the
additional factor $\mathcal{M}(\ell_1,\ell_2)$. For $\ell_2\sim p_4$ and $\ell_1\sim p_6$ this factor can be further simplified leading, respectively, to
\begin{align}  \label{M} 
\mathcal{M}_4 &=  \frac1{x_{14}^4}\frac{\vev{12}\vev{23}\vev{34}}{[12][23][34]}\frac{\bra{1}x_{14}|4]}{[1|x_{14}\ket{4}}\,,
\qqqquad \mathcal{M}_6 =  \frac1{x_{14}^4}\frac{\vev{61}\vev{12}\vev{23}}{[61][12][23]}\frac{\bra{3}x_{36}|6]}{[3|x_{36}\ket{6}}\,.
\end{align}
Combining these relations with \re{Qbar-MHV} (evaluated for $j=3$ and $n=6$) we obtain
\begin{align}
\bar Q^A_{\dot\alpha} \bigg( {\rm Disc}_{x_{14}^2}   \mathcal{A}_6^{\rm NMHV;1}\bigg) &= - 2\pi i c_\Gamma \mathcal{A}_6^{\rm MHV;0} \delta^{(4)}(\eta_1[23]+\eta_2[31]+\eta_3[12])
\\  \notag
&\times
\Bigg[\frac{\tilde\lambda_{4\dot\alpha}\,\Xi_{514}^A \vev{61} \mathcal{M}_4}{\vev{45}\bra{6}x_{64}|4][4|x_{41}\ket{1}}
-  \frac{\tilde\lambda_{6\dot\alpha}\,\Xi_{146}^A \vev{34} \mathcal{M}_6}{\vev{61}\bra{3}x_{36}|6][6|x_{64}\ket{4}}
  \Bigg]+ (i\to i+3)\,.
\end{align}
Using the definition \re{Xi} we replace $\Xi-$functions by their explicit expressions
\begin{align}\label{combi}
\frac{\Xi_{514}^A}{\vev{45}\vev{56}} = \frac{\Xi_{146}^A}{\vev{45}\vev{61}} = -\lr{\eta_4[56]+\eta_5[46]+\eta_6[45]}^A
\end{align}
and, finally, obtain the expression \p{Qb-NMHV} for the $\bar Q-$anomaly of one-loop $n=6$ NMHV superamplitude.

Next, we turn to the anomaly of the ratio function.
It follows from the definition \p{mhvfac} that
\begin{align}
\mathcal{A}_6^{\rm NMHV;1}=R_6^{\rm NMHV;1} \mathcal{A}_6^{\rm MHV;0}
+  R_6^{\rm NMHV;0} \mathcal{A}_6^{\rm MHV;1}\,.
\end{align}
We apply $\bar Q^A_{\dot\alpha}$ to both sides of this relation and take into account
$\bar Q-$invariance of tree amplitudes to get
\begin{align}
\lr{\bar Q^A_{\dot\alpha} R_6^{\rm NMHV;1}} \mathcal{A}_6^{\rm MHV;0}
=
\bar Q^A_{\dot\alpha} \mathcal{A}_6^{\rm NMHV;1} -R_6^{\rm NMHV;0} \lr{\bar Q^A_{\dot\alpha} \mathcal{A}_6^{\rm MHV;1}} \,.
\end{align}
Finally, we take discontinuity with respect to $x_{14}^2$ in the both sides of this relation, substitute \re{Qb-NMHV} and \re{Qb-MHV6} and replace the tree-level ratio function $R_6^{\rm NMHV;0}$  by its explicit
expression, Eqs.~\re{NMHV-67} and \re{R6-ini}, to get
\begin{align}  \notag
\bar Q^A_{\dot\alpha} \bigg( {\rm Disc}_{x_{14}^2} & {R}_6^{\rm NMHV;1} \bigg) = 2\pi i c_{\Gamma}
\lr{\eta_4 [56] + \eta_5[64]+\eta_6[45]}^A\delta^{(4)}(\eta_1[23]+\eta_2[31]+\eta_3[12])
\\
&\times \label{Qbar-anom}
\frac{\vev{61}\vev{12}\vev{23}\vev{34}}{x_{14}^4 [12][23]}
\Bigg(
   \frac{\tilde\lambda_{4\dot\alpha}(1+\xi_4)}{[1|x_{14}\ket{4} [34][45]}
+
 \frac{\tilde\lambda_{6\dot\alpha}(1+\xi_6)}{[3|x_{36}\ket{6}[56][61]}
\Bigg)
+ (i\to i+3)\,.
\end{align}
Here the notation was introduced for
\begin{align}
\xi_4 &= \frac{x_{14}^2 [34]\vev{61}}{\bra{1} x_{14} |4] \bra{6} x_{63} |3]}\,,\qqqquad \xi_6 = \frac{x_{14}^2 \vev{34} [61]}{\bra{4} x_{41} |1] \bra{3} x_{36} |6]} \,,
\end{align}
such that  $\xi_6/\xi_4 = \mathcal{M}_4/\mathcal{M}_6$ (see Eq.~\re{M}).
Notice that $\xi_4$ and $\xi_6$ are invariant under dual conformal transformations
and going through some algebra we find
\begin{align}
\xi=\xi_4 =\xi_6 = \frac{x_{14}^2 x_{26}^2 x_{35}^2}{x_{13}^2 x_{25}^2x_{46}^2 - x_{14}^2 x_{25}^2 x_{36}^2 + x_{15}^2 x_{24}^2 x_{36}^2-x_{14}^2 x_{26}^2 x_{35}^2}\,.
\end{align}
In addition, $\xi$ is invariant under shift of indices $i\mapsto i+3$ and, therefore,
$(1+\xi)$ can be factored out in the right-hand side of \re{Qbar-anom}. Then, the
remaining expression is proportional to the coefficient in front of $\mathcal{A}_6^{\rm MHV;0}$ in \re{Qb-NMHV}. In this way,
we arrive at the relation \p{power}.

\end{document}